\documentclass[aps,prb,superscriptaddress,twocolumn,showpacs]{revtex4} 
\usepackage{amsmath}
\usepackage{amssymb}
\usepackage{dsfont}
\usepackage[all,arc,knot,web]{xy}
\SelectTips{lu}{12}
\usepackage{graphicx}
\usepackage{pstricks}
\usepackage{scalefnt}
\usepackage{color}
\usepackage{subfigure}
\usepackage{ifpdf}
\definecolor{Blue}{rgb}{0.3,0.3,1}
\usepackage{pgffor}
\usepackage{verbatim}
\usepackage{calc} 
\newcommand{\arl}{\ar@{-}|@{>}}
\newcommand{\arr}{\ar@{-}|@{<}}
\newcommand{\aline}{\ar@{-}}
\newcommand{\rmd}{\mathrm{d}}

\newcommand{\rmv}{\mathrm{v}}
\newcommand{\unit}{\mathbf{1}}
\newcommand{\Z}{\mathbb{Z}}
\newcommand{\scC}{\mathcal{C}}
\newcommand{\scA}{\mathcal{A}}
\newcommand{\scH}{\mathcal{H}}

\newcommand{\ds}[1]{\mathbb{#1}}
\newcommand{\Q}{\mathcal{Q}}
\newcommand{\B}{\mathcal{B}}
\newcommand{\Bt}{\tilde{\mathcal{B}}}

\newcommand{\e}{\mathrm{e}}
\newcommand{\ii}{\mathrm{i}}
\newcommand{\dual}[1]{{#1}^*}

\newcommand{\ket}[1]{\left|{#1}\right\rangle}
\newcommand{\bra}[1]{\left\langle{#1}\right|}
\newcommand{\bket}[1]{\Biggl|\;{\bmm #1 \emm}\;\Biggr\rangle}

\newcommand{\bpm}{\begin{pmatrix}}
\newcommand{\epm}{\end{pmatrix}}
\newcommand{\bmm}{\begin{matrix}}
\newcommand{\emm}{\end{matrix}}

\def\objectstyle{\scriptstyle}
\def\labelstyle{\scriptstyle}

\newcommand{\dyonpair}[3]{
\bmm\xy
0;/r0.2pc/:;
{\ar@{=}@`{(4,2), (8,-1)} (0,0)*@{*};(12,0)*@{*}};
(-2,0)*{\scriptstyle #2};
(14,0)*{\scriptstyle #3};
(7,2.2)*{\scriptstyle #1};
\endxy\emm
}

\newcommand{\BareHpart}{
	\bmm\xy
	0;/r0.1pc/:;
	(-10,-10)*{}="p1";
	(-5,0)*{}="p2";
	(-10,10)*{}="p3";
	(10,-10)*{}="p4";
	(5,0)*{}="p5";
	(10,10)*{}="p6";
	"p1";"p2" **\dir{-} ;
	"p2";"p3" **\dir{-} ;
	"p4";"p5" **\dir{-} ;
	"p5";"p6" **\dir{-} ;
	"p5";"p2" **\dir{-} ;
	\endxy\emm
}
\newcommand{\Hpart}[5]{
	\def\jjA{#1}
	\def\jjB{#2}
	\def\jjC{#3}
	\def\jjD{#4}
	\def\jjE{#5}
	\HpartExtended
}
\newcommand{\HpartExtended}[5]{
	\bmm\xy
	0;/r0.11pc/:;
	(-10,-10)*{}="p1";
	(-5,0)*{}="p2";
	(-10,10)*{}="p3";
	(10,-10)*{}="p4";
	(5,0)*{}="p5";
	(10,10)*{}="p6";
	"p1";"p2" **\dir{-} ?(0.6)*\dir{#1}+(3,-2) *{\scriptstyle \jjA};
	"p4";"p5" **\dir{-} ?(0.6)*\dir{#2}+(-3,-2) *{\scriptstyle \jjB};
	"p5";"p2" **\dir{-} ?(0.5)*\dir{#3}+(0,3.5) *{\scriptstyle \jjC};
	"p2";"p3" **\dir{-} ?(0.6)*\dir{#4}+(3,2) *{\scriptstyle \jjD};
	"p5";"p6" **\dir{-} ?(0.6)*\dir{#5}+(-3,2) *{\scriptstyle \jjE};
	\endxy\emm
}

\newcommand{\BareRotatedHpart}{
	\bmm\xy
	0;/r0.1pc/:;
	(-10,-10)*{}="p1";
	(10,-10)*{}="p2";
	(0,-5)*{}="p3";
	(0,5)*{}="p4";
	(-10,10)*{}="p5";
	(10,10)*{}="p6";
	"p1";"p3" **\dir{-} ;
	"p2";"p3" **\dir{-} ;
	"p3";"p4" **\dir{-} ;
	"p4";"p5" **\dir{-} ;
	"p4";"p6" **\dir{-} ;
	\endxy\emm
}
\newcommand{\RotatedHpart}[5]{
	\def\jjA{#1}
	\def\jjB{#2}
	\def\jjC{#3}
	\def\jjD{#4}
	\def\jjE{#5}
	\RotatedHpartExtended
}
\newcommand{\RotatedHpartExtended}[5]{
	\bmm\xy
	0;/r0.1pc/:;
	(-10,-10)*{}="p1";
	(10,-10)*{}="p2";
	(0,-5)*{}="p3";
	(0,5)*{}="p4";
	(-10,10)*{}="p5";
	(10,10)*{}="p6";
	"p1";"p3" **\dir{-} ?(0.55)*\dir{#1}+(-3,3) *{\scriptstyle \jjA};
	"p2";"p3" **\dir{-} ?(0.55)*\dir{#2}+(3,3) *{\scriptstyle \jjB};
	"p3";"p4" **\dir{-} ?(0.55)*\dir{#3}+(3,-1) *{\scriptstyle \jjC};
	"p4";"p5" **\dir{-} ?(0.55)*\dir{#4}+(-2,-3) *{\scriptstyle \jjD};
	"p4";"p6" **\dir{-} ?(0.55)*\dir{#5}+(2,-3) *{\scriptstyle \jjE};
	\endxy\emm
}

  \newcommand{\BareYpart}{
  	\bmm\xy
  	0;/r0.13pc/:;
  	(-10,10)*{}="p1";
  	(10,10)*{}="p2";
  	(0,2)*{}="p3";
  	(0,-10)*{}="p4";
  	"p1";"p3" **\dir{-} ;
  	"p2";"p3" **\dir{-} ;
  	"p3";"p4" **\dir{-} ;
  	\endxy\emm
  }
  
  \newcommand{\Ypart}[6]{
  	\bmm\xy
  	0;/r0.12pc/:;
  	(-10,10)*{}="p1";
  	(10,10)*{}="p2";
  	(0,2)*{}="p3";
  	(0,-10)*{}="p4";
  	"p4";"p3" **\dir{-} ?(0.5)*\dir{#4}+(4,0) *{\scriptstyle #1};
  	"p3";"p2" **\dir{-} ?(0.6)*\dir{#6}+(2,-3) *{\scriptstyle #3};
  	"p3";"p1" **\dir{-} ?(0.6)*\dir{#5}+(-1,-3) *{\scriptstyle #2};
  	\endxy\emm
  }
  
  \newcommand{\BareTriangleYpart}{
  	\bmm\xy
  	0;/r0.1pc/:;
  	(-10,10)*{}="p1";
  	(10,10)*{}="p2";
  	(0,-10)*{}="p3";
  	(-4,5)*{}="p4";
  	(4,5)*{}="p5";
  	(0,-3)*{}="p6";
  	"p1";"p4" **\dir{-} ;
  	"p2";"p5" **\dir{-} ;
  	"p3";"p6" **\dir{-} ;
  	"p4";"p5" **\dir{-} ;
  	"p5";"p6" **\dir{-} ;
  	"p6";"p4" **\dir{-} ;
  	\endxy\emm
  }
  \newcommand{\TriangleYpart}[6]{
  	\def\jjA{#1}
  	\def\jjB{#2}
  	\def\jjC{#3}
  	\def\jjD{#4}
  	\def\jjE{#5}
  	\def\jjF{#6}
  	\TriangleYpartExtended
  }
  \newcommand{\TriangleYpartExtended}[6]{
  	\bmm\xy
  	0;/r0.12pc/:;
  	(-10,10)*{}="p1";
  	(10,10)*{}="p2";
  	(0,-10)*{}="p3";
  	(-4,5)*{}="p4";
  	(4,5)*{}="p5";
  	(0,-3)*{}="p6";
  	"p4";"p1" **\dir{-} ?(0.6)*\dir{#5}+(-3,-1) *{\scriptstyle \jjE};
  	"p5";"p2" **\dir{-} ?(0.6)*\dir{#6}+(3.5,-1) *{\scriptstyle \jjF};
  	"p3";"p6" **\dir{-} ?(0.6)*\dir{#1}+(3,-1) *{\scriptstyle \jjA};
  	"p4";"p5" **\dir{-} ?(0.5)*\dir{#4}+(0,3.5) *{\scriptstyle \jjD};
  	"p6";"p5" **\dir{-} ?(0.5)*\dir{#3}+(3.5,0) *{\scriptstyle \jjC};
  	"p6";"p4" **\dir{-} ?(0.5)*\dir{#2}+(-3,0) *{\scriptstyle \jjB};
  	\endxy\emm
  }
   \newcommand{\TriangulationYpartLarge}[6]{
   	\def\jjA{#1}
   	\def\jjB{#2}
   	\def\jjC{#3}
   	\def\jjD{#4}
   	\def\jjE{#5}
   	\def\jjF{#6}
   	\TriangulationYpartExtendedLarge
   }
   \newcommand{\TriangulationYpartExtendedLarge}[6]{
   	\bmm\xy
   	0;/r0.2pc/:;
   	(-10,10)*{}="p1";
   	(10,10)*{}="p2";
   	(0,-10)*{}="p3";
   	(-4,5)*{}="p4";
   	(4,5)*{}="p5";
   	(0,-3)*{}="p6";
   	(13,-4)*{}="v1";
   	(-13,-4)*{}="v2";
   	(0,13)*{}="v3";
   	(0,1)*{}="v4";
   	"p4";"p1" **\dir{.} ;
   	"p5";"p2" **\dir{.} ;
   	"p3";"p6" **\dir{.} ;
   	"p4";"p5" **\dir{.} ;
   	"p6";"p5" **\dir{.} ;
   	"p6";"p4" **\dir{.} ;
   	"v2";"v3" **\dir{-} ?(0.5)*\dir{#5}+(-3.5,0)  *{\scriptstyle \jjE};
   	"v3";"v1" **\dir{-} ?(0.5)*\dir{#6}+(3.5,0)  *{\scriptstyle \jjF};
   	"v2";"v1" **\dir{-} ?(0.6)*\dir{#1}+(0,-3) *{\scriptstyle \jjA};
   	"v3";"v4" **\dir{-} ?(0.5)*\dir{#4}+(2,0)  *{\scriptstyle \jjD};
   	"v4";"v1" **\dir{-} ?(0.5)*\dir{#3}+(-1,3)  *{\scriptstyle \jjC};
   	"v2";"v4" **\dir{-} ?(0.5)*\dir{#2}+(1,3)  *{\scriptstyle \jjB};   
   	\endxy\emm
   }
   \newcommand{\TriangulationYpart}[6]{
   	\def\jjA{#1}
   	\def\jjB{#2}
   	\def\jjC{#3}
   	\def\jjD{#4}
   	\def\jjE{#5}
   	\def\jjF{#6}
   	\TriangulationYpartExtended
   }
   \newcommand{\TriangulationYpartExtended}[6]{
   	\bmm\xy
   	0;/r0.13pc/:;
   	(13,-4)*{}="v1";
   	(-13,-4)*{}="v2";
   	(0,13)*{}="v3";
   	(0,1)*{}="v4";
   	"v2";"v3" **\dir{-} ?(0.5)*\dir{#5}+(-3.5,0)  *{\scriptstyle \jjE};
   	"v3";"v1" **\dir{-} ?(0.5)*\dir{#6}+(3.5,0)  *{\scriptstyle \jjF};
   	"v2";"v1" **\dir{-} ?(0.6)*\dir{#1}+(0,-3) *{\scriptstyle \jjA};
   	"v3";"v4" **\dir{-} ?(0.5)*\dir{#4}+(2.5,-1)  *{\scriptstyle \jjD};
   	"v4";"v1" **\dir{-} ?(0.5)*\dir{#3}+(-1,3)  *{\scriptstyle \jjC};
   	"v2";"v4" **\dir{-} ?(0.5)*\dir{#2}+(1.5,3.5)  *{\scriptstyle \jjB};   
   	\endxy\emm
   }
   
   \newcommand{\CorbodismYpart}[9]{
   	\def\jjA{#1}
   	\def\jjB{#2}
   	\def\jjC{#3}
   	\def\jjD{#4}
   	\def\jjE{#5}
   	\def\jjF{#6}
   	\def\jjBB{#7}
   	\def\jjCC{#8}
   	\def\jjDD{#9}
   	\CorbodismYpartExtended
   }
   \newcommand{\CorbodismYpartExtended}[6]{
   	\bmm\xy
   	0;/r0.25pc/:;
   	(13,-4)*{}="v1";
   	(-13,-4)*{}="v2";
   	(-3,13)*{}="v3";
   	(0,1)*{}="v4";
   	(5,14)*{}="v5";
   	"v2";"v3" **\dir{-} ?(0.5)*\dir{#5}+(-2.5,0)  *{\scriptstyle \jjE};
   	{\aline@{-}|(0.21){\hole}|(0.36){\hole}|(0.5)@{#6}^{\jjF} "v3";"v1"};
   	"v2";"v1" **\dir{-} ?(0.6)*\dir{#1}+(-1,2) *{\scriptstyle \jjA};
   	{\aline@{-}|(0.43){\hole}|(0.63)@{#4}^>>>>>>{\jjD} "v3";"v4" };   	
   	"v4";"v1" **\dir{-} ?(0.5)*\dir{#3}+(-1,3)  *{\scriptstyle \jjC};
   	"v2";"v4" **\dir{-} ?(0.5)*\dir{#2}+(1.5,2.5)  *{\scriptstyle \jjB};
   	"v3";"v5" **\dir{-} ?(0.5)*\dir{#4}+(-1,2)  *{\scriptstyle \jjDD};
   	"v5";"v1" **\dir{-} ?(0.5)*\dir{#3}+(1,2)  *{\scriptstyle \jjCC};
   	"v2";"v5" **\dir{-} ?(0.5)*\dir{#2}+(0,2.5)  *{\scriptstyle \jjBB};
   	"v4";"v5" **\dir{-} ?(0.6)*\dir{>}+(1.5,0.5)  *{s};
   	\endxy\emm
   }

\newcommand{\NewVertex}
{
\xy
0;<6mm,0mm>:
<0mm,8mm>::
0="o";
a(150)="l";
a(30)="r";
a(270)="b";
(0,-0.25)="i";
(0.7,-0.25)*+{q_1}="f";
(0,-0.75)="i1";
(0.7,-0.75)*+{q_2}="f1";
\xygraph{
"l"-"o"|@{>}^{i}
"r"-"o"|@{>}_{j}
"b"-"o"|@{>}|(0.1)@{>}|(0.9)@{>}^(0.1){k_2}^{l}^(0.9){k_1}
"i":"f"
"i1":"f1"
}
\endxy}

\newcommand{\NewPlaquette}{
\xy
0;<7mm,0mm>:
<0mm,8mm>::
(0,0.4)="u";
(0,-0.4)="d";
{\arl^<<{l_1} a(150)+a(150)+"u";a(150)+"u"};
a(90)+a(90)+"u";a(90)+"u"**@{-}?(0.4)="pt"?(0.25)*@{>}?(0.2)+/_/*{l_6}?(0.7)*@{>}+/_/*{k_6};
{\ar "pt";"pt"+(0.7,0)*+{q_6}};
{\arl_<<{l_5} a(30)+a(30)+"u";a(30)+"u"};
{\arl^<<{l_4} a(330)+a(330);a(330)};
a(270)+a(270);a(270)**@{-}?(0.4)="pt"?(0.25)*@{>}?(0.2)+/^/*{l_3}?(0.7)*@{>}+/^/*{k_3};
{\ar "pt";"pt"+(0.7,0)*+{q_3}};
{\arl_<<{l_2} a(210)+a(210);a(210)};
{\arr^{j_1} a(150)+"u";a(90)+"u"};
{\arr^{j_6} a(90)+"u";a(30)+"u"};
{a(30)+"u";a(330)**@{-}?(0.25)="pt25"?(0.75)="pt75"
    ?(0.1)*@{<}?(0.5)*@{<}?(0.85)*@{<}
    ?(0.15)+/_/*{k_5}?(0.5)+/_/*{j_5}?(0.89)+/_/*{k_4}};
{\ar "pt25";"pt25"+(0.7,0)*+{q_5}};
{\ar "pt75";"pt75"+(0.7,0)*+{q_4}};
{\arr^{j_4} a(330);a(270)};
{\arr^{j_3} a(270);a(210)};
{a(210);a(150)+"u"**@{-}?(0.25)="pt25"?(0.75)="pt75"
    ?(0.1)*@{<}?(0.5)*@{<}?(0.85)*@{<}
    ?(0.15)+/^/*{k_2}?(0.5)+/^/*{j_2}?(0.89)+/^/*{k_1}};
{\ar "pt25";"pt25"+(0.7,0)*+{q_2}};
{\ar "pt75";"pt75"+(0.7,0)*+{q_1}};
\endxy
}

\newcommand{\NewPlaquetteMoveTailA}{
	\xy
	0;<6mm,0mm>:
	<0mm,7mm>::
	(0,0.4)="u";
	(0,-0.4)="d";
	{\arl^<<{l_1} a(150)+a(150)+"u";a(150)+"u"};
	a(90)+a(90)+"u";a(90)+"u"**@{-}?(0.4)="pt"?(0.25)*@{>}?(0.2)+/_/*{l_6}?(0.7)*@{>}+/_/*{k_6};
	{\ar "pt";"pt"+(0.7,0)*+{q_6}};
	a(270)+a(270);a(270)**@{-}?(0.4)="pt"?(0.25)*@{>}?(0.2)+/^/*{l_3}?(0.7)*@{>}+/^/*{k_3};
	{\ar "pt";"pt"+(0.7,0)*+{q_3}};
	{\arl_<<{l_2} a(210)+a(210);a(210)};
	{\arr^{j_1} a(150)+"u";a(90)+"u"};
	{\arr^{j_6} a(90)+"u";a(30)+"u"};
	{a(30)+"u";a(330)**@{.}?(0.5)*@{<}};
	{\arr^{j_4} a(330);a(270)};
	{\arr^{j_3} a(270);a(210)};
	{a(210);a(150)+"u"**@{-}?(0.25)="pt25"?(0.75)="pt75"
		?(0.1)*@{<}?(0.5)*@{<}?(0.85)*@{<}
        ?(0.15)+/^/*{k_2}?(0.5)+/^/*{j_2}?(0.89)+/^/*{k_1}};
	{\ar "pt25";"pt25"+(0.7,0)*+{q_2}};
	{\ar "pt75";"pt75"+(0.7,0)*+{q_1}};
	\endxy
}

\newcommand{\NewPlaquetteMoveTailB}{
	\xy
	0;<6mm,0mm>:
	<0mm,7mm>::
	(0,0.4)="u";
	(0,-0.4)="d";
	{\arl^<<{l_1} a(150)+a(150)+"u";a(150)+"u"};
	a(90)+a(90)+"u";a(90)+"u"**@{-}?(0.4)="pt"?(0.25)*@{>}?(0.2)+/_/*{l_6}?(0.7)*@{>}+/_/*{k_6};
	{\ar "pt";"pt"+(0.7,0)*+{q_6}};
	a(270)+a(270);a(270)**@{-}?(0.4)="pt"?(0.25)*@{>}?(0.2)+/^/*{l_3}?(0.7)*@{>}+/^/*{k_3};
	{\ar "pt";"pt"+(0.7,0)*+{q_3}};
	{\arl_<<{l_2} a(210)+a(210);a(210)};
    a(90)+"u";a(150)+"u"**@{-}?(0.4)*@{>}+/_/*{j_1}?(0.6)="pt"?(0.8)*@{>}+/_/*{k'_1};
    	{\ar "pt";"pt"+(0.35,-0.61)*+{q_1}};
	{\arr^{j_6} a(90)+"u";a(30)+"u"};
	{a(30)+"u";a(330)**@{.}?(0.5)*@{<}};	
	{\arr^{j_4} a(330);a(270)};
	{\arr^{j_3} a(270);a(210)};
	{a(210);a(150)+"u"**@{-}?(0.25)="pt25"?(0.75)="pt75"
		?(0.1)*@{<}?(0.6)*@{<}
		?(0.15)+/^/*{k_2}?(0.6)+/^/*{j_2}};
	{\ar "pt25";"pt25"+(0.7,0)*+{q_2}};
	\endxy
}

\newcommand{\NewPlaquetteOneTail}{
	\xy
	0;<5mm,0mm>:
	<0mm,6mm>::
	(0,0.4)="u";
	(0,-0.4)="d";
	{\arl_<<{l_1} a(150)+a(150)+"u";a(150)+"u"};
	a(90)+a(90)+"u";a(90)+"u"**@{-}?(0.4)="pt"?(0.5)*@{>}?(0.5)+/^/*{l_6};
	{\arl^<<{l_5} a(30)+a(30)+"u";a(30)+"u"};
	{\arl_<<{l_4} a(330)+a(330);a(330)};
	a(270)+a(270);a(270)**@{-}?(0.4)="pt"?(0.5)*@{>}?(0.5)+/_/*{l_3};
	{\arl^<<{l_2} a(210)+a(210);a(210)};
	{\arr^{j_1} a(150)+"u";a(90)+"u"};
	{\arr^{j_6} a(90)+"u";a(30)+"u"};
	{a(30)+"u";a(330)**@{-}?(0.25)="pt25"?(0.75)="pt75"
		?(0.5)*@{<}+/^/*{j_5}};
	{\arr^{j_4} a(330);a(270)};
	{\arr^{j_3} a(270);a(210)};
	{a(210);a(150)+"u"**@{-}?(0.25)="pt25"?(0.5)="pt75"
		?(0.25)*@{<}?(0.75)*@{<}
		?(0.25)+/^/*{j_2}?(0.75)+/^/*{k}};
	{\ar "pt75";"pt75"+(0.9,0)*+{q}};
	\endxy
}

\newcommand{\NewPlaquetteOneTailPrime}{
	\xy
	0;<5mm,0mm>:
	<0mm,6mm>::
	(0,0.4)="u";
	(0,-0.4)="d";
	{\arl_<<{l_1} a(150)+a(150)+"u";a(150)+"u"};
	a(90)+a(90)+"u";a(90)+"u"**@{-}?(0.4)="pt"?(0.5)*@{>}?(0.5)+/^/*{l_6};
	{\arl^<<{l_5} a(30)+a(30)+"u";a(30)+"u"};
	{\arl_<<{l_4} a(330)+a(330);a(330)};
	a(270)+a(270);a(270)**@{-}?(0.4)="pt"?(0.5)*@{>}?(0.5)+/_/*{l_3};
	{\arl^<<{l_2} a(210)+a(210);a(210)};
	{\arr^{j'_1} a(150)+"u";a(90)+"u"};
	{\arr^{j'_6} a(90)+"u";a(30)+"u"};
	{a(30)+"u";a(330)**@{-}?(0.25)="pt25"?(0.75)="pt75"
		?(0.5)*@{<}+/^/*{j'_5}};
	{\arr^{j'_4} a(330);a(270)};
	{\arr^{j'_3} a(270);a(210)};
	{a(210);a(150)+"u"**@{-}?(0.25)="pt25"?(0.5)="pt75"
		?(0.25)*@{<}?(0.75)*@{<}
		?(0.25)+/^/*{j'_2}?(0.75)+/^/*{k'}};
	{\ar "pt75";"pt75"+(0.9,0)*+{q'}};
	\endxy
}

\newcommand{\PlaquetteLoop}{
	\xy
	0;<6mm,0mm>:
	<0mm,7mm>::
	(0,0.4)="u";
	(0,-0.4)="d";
	{\aline a(150)+a(150)+"u";a(150)+"u"};
	a(90)+a(90)+"u";a(90)+"u"**@{-};
	{\aline a(30)+a(30)+"u";a(30)+"u"};
	{\aline a(330)+a(330);a(330)};
	a(270)+a(270);a(270)**@{-};
	{\aline a(210)+a(210);a(210)};
	{\aline a(150)+"u";a(90)+"u"};
	{\aline a(90)+"u";a(30)+"u"};
	{a(30)+"u";a(330)**@{-}};
	{\aline a(330);a(270)};
	{\aline a(270);a(210)};
	{a(210);a(150)+"u"**@{-}};
    {\ar@{-}@`{(0.6928, 0.72), (0.4, 1.0128), 
		(0., 1.12), (-0.4, 1.0128), 
		(-0.6928, 0.72), (-0.8,0.16), (-0.6928, -0.4), (-0.4, -0.6928), 
		(0., -0.8), (0.4, -0.6928), 
		(0.6928, -0.4)}|(0.05)*\dir{<} (0.735,0.2);(0.735,0.2)};
    (0.35,0.55)*{s};
	\endxy
}

\newcommand{\PlaquetteOneTailLoop}{
	\xy
	0;<7mm,0mm>:
	<0mm,7mm>::
	(0,0.4)="u";
	(0,-0.4)="d";
	{\aline a(150)+a(150)+"u";a(150)+"u"};
	a(90)+a(90)+"u";a(90)+"u"**@{-};
	{\aline a(30)+a(30)+"u";a(30)+"u"};
	{\aline a(330)+a(330);a(330)};
	a(270)+a(270);a(270)**@{-};
	{\aline a(210)+a(210);a(210)};
	{\aline a(150)+"u";a(90)+"u"};
	{\aline a(90)+"u";a(30)+"u"};
	{a(30)+"u";a(330)**@{-}};
	{\aline a(330);a(270)};
	{\aline a(270);a(210)};
	{a(210);a(150)+"u"**@{-}?(0.5)="pt75"};
	"pt75";"pt75"+(0.8,0)**@{-}?(0.18)*@{>}+(0,0.2)*{q}?(0.33)="pt33"?(0.5)*@{>}+(0.05,-0.2)*{u}?(0.66)="pt66"?(1)*@{>}+(0.1,0.25)*{q'};
    {\ar@{-}@`{	(-0.6928, -0.4), (-0.4, -0.6928), 
	    		(0., -0.8), (0.4, -0.6928), 
	    		(0.6928, -0.4),(0.6928, 0.72), (0.4, 1.0128), 
	    		(0., 1.12), (-0.4, 1.0128), 
	    		(-0.6928, 0.72)}|(0.4)*\dir{<} "pt33";"pt66"};
	(0.3,-0.5)*{s};
	\endxy
}

\newcommand{\PlaquetteOneTailTwist}{
	\xy
	0;<7mm,0mm>:
	<0mm,7mm>::
	(0,0.4)="u";
	(0,-0.4)="d";
	{\aline a(150)+a(150)+"u";a(150)+"u"};
	a(90)+a(90)+"u";a(90)+"u"**@{-};
	{\aline a(30)+a(30)+"u";a(30)+"u"};
	{\aline a(330)+a(330);a(330)};
	a(270)+a(270);a(270)**@{-};
	{\aline a(210)+a(210);a(210)};
	{\aline a(150)+"u";a(90)+"u"};
	{\aline a(90)+"u";a(30)+"u"};
	{a(30)+"u";a(330)**@{-}};
	{\aline a(330);a(270)};
	{\aline a(270);a(210)};
	{a(210);a(150)+"u"**@{-}?(0.5)="pt75"};
	{\ar@{-}@`{	(-0.6,0.16),(-0.6928, -0.4), (-0.4, -0.6928), 
			(0., -0.8), (0.4, -0.6928), 
			(0.6928, -0.4),(0.6928, 0.72), (0.4, 1.0128), 
			(0., 1.12), (-0.4, 1.0128), 
			(-0.6928, 0.72),(-0.8,0.16)}|(1)*\dir{>} "pt75";"pt75"+(0.8,0)};
	(0,0.4)*{q};
	\endxy
}

\newcommand{\NewPlaquettePrime}{
\xy
0;<7mm,0mm>:
<0mm,8mm>::
(0,0.4)="u";
(0,-0.4)="d";
{\arl^<<{l_1} a(150)+a(150)+"u";a(150)+"u"};
a(90)+a(90)+"u";a(90)+"u"**@{-}?(0.4)="pt"?(0.2)*@{>}?(0.2)+/_/*{l_6}?(0.7)*@{>}+/_/*{k_6};
{\ar "pt";"pt"+(0.7,0)*+{q_6}};
{\arl_<<{l_5} a(30)+a(30)+"u";a(30)+"u"};
{\arl^<<{l_4} a(330)+a(330);a(330)};
a(270)+a(270);a(270)**@{-}?(0.4)="pt"?(0.2)*@{>}?(0.2)+/^/*{l_3}?(0.7)*@{>}+/^/*{k_3};
{\ar "pt";"pt"+(0.7,0)*+{q_3}};
{\arl_<<{l_2} a(210)+a(210);a(210)};
{\arr^{j'_1} a(150)+"u";a(90)+"u"};
{\arr^{j'_6} a(90)+"u";a(30)+"u"};
{a(30)+"u";a(330)**@{-}?(0.25)="pt25"?(0.75)="pt75"
    ?(0.1)*@{<}?(0.5)*@{<}?(0.85)*@{<}
    ?(0.15)+/_/*{k'_5}?(0.5)+/_/*{j'_5}?(0.88)+/_/*{k'_4}};
{\ar "pt25";"pt25"+(0.7,0)*+{q_5}};
{\ar "pt75";"pt75"+(0.7,0)*+{q_4}};
{\arr^{j'_4} a(330);a(270)};
{\arr^{j'_3} a(270);a(210)};
{a(210);a(150)+"u"**@{-}?(0.25)="pt25"?(0.75)="pt75"
    ?(0.1)*@{<}?(0.5)*@{<}?(0.85)*@{<}
    ?(0.15)+/^/*{k'_2}?(0.5)+/^/*{j'_2}?(0.88)+/^/*{k'_1}};
{\ar "pt25";"pt25"+(0.7,0)*+{q_2}};
{\ar "pt75";"pt75"+(0.7,0)*+{q_1}};
\endxy
}

\newcommand{\BpsPrime}{
	\xy
	0;<7mm,0mm>:
	<0mm,8mm>::
	(0,0.4)="u";
	(0,-0.4)="d";
	{\arl^<<{l_1} a(150)+a(150)+"u";a(150)+"u"};
	a(90)+a(90)+"u";a(90)+"u"**@{-}?(0.4)="pt"?(0.2)*@{>}?(0.2)+/_/*{l_6}?(0.7)*@{>}+/_/*{k_6};
	{\ar "pt";"pt"+(0.7,0)*+{q_6}};
	{\arl_<<{l_5} a(30)+a(30)+"u";a(30)+"u"};
	{\arl^<<{l_4} a(330)+a(330);a(330)};
	a(270)+a(270);a(270)**@{-}?(0.4)="pt"?(0.2)*@{>}?(0.2)+/^/*{l_3}?(0.7)*@{>}+/^/*{k_3};
	{\ar "pt";"pt"+(0.7,0)*+{q_3}};
	{\arl_<<{l_2} a(210)+a(210);a(210)};
	{\arr^{j'_1} a(150)+"u";a(90)+"u"};
	{\arr^{j'_6} a(90)+"u";a(30)+"u"};
	{a(30)+"u";a(330)**@{-}?(0.25)="pt25"?(0.75)="pt75"
		?(0.1)*@{<}?(0.5)*@{<}?(0.85)*@{<}
		?(0.15)+/_/*{k'_5}?(0.5)+/_/*{j'_5}?(0.88)+/_/*{k'_4}};
	{\ar "pt25";"pt25"+(0.7,0)*+{q_5}};
	{\ar "pt75";"pt75"+(0.7,0)*+{q_4}};
	{\arr^{j'_4} a(330);a(270)};
	{\arr^{j'_3} a(270);a(210)};
	{a(210);a(150)+"u"**@{-}?(0.25)="pt25"?(0.75)="pt75"
		?(0.1)*@{<}?(0.5)*@{<}?(0.85)*@{<}
		?(0.15)+/^/*{j'_2}?(0.5)+/^/*{j'_2}?(0.88)+/^/*{j'_2}};
	{\ar@{.} "pt25";"pt25"+(0.7,0)*+{0}};
	{\ar@{.} "pt75";"pt75"+(0.7,0)*+{0}};
	\endxy
}

\newcommand{\NewPlaquettePrimePrime}{
\xy
0;<7mm,0mm>:
<0mm,8mm>::
(0,0.4)="u";
(0,-0.4)="d";
{\arl^<<{l_1} a(150)+a(150)+"u";a(150)+"u"};
a(90)+a(90)+"u";a(90)+"u"**@{-}?(0.4)="pt"?(0.2)*@{>}?(0.2)+/_/*{l_6}?(0.7)*@{>}+/_/*{k_6};
{\ar "pt";"pt"+(0.7,0)*+{q_6}};
{\arl_<<{l_5} a(30)+a(30)+"u";a(30)+"u"};
{\arl^<<{l_4} a(330)+a(330);a(330)};
a(270)+a(270);a(270)**@{-}?(0.4)="pt"?(0.2)*@{>}?(0.2)+/^/*{l_3}?(0.7)*@{>}+/^/*{k_3};
{\ar "pt";"pt"+(0.7,0)*+{q_3}};
{\arl_<<{l_2} a(210)+a(210);a(210)};
{\arr^{j'_1} a(150)+"u";a(90)+"u"};
{\arr^{j'_6} a(90)+"u";a(30)+"u"};
{a(30)+"u";a(330)**@{-}?(0.25)="pt25"?(0.75)="pt75"
    ?(0.1)*@{<}?(0.5)*@{<}?(0.85)*@{<}
    ?(0.15)+/_/*{k'_5}?(0.5)+/_/*{j'_5}?(0.88)+/_/*{k'_4}};
{\ar "pt25";"pt25"+(0.7,0)*+{q_5}};
{\ar "pt75";"pt75"+(0.7,0)*+{q_4}};
{\arr^{j'_4} a(330);a(270)};
{\arr^{j'_3} a(270);a(210)};
{a(210);a(150)+"u"**@{-}?(0.25)="pt25"?(0.75)="pt75"
    ?(0.1)*@{<}?(0.5)*@{<}?(0.85)*@{<}
    ?(0.15)+/^/*{k'_2}?(0.5)+/^/*{k_1}?(0.88)+/^/*{k'_1}};
{\ar "pt25";"pt25"+(0.7,0)*+{q_2}};
{\ar "pt75";"pt75"+(0.7,0)*+{q_1}};
\endxy
}

\newcommand{\ChargePairNewPlaquette}{
\xy
0;<7mm,0mm>:
<0mm,8mm>::
(0,0.4)="u";
(0,-0.4)="d";
{\arl^<<{l_1} a(150)+a(150)+"u";a(150)+"u"};
a(90)+a(90)+"u";a(90)+"u"**@{-}?(0.4)="pt"?(0.25)*@{>}?(0.2)+/_/*{l_6}?(0.7)*@{>}+/_/*{k_6};
{\ar "pt";"pt"+(0.7,0)*+{q_6}};
{\arl_<<{l_5} a(30)+a(30)+"u";a(30)+"u"};
{\arl^<<{l_4} a(330)+a(330);a(330)};
a(270)+a(270);a(270)**@{-}?(0.4)="pt"?(0.25)*@{>}?(0.2)+/^/*{l_3}?(0.7)*@{>}+/^/*{k_3};
{\ar "pt";"pt"+(0.7,0)*+{q_3}};
{\arl_<<{l_2} a(210)+a(210);a(210)};
{\arr^{j_1} a(150)+"u";a(90)+"u"};
{\arr^{j_6} a(90)+"u";a(30)+"u"};
{a(30)+"u";a(330)**@{-}?(0.25)="pt25"?(0.75)="pt75"
    ?(0.1)*@{<}?(0.5)*@{<}?(0.85)*@{<}
    ?(0.15)+/_/*{k_5}?(0.5)+/_/*{j_5}?(0.89)+/_/*{k_4}};
{\ar "pt25";"pt25"+(0.7,0)*+{q_5}};
{\ar "pt75";"pt75"+(0.7,0)*+{q_4}};
{\arr^{j_4} a(330);a(270)};
{\arr^{j_3} a(270);a(210)};
{a(210);a(150)+"u"**@{-}?(0.25)="pt25"?(0.75)="pt75"
    ?(0.1)*@{<}?(0.5)*@{<}?(0.85)*@{<}
    ?(0.15)+/^/*{k_2}?(0.5)+/^/*{j'_2}?(0.89)+/^/*{k_1}};
{\ar "pt25";"pt25"+(0.7,0)*+{q'_2}};
{\ar "pt75";"pt75"+(0.7,0)*+{q'_1}};
\endxy
}

\newcommand{\ChargeHoppingNewPlaquette}{
\xy
0;<7mm,0mm>:
<0mm,8mm>::
(0,0.4)="u";
(0,-0.4)="d";
{\arl^<<{l_1} a(150)+a(150)+"u";a(150)+"u"};
a(90)+a(90)+"u";a(90)+"u"**@{-}?(0.4)="pt"?(0.25)*@{>}?(0.2)+/_/*{l_6}?(0.7)*@{>}+/_/*{k_6};
{\ar "pt";"pt"+(0.7,0)*+{q_6}};
{\arl_<<{l_5} a(30)+a(30)+"u";a(30)+"u"};
{\arl^<<{l_4} a(330)+a(330);a(330)};
a(270)+a(270);a(270)**@{-}?(0.4)="pt"?(0.25)*@{>}?(0.2)+/^/*{l_3}?(0.7)*@{>}+/^/*{k_3};
{\ar "pt";"pt"+(0.7,0)*+{q_3}};
{\arl_<<{l_2} a(210)+a(210);a(210)};
{\arr^{j_1} a(150)+"u";a(90)+"u"};
{\arr^{j_6} a(90)+"u";a(30)+"u"};
{a(30)+"u";a(330)**@{-}?(0.25)="pt25"?(0.75)="pt75"
    ?(0.1)*@{<}?(0.5)*@{<}?(0.85)*@{<}
    ?(0.15)+/_/*{k_5}?(0.5)+/_/*{j_5}?(0.89)+/_/*{k_4}};
{\ar "pt25";"pt25"+(0.7,0)*+{q_5}};
{\ar "pt75";"pt75"+(0.7,0)*+{q_4}};
{\arr^{j_4} a(330);a(270)};
{\arr^{j_3} a(270);a(210)};
{a(210);a(150)+"u"**@{-}?(0.25)="pt25"?(0.75)="pt75"
    ?(0.1)*@{<}?(0.5)*@{<}?(0.85)*@{<}
    ?(0.15)+/^/*{k_2}?(0.5)+/^/*{k_2}?(0.89)+/^/*{k_1}};
{\ar@{.>} "pt25";"pt25"+(0.65,0)*+{0}};
{\ar "pt75";"pt75"+(0.7,0)*+{q'_1}};
\endxy
}

\newcommand{\ChargeqqHoppingNewPlaquette}{
	\xy
	0;<7mm,0mm>:
	<0mm,8mm>::
	(0,0.4)="u";
	(0,-0.4)="d";
	{\arl^<<{l_1} a(150)+a(150)+"u";a(150)+"u"};
	a(90)+a(90)+"u";a(90)+"u"**@{-}?(0.4)="pt"?(0.25)*@{>}?(0.2)+/_/*{l_6}?(0.7)*@{>}+/_/*{k_6};
	{\ar "pt";"pt"+(0.7,0)*+{q_6}};
	{\arl_<<{l_5} a(30)+a(30)+"u";a(30)+"u"};
	{\arl^<<{l_4} a(330)+a(330);a(330)};
	a(270)+a(270);a(270)**@{-}?(0.4)="pt"?(0.25)*@{>}?(0.2)+/^/*{l_3}?(0.7)*@{>}+/^/*{k_3};
	{\ar "pt";"pt"+(0.7,0)*+{q_3}};
	{\arl_<<{l_2} a(210)+a(210);a(210)};
	{\arr^{j_1} a(150)+"u";a(90)+"u"};
	{\arr^{j_6} a(90)+"u";a(30)+"u"};
	{a(30)+"u";a(330)**@{-}?(0.25)="pt25"?(0.75)="pt75"
		?(0.1)*@{<}?(0.5)*@{<}?(0.85)*@{<}
		?(0.15)+/_/*{k_5}?(0.5)+/_/*{j_5}?(0.89)+/_/*{k_4}};
	{\ar "pt25";"pt25"+(0.7,0)*+{q_5}};
	{\ar "pt75";"pt75"+(0.7,0)*+{q_4}};
	{\arr^{j_4} a(330);a(270)};
	{\arr^{j_3} a(270);a(210)};
	{a(210);a(150)+"u"**@{-}?(0.25)="pt25"?(0.75)="pt75"
		?(0.1)*@{<}?(0.5)*@{<}?(0.85)*@{<}
		?(0.15)+/^/*{k_2}?(0.5)+/^/*{k_2}?(0.89)+/^/*{k_1}};
	{\ar@{.>} "pt25";"pt25"+(0.65,0)*+{0}};
	{\ar@{.>} "pt75";"pt75"+(0.65,0)*+{0}};
	\endxy
}

\newcommand{\NewPlaquetteEffective}{
\xy
0;<7mm,0mm>:
<0mm,8mm>::
(0,0.4)="u";
(0,-0.4)="d";
{\arl^<<{l_1} a(150)+a(150)+"u";a(150)+"u"};
a(90)+a(90)+"u";a(90)+"u"**@{-}?(0.4)="pt"?(0.2)*@{>}?(0.2)+/_/*{l_6}?(0.7)*@{>}+/_/*{k_6};
{\ar "pt";"pt"+(0.7,0)*+{q_6}};
{\arl_<<{l_5} a(30)+a(30)+"u";a(30)+"u"};
{\arl^<<{l_4} a(330)+a(330);a(330)};
a(270)+a(270);a(270)**@{-}?(0.4)="pt"?(0.2)*@{>}?(0.2)+/^/*{l_3}?(0.7)*@{>}+/^/*{k_3};
{\ar "pt";"pt"+(0.7,0)*+{q_3}};
{\arl_<<{l_2} a(210)+a(210);a(210)};
{\arr^{j_1} a(150)+"u";a(90)+"u"};
{\arr^{j_6} a(90)+"u";a(30)+"u"};
{a(30)+"u";a(330)**@{-}?(0.25)="pt25"?(0.75)="pt75"
    ?(0.1)*@{<}?(0.5)*@{<}?(0.85)*@{<}
    ?(0.15)+/_/*{k_5}?(0.5)+/_/*{j_5}?(0.88)+/_/*{k_4}};
{\ar "pt25";"pt25"+(0.7,0)*+{q_5}};
{\ar "pt75";"pt75"+(0.7,0)*+{q_4}};
{\arr^{j_4} a(330);a(270)};
{\arr^{j_3} a(270);a(210)};
{a(210);a(150)+"u"**@{-}?(0.25)="pt25"?(0.75)="pt75"
    ?(0.1)*@{<}?(0.5)*@{<}?(0.85)*@{<}
    ?(0.15)+/_/*{k_2}?(0.5)+/_/*{j_2}?(0.88)+/_/*{k_1}};
{\ar "pt25";"pt25"+(-0.7,0)*+{q_2}};
{\ar "pt75";"pt75"+(-0.7,0)*+{q_1}};
\endxy
}

\newcommand{\WilsonLoopPlaquette}{
\xy
0;<7mm,0mm>:
<0mm,8mm>::
(0,0.4)="u";
(0,-0.4)="d";
{\arl^<<{l_1} a(150)+a(150)+"u";a(150)+"u"};
a(90)+a(90)+"u";a(90)+"u"**@{-}?(0.4)="pt"?(0.25)*@{>}?(0.2)+/_/*{l_6}?(0.7)*@{>}+/_/*{k_6};
{\ar "pt";"pt"+(0.7,0)*+{q_6}};
{\arl_<<{l_5} a(30)+a(30)+"u";a(30)+"u"};
{\arl^<<{l_4} a(330)+a(330);a(330)};
a(270)+a(270);a(270)**@{-}?(0.4)="pt"?(0.25)*@{>}?(0.2)+/^/*{l_3}?(0.7)*@{>}+/^/*{k_3};
{\ar "pt";"pt"+(0.7,0)*+{q_3}};
{\arl_<<{l_2} a(210)+a(210);a(210)};
{\arr^{j_1} a(150)+"u";a(90)+"u"};
{\arr^{j_6} a(90)+"u";a(30)+"u"};
{a(30)+"u";a(330)**@{-}?(0.25)="pt25"?(0.75)="pt75"
    ?(0.1)*@{<}?(0.5)*@{<}?(0.85)*@{<}
    ?(0.12)+/^/*{\scriptscriptstyle k_5}?(0.5)+/^/*{\scriptscriptstyle j_5}?(0.89)+/^/*{\scriptscriptstyle k_4}};
{\ar "pt25";"pt25"+(0.7,0)*+{q_5}};
{\ar "pt75";"pt75"+(0.7,0)*+{q_4}};
{\arr^{j_4} a(330);a(270)};
{\arr^{j_3} a(270);a(210)};
{a(210);a(150)+"u"**@{-}?(0.25)="pt25"?(0.75)="pt75"
    ?(0.1)*@{<}?(0.5)*@{<}?(0.85)*@{<}
    ?(0.15)+/^/*{k_2}?(0.5)+/^/*{j_2}?(0.89)+/^/*{k_1}};
{\ar "pt25";"pt25"+(0.7,0)*+{q_2}};
{\ar "pt75";"pt75"+(0.7,0)*+{q_1}};
{\ar@{-}@`{(0.6928, 0.72), (0.4, 1.0128), 
 (0., 1.12), (-0.4, 1.0128), 
 (-0.6928, 0.72), (-0.8,0.16), (-0.6928, -0.4), (-0.4, -0.6928), 
 (0., -0.8), (0.4, -0.6928), 
 (0.6928, -0.4)}|(0.05)*\dir{>}|(0.42)\hole|(0.55)\hole (0.735,0.2);(0.735,0.2)};
(0.45,0.65)*{s};
\endxy
}

\newcommand{\TwistFluxonPlaquette}{
\xy
0;<7mm,0mm>:
<0mm,8mm>::
(0,0.4)="u";
(0,-0.4)="d";
{\arl^<<{l_1} a(150)+a(150)+"u";a(150)+"u"};
a(90)+a(90)+"u";a(90)+"u"**@{-}?(0.4)="pt"?(0.25)*@{>}?(0.2)+/_/*{l_6}?(0.7)*@{>}+/_/*{k_6};
{\ar "pt";"pt"+(0.7,0)*+{q_6}};
{\arl_<<{l_5} a(30)+a(30)+"u";a(30)+"u"};
{\arl^<<{l_4} a(330)+a(330);a(330)};
a(270)+a(270);a(270)**@{-}?(0.4)="pt"?(0.25)*@{>}?(0.2)+/^/*{l_3}?(0.7)*@{>}+/^/*{k_3};
{\ar "pt";"pt"+(0.7,0)*+{q_3}};
{\arl_<<{l_2} a(210)+a(210);a(210)};
{\arr^{j_1} a(150)+"u";a(90)+"u"};
{\arr^{j_6} a(90)+"u";a(30)+"u"};
{a(30)+"u";a(330)**@{-}?(0.25)="pt25"?(0.75)="pt75"
    ?(0.1)*@{<}?(0.5)*@{<}?(0.85)*@{<}
    ?(0.15)+/^/*{\scriptscriptstyle k_5}?(0.5)+/^/*{\scriptscriptstyle j_5}?(0.89)+/^/*{\scriptscriptstyle k_4}};
{\ar "pt25";"pt25"+(0.7,0)*+{q_5}};
{\ar "pt75";"pt75"+(0.7,0)*+{q_4}};
{\arr^{j_4} a(330);a(270)};
{\arr^{j_3} a(270);a(210)};
{a(210);a(150)+"u"**@{-}?(0.25)="pt25"?(0.75)="pt75"
    ?(0.1)*@{<}?(0.5)*@{<}?(0.85)*@{<}
    ?(0.15)+/^/*{k_2}?(0.5)+/^/*{j_2}?(0.89)+/^/*{k_1}};
{\ar "pt25";"pt25"+(0.7,0)*+{q_2}};
{\ar@{-} "pt75";"pt75"+(0.2,0)};
{\ar@`{"pt75"+(0.3,-0.3),(-0.8,0.16), (-0.6928, -0.4), (-0.4, -0.6928), 
 (0., -0.8), (0.4, -0.6928), 
 (0.6928, -0.4), (0.8,0.16), (0.6928, 0.72), (0.4, 1.0128), 
 (0., 1.12), (-0.4, 1.0128), 
 (-0.74,0.6)}|(0.13)\hole
 "pt75"+(0.2,0);"pt75"+(0.5,0)};
(-0.1,0.6)*{q_1};
\endxy
}

\newcommand{\TwistDyonPlaquette}{
\xy
0;<7mm,0mm>:
<0mm,8mm>::
(0,0.4)="u";
(0,-0.4)="d";
{\ar^<<{l_1}@{-}|(0.35)@{>}|(0.65)\hole a(150)+a(150)+"u";a(150)+"u"};
a(90)+a(90)+"u";a(90)+"u"**@{-}?(0.4)="pt"?(0.25)*@{>}?(0.2)+/_/*{l_6}?(0.7)*@{>}+/_/*{k_6};
{\ar "pt";"pt"+(0.7,0)*+{q_6}};
{\arl_<<{l_5} a(30)+a(30)+"u";a(30)+"u"};
{\arl^<<{l_4} a(330)+a(330);a(330)};
a(270)+a(270);a(270)**@{-}?(0.4)="pt"?(0.25)*@{>}?(0.2)+/^/*{l_3}?(0.7)*@{>}+/^/*{k_3};
{\ar "pt";"pt"+(0.7,0)*+{q_3}};
{\arl_<<{l_2} a(210)+a(210);a(210)};
{\ar^{j_1}@{-}|(0.33)\hole|(0.6)@{<} a(150)+"u";a(90)+"u"};
{\arr^{j_6} a(90)+"u";a(30)+"u"};
{a(30)+"u";a(330)**@{-}?(0.25)="pt25"?(0.75)="pt75"
    ?(0.1)*@{<}?(0.5)*@{<}?(0.85)*@{<}
    ?(0.15)+/^/*{\scriptscriptstyle k_5}?(0.5)+/^/*{\scriptscriptstyle j_5}?(0.89)+/^/*{\scriptscriptstyle k_4}};
{\ar "pt25";"pt25"+(0.7,0)*+{q_5}};
{\ar "pt75";"pt75"+(0.7,0)*+{q_4}};
{\arr^{j_4} a(330);a(270)};
{\arr^{j_3} a(270);a(210)};
{a(210);a(150)+"u"**{}?(0.25)="pt25"?(0.70)="pt75"
    ?(0.1)*@{<}?(0.5)*@{<}?(0.85)*{}
    ?(0.15)+/^/*{k_2}?(0.5)+/^/*{j_2}?(0.83)+/^/*{k_1}};
{\ar@{-}|(0.86)\hole a(210);a(150)+"u"};
{\ar "pt25";"pt25"+(0.7,0)*+{q_2}};
{\ar@{-} "pt75";"pt75"+(0.2,0)};
{\ar@`{"pt75"+(0.3,-0.3),(-0.8,0.16), (-0.6928, -0.4), (-0.4, -0.6928), 
 (0., -0.8), (0.4, -0.6928), 
 (0.6928, -0.4), (0.8,0.16), (0.6928, 0.72), (0.4, 1.0128), 
 (0., 1.12), (-0.3, 1.1), 
 (-0.6,1.), (-1.1,1.3), (-1.35,0.85), (-0.8,0.65), (-0.6,0.75)}|(0.1)\hole
  "pt75"+(0.2,0);"pt75"+(0.5,0)};
(-0.05,0.55)*{q_1};
\endxy
}

\newcommand{\GroundStatePlaquette}{
\xy
0;<7mm,0mm>:
<0mm,8mm>::
(0,0.4)="u";
(0,-0.4)="d";
{\arl^<<{l_1} a(150)+a(150)+"u";a(150)+"u"};
a(90)+a(90)+"u";a(90)+"u"**@{-}?(0.4)="pt"?(0.25)*@{>}?(0.2)+/_/*{l_6}?(0.7)*@{>}+/_/*{l_6}; 
{\ar@{.>} "pt";"pt"+(0.7,0)*+{0}}; 
{\arl_<<{l_5} a(30)+a(30)+"u";a(30)+"u"};
{\arl^<<{l_4} a(330)+a(330);a(330)};
a(270)+a(270);a(270)**@{-}?(0.4)="pt"?(0.25)*@{>}?(0.2)+/^/*{l_3}?(0.7)*@{>}+/^/*{l_3}; 
{\ar@{.>} "pt";"pt"+(0.7,0)*+{0}}; 
{\arl_<<{l_2} a(210)+a(210);a(210)};
{\arr^{j_1} a(150)+"u";a(90)+"u"};
{\arr^{j_6} a(90)+"u";a(30)+"u"};
{a(30)+"u";a(330)**@{-}?(0.25)="pt25"?(0.75)="pt75"
    ?(0.1)*@{<}?(0.5)*@{<}?(0.85)*@{<}
    ?(0.15)+/_/*{j_5}?(0.5)+/_/*{j_5}?(0.89)+/_/*{j_5}}; 
{\ar@{.>} "pt25";"pt25"+(0.7,0)*+{0}}; 
{\ar@{.>} "pt75";"pt75"+(0.7,0)*+{0}}; 
{\arr^{j_4} a(330);a(270)};
{\arr^{j_3} a(270);a(210)};
{a(210);a(150)+"u"**@{-}?(0.25)="pt25"?(0.75)="pt75"
    ?(0.1)*@{<}?(0.5)*@{<}?(0.85)*@{<}
    ?(0.15)+/^/*{j_2}?(0.5)+/^/*{j_2}?(0.89)+/^/*{j_2}}; 
{\ar@{.>} "pt25";"pt25"+(0.7,0)*+{0}}; 
{\ar@{.>} "pt75";"pt75"+(0.7,0)*+{0}}; 
\endxy
}

\newcommand{\ChargePairPlaquette}{
\xy
0;<7mm,0mm>:
<0mm,8mm>::
(0,0.4)="u";
(0,-0.4)="d";
{\arl^<<{l_1} a(150)+a(150)+"u";a(150)+"u"};
a(90)+a(90)+"u";a(90)+"u"**@{-}?(0.4)="pt"?(0.25)*@{>}?(0.2)+/_/*{l_6}?(0.7)*@{>}+/_/*{l_6}; 
{\ar@{.>} "pt";"pt"+(0.7,0)*+{0}}; 
{\arl_<<{l_5} a(30)+a(30)+"u";a(30)+"u"};
{\arl^<<{l_4} a(330)+a(330);a(330)};
a(270)+a(270);a(270)**@{-}?(0.4)="pt"?(0.25)*@{>}?(0.2)+/^/*{l_3}?(0.7)*@{>}+/^/*{l_3}; 
{\ar@{.>} "pt";"pt"+(0.7,0)*+{0}}; 
{\arl_<<{l_2} a(210)+a(210);a(210)};
{\arr^{j_1} a(150)+"u";a(90)+"u"};
{\arr^{j_6} a(90)+"u";a(30)+"u"};
{a(30)+"u";a(330)**@{-}?(0.25)="pt25"?(0.75)="pt75"
    ?(0.1)*@{<}?(0.5)*@{<}?(0.85)*@{<}
    ?(0.15)+/_/*{j_5}?(0.5)+/_/*{j_5}?(0.89)+/_/*{j_5}}; 
{\ar@{.>} "pt25";"pt25"+(0.7,0)*+{0}}; 
{\ar@{.>} "pt75";"pt75"+(0.7,0)*+{0}}; 
{\arr^{j_4} a(330);a(270)};
{\arr^{j_3} a(270);a(210)};
{a(210);a(150)+"u"**@{-}?(0.25)="pt25"?(0.75)="pt75"
    ?(0.1)*@{<}?(0.5)*@{<}?(0.85)*@{<}
    ?(0.15)+/^/*{j_2}?(0.5)+/^/*{j'_2}?(0.89)+/^/*{j_2}}; 
{\ar "pt25";"pt25"+(0.7,0)*+{q^*}}; 
{\ar "pt75";"pt75"+(0.7,0)*+{q{\;\,}}}; 
\endxy
}

\newcommand{\CrossedGroundStatePlaquette}{
\xy
0;<7mm,0mm>:
<0mm,8mm>::
(0,0.4)="u";
(0,-0.4)="d";
{\arl^<<{l_1} a(150)+a(150)+"u";a(150)+"u"};
a(90)+a(90)+"u";a(90)+"u"**@{-}?(0.4)="pt"?(0.25)*@{>}?(0.2)+/_/*{l_6}?(0.7)*@{>}+/_/*{l_6}; 
{\ar@{.>} "pt";"pt"+(0.7,0)*+{0}}; 
{\arl_<<{l_5} a(30)+a(30)+"u";a(30)+"u"};
{\arl^<<{l_4} a(330)+a(330);a(330)};
a(270)+a(270);a(270)**@{-}?(0.4)="pt"?(0.25)*@{>}?(0.2)+/^/*{l_3}?(0.7)*@{>}+/^/*{l_3}; 
{\ar@{.>} "pt";"pt"+(0.7,0)*+{0}}; 
{\arl_<<{l_2} a(210)+a(210);a(210)};
{\arr^{j_1} a(150)+"u";a(90)+"u"};
{\arr^{j_6} a(90)+"u";a(30)+"u"};
{a(30)+"u";a(330)**@{-}?(0.25)="pt25"?(0.75)="pt75"
    ?(0.1)*@{<}?(0.5)*@{<}?(0.85)*@{<}
    ?(0.15)+/_/*{j_5}?(0.5)+/_/*{j_5}?(0.89)+/_/*{j_5}}; 
{\ar@{.>} "pt25";"pt25"+(0.7,0)*+{0}}; 
{\ar@{.>} "pt75";"pt75"+(0.7,0)*+{0}}; 
{\arr^{j_4} a(330);a(270)};
{\arr^{j_3} a(270);a(210)};
{a(210);a(150)+"u"**@{-}?(0.25)="pt25"?(0.75)="pt75"
    ?(0.1)*@{<}?(0.5)*@{<}?(0.85)*@{<}
    ?(0.15)+/_/*{j_2}?(0.5)+/^/*{j_2}?(0.89)+/^/*{j_2}}; 
{\ar@{.>} "pt25";"pt25"+(-0.7,0)*+{0}}; 
{\ar@{.>} "pt75";"pt75"+(0.7,0)*+{0}}; 
\endxy
}

\newcommand{\CrossedChargePairPlaquette}{
\xy
0;<7mm,0mm>:
<0mm,8mm>::
(0,0.4)="u";
(0,-0.4)="d";
{\arl^<<{l_1} a(150)+a(150)+"u";a(150)+"u"};
a(90)+a(90)+"u";a(90)+"u"**@{-}?(0.4)="pt"?(0.25)*@{>}?(0.2)+/_/*{l_6}?(0.7)*@{>}+/_/*{l_6}; 
{\ar@{.>} "pt";"pt"+(0.7,0)*+{0}}; 
{\arl_<<{l_5} a(30)+a(30)+"u";a(30)+"u"};
{\arl^<<{l_4} a(330)+a(330);a(330)};
a(270)+a(270);a(270)**@{-}?(0.4)="pt"?(0.25)*@{>}?(0.2)+/^/*{l_3}?(0.7)*@{>}+/^/*{l_3}; 
{\ar@{.>} "pt";"pt"+(0.7,0)*+{0}}; 
{\arl_<<{l_2} a(210)+a(210);a(210)};
{\arr^{j_1} a(150)+"u";a(90)+"u"};
{\arr^{j_6} a(90)+"u";a(30)+"u"};
{a(30)+"u";a(330)**@{-}?(0.25)="pt25"?(0.75)="pt75"
    ?(0.1)*@{<}?(0.5)*@{<}?(0.85)*@{<}
    ?(0.15)+/_/*{j_5}?(0.5)+/_/*{j_5}?(0.89)+/_/*{j_5}}; 
{\ar@{.>} "pt25";"pt25"+(0.7,0)*+{0}}; 
{\ar@{.>} "pt75";"pt75"+(0.7,0)*+{0}}; 
{\arr^{j_4} a(330);a(270)};
{\arr^{j_3} a(270);a(210)};
{a(210);a(150)+"u"**@{-}?(0.25)="pt25"?(0.75)="pt75"
    ?(0.1)*@{<}?(0.5)*@{<}?(0.85)*@{<}
    ?(0.15)+/_/*{j_2}?(0.5)+/^/*{j'_2}?(0.89)+/^/*{j_2}}; 
{\ar "pt25";"pt25"+(-0.7,0)*+{q^*}}; 
{\ar "pt75";"pt75"+(0.7,0)*+{q{\;\,}}}; 
\endxy
}

\newcommand{\CrossedDyonPairPlaquette}{
\xy
0;<7mm,0mm>:
<0mm,8mm>::
(0,0.4)="u";
(0,-0.4)="d";
{\arl^<<{l_1} a(150)+a(150)+"u";a(150)+"u"};
a(90)+a(90)+"u";a(90)+"u"**@{-}?(0.4)="pt"?(0.25)*@{>}?(0.2)+/_/*{l_6}?(0.7)*@{>}+/_/*{l_6}; 
{\ar@{.>} "pt";"pt"+(0.7,0)*+{0}}; 
{\arl_<<{l_5} a(30)+a(30)+"u";a(30)+"u"};
{\arl^<<{l_4} a(330)+a(330);a(330)};
a(270)+a(270);a(270)**@{-}?(0.4)="pt"?(0.25)*@{>}?(0.2)+/^/*{l_3}?(0.7)*@{>}+/^/*{l_3}; 
{\ar@{.>} "pt";"pt"+(0.7,0)*+{0}}; 
{\arl_<<{l_2} a(210)+a(210);a(210)};
{\arr^{j_1} a(150)+"u";a(90)+"u"};
{\arr^{j_6} a(90)+"u";a(30)+"u"};
{a(30)+"u";a(330)**@{-}?(0.25)="pt25"?(0.75)="pt75"
    ?(0.1)*@{<}?(0.5)*@{<}?(0.85)*@{<}
    ?(0.15)+/_/*{j_5}?(0.5)+/_/*{j_5}?(0.89)+/_/*{j_5}}; 
{\ar@{.>} "pt25";"pt25"+(0.7,0)*+{0}}; 
{\ar@{.>} "pt75";"pt75"+(0.7,0)*+{0}}; 
{\arr^{j_4} a(330);a(270)};
{\arr^{j_3} a(270);a(210)};
{a(210);a(150)+"u"**@{-}?(0.25)="pt25"?(0.75)="pt75"
    ?(0.1)*@{<}?(0.5)*@{<}?(0.85)*@{<}
    ?(0.15)+/_/*{j_2}?(0.5)+/^/*{j'_2}?(0.89)+/^/*{j_2}}; 
{\ar "pt25";"pt25"+(-0.7,0)*+{p{\;}}}; 
{\ar "pt75";"pt75"+(0.7,0)*+{q{\;}}}; 
\endxy
}

\newcommand{\GroundStateTailedTwoPlaquette}{
	\xy
	0;<7mm,0mm>:
	<0mm,8mm>::
	(0,0.4)="u";
	(0,-0.4)="d";
	{\arl^{j_7} a(150)+a(150)+"u";a(150)+"u"};
	a(90)+a(90)+"u";a(90)+"u"**@{-}?(0.4)="pt"?(0.25)*@{>}?(0.2)+/_/*{l_6}?(0.7)*@{>}+/_/*{l_6}; 
	{\ar@{.>} "pt";"pt"+(0.7,0)*+{0}}; 
	{\arl_<<{l_5} a(30)+a(30)+"u";a(30)+"u"};
	{\arr^<<{l_4} a(330)+a(330);a(330)};
	a(270)+a(270);a(270)**@{-}?(0.4)="pt"?(0.25)*@{>}?(0.2)+/^/*{l_3}?(0.7)*@{>}+/^/*{l_3}; 
	{\ar@{.>} "pt";"pt"+(0.7,0)*+{0}}; 
	{\arl_{j_{11}} a(210)+a(210);a(210)};
	{\arr^{j_1} a(150)+"u";a(90)+"u"};
	{\arr^{j_6} a(90)+"u";a(30)+"u"};
	{a(30)+"u";a(330)**@{-}?(0.25)="pt25"?(0.75)="pt75"
		?(0.1)*@{<}?(0.5)*@{<}?(0.85)*@{<}
		?(0.15)+/_/*{j_5}?(0.5)+/_/*{j_5}?(0.89)+/_/*{j_5}}; 
	{\ar@{.>} "pt25";"pt25"+(0.7,0)*+{0}}; 
	{\ar@{.>} "pt75";"pt75"+(0.7,0)*+{0}}; 
	{\arr^{j_4} a(330);a(270)};
	{\arr^{j_3} a(270);a(210)};
	{a(210);a(150)+"u"**@{-}?(0.25)="pt25"?(0.75)="pt75"
		?(0.1)*@{>}?(0.5)*@{>}?(0.85)*@{>}
		?(0.15)+/^/*{j_2}?(0.5)+/^/*{j_2}?(0.89)+/^/*{j_2}}; 
	{\ar@{.>} "pt25";"pt25"+(0.7,0)*+{0}}; 
	{\ar@{.>} "pt75";"pt75"+(0.7,0)*+{0}}; 
	{\arr^{j_8} a(150)+a(150)+"u";a(150)+a(150)+a(210)+"u"};
	{\arr_{j_{10}} a(210)+a(210)+a(150);a(210)+a(210)};
	{a(150)+a(150)+a(210)+"u";a(210)+a(210)+a(150)**@{-}?(0.25)="pt25"?(0.75)="pt75"
		?(0.1)*@{<}?(0.5)*@{<}?(0.85)*@{<}
		?(0.15)+/_/*{j_9}?(0.5)+/_/*{j_9}?(0.89)+/_/*{j_9}}; 
	{\ar@{.>} "pt25";"pt25"+(0.7,0)*+{0}}; 
	{\ar@{.>} "pt75";"pt75"+(0.7,0)*+{0}}; 
	a(150)+a(150)+a(90)+"u";a(150)+a(150)+"u"**@{-}?(0.4)="pt"?(0.25)*@{>}?(0.2)+/_/*{l_1}?(0.7)*@{>}+/_/*{l_1}; 
	{\ar@{.>} "pt";"pt"+(0.7,0)*+{0}}; 
	a(210)+a(210)+a(270);a(210)+a(210)**@{-}?(0.4)="pt"?(0.25)*@{>}?(0.2)+/^/*{l_2}?(0.7)*@{>}+/^/*{l_2}; 
	{\ar@{.>} "pt";"pt"+(0.7,0)*+{0}}; 
	{\arl^<<{l_8} a(150)+a(210)+a(150)+a(150)+"u";a(150)+a(210)+a(150)+"u"};
	{\arl_<<{l_9} a(210)+a(210)+a(210)+a(150);a(210)+a(210)+a(150)};
	\endxy
}

\newcommand{\DyonTwoPlaquette}{
	\xy
	0;<7mm,0mm>:
	<0mm,8mm>::
	(0,0.4)="u";
	(0,-0.4)="d";
	{\arl^{j_7} a(150)+a(150)+"u";a(150)+"u"};
	a(90)+a(90)+"u";a(90)+"u"**@{-}?(0.4)="pt"?(0.25)*@{>}?(0.2)+/_/*{l_6}?(0.7)*@{>}+/_/*{l_6}; 
	{\ar@{.>} "pt";"pt"+(0.7,0)*+{0}}; 
	{\arl_<<{l_5} a(30)+a(30)+"u";a(30)+"u"};
	{\arr^<<{l_4} a(330)+a(330);a(330)};
	a(270)+a(270);a(270)**@{-}?(0.4)="pt"?(0.25)*@{>}?(0.2)+/^/*{l_3}?(0.7)*@{>}+/^/*{l_3}; 
	{\ar@{.>} "pt";"pt"+(0.7,0)*+{0}}; 
	{\arl_{j_{11}} a(210)+a(210);a(210)};
	{\arr^{j_1} a(150)+"u";a(90)+"u"};
	{\arr^{j_6} a(90)+"u";a(30)+"u"};
	{a(30)+"u";a(330)**@{-}?(0.25)="pt25"?(0.75)="pt75"
		?(0.1)*@{<}?(0.5)*@{<}?(0.85)*@{<}
		?(0.15)+/_/*{j_5}?(0.5)+/_/*{j_5}?(0.89)+/_/*{j_5}}; 
	{\ar@{.>} "pt25";"pt25"+(0.7,0)*+{0}}; 
	{\ar@{.>} "pt75";"pt75"+(0.7,0)*+{0}}; 
	{\arr^{j_4} a(330);a(270)};
	{\arr^{j_3} a(270);a(210)};
	{a(210);a(150)+"u"**@{-}?(0.25)="pt25"?(0.75)="pt75"
		?(0.1)*@{>}?(0.5)*@{>}?(0.85)*@{>}
		?(0.15)+/^/*{j_2}?(0.5)+/^/*{j'_2}?(0.89)+/^/*{j_2}}; 
	{\ar@{->} "pt25";"pt25"+(-0.7,0)*+{p}}; 
	{\ar@{->} "pt75";"pt75"+(0.7,0)*+{q^*}}; 
	{\arr^{j_8} a(150)+a(150)+"u";a(150)+a(150)+a(210)+"u"};
	{\arr_{j_{10}} a(210)+a(210)+a(150);a(210)+a(210)};
	{a(150)+a(150)+a(210)+"u";a(210)+a(210)+a(150)**@{-}?(0.25)="pt25"?(0.75)="pt75"
		?(0.1)*@{<}?(0.5)*@{<}?(0.85)*@{<}
		?(0.15)+/_/*{j_9}?(0.5)+/_/*{j_9}?(0.89)+/_/*{j_9}}; 
	{\ar@{.>} "pt25";"pt25"+(0.7,0)*+{0}}; 
	{\ar@{.>} "pt75";"pt75"+(0.7,0)*+{0}}; 
	a(150)+a(150)+a(90)+"u";a(150)+a(150)+"u"**@{-}?(0.4)="pt"?(0.25)*@{>}?(0.2)+/_/*{l_1}?(0.7)*@{>}+/_/*{l_1}; 
	{\ar@{.>} "pt";"pt"+(0.7,0)*+{0}}; 
	a(210)+a(210)+a(270);a(210)+a(210)**@{-}?(0.4)="pt"?(0.25)*@{>}?(0.2)+/^/*{l_2}?(0.7)*@{>}+/^/*{l_2}; 
	{\ar@{.>} "pt";"pt"+(0.7,0)*+{0}}; 
	{\arl^<<{l_8} a(150)+a(210)+a(150)+a(150)+"u";a(150)+a(210)+a(150)+"u"};
	{\arl_<<{l_9} a(210)+a(210)+a(210)+a(150);a(210)+a(210)+a(150)};
	\endxy
}

\newcommand{\TwoDyonTwoPlaquette}{
	\xy
	0;<7mm,0mm>:
	<0mm,8mm>::
	(0,0.4)="u";
	(0,-0.4)="d";
	{\arl^{j_7} a(150)+a(150)+"u";a(150)+"u"};
	a(90)+a(90)+"u";a(90)+"u"**@{-}?(0.4)="pt"?(0.25)*@{>}?(0.2)+/_/*{l_6}?(0.7)*@{>}+/_/*{l_6}; 
	{\ar@{.>} "pt";"pt"+(0.7,0)*+{0}}; 
	{\arl_<<{l_5} a(30)+a(30)+"u";a(30)+"u"};
	{\arr^<<{l_4} a(330)+a(330);a(330)};
	a(270)+a(270);a(270)**@{-}?(0.4)="pt"?(0.25)*@{>}?(0.2)+/^/*{l_3}?(0.7)*@{>}+/^/*{l_3}; 
	{\ar@{.>} "pt";"pt"+(0.7,0)*+{0}}; 
	{\arl_{j_{11}} a(210)+a(210);a(210)};
	{\arr^{j_1} a(150)+"u";a(90)+"u"};
	{\arr^{j_6} a(90)+"u";a(30)+"u"};
	{a(30)+"u";a(330)**@{-}?(0.25)="pt25"?(0.75)="pt75"
		?(0.1)*@{<}?(0.5)*@{<}?(0.85)*@{<}
		?(0.15)+/_/*{j_5}?(0.5)+/_/*{j_5}?(0.89)+/_/*{j_5}}; 
	{\ar@{.>} "pt25";"pt25"+(0.7,0)*+{0}}; 
	{\ar@{.>} "pt75";"pt75"+(0.7,0)*+{0}}; 
	{\arr^{j_4} a(330);a(270)};
	{\arr^{j_3} a(270);a(210)};
	{a(210);a(150)+"u"**@{-}?(0.25)="pt25"?(0.75)="pt75"
		?(0.1)*@{<}?(0.5)*@{<}?(0.85)*@{<}
		?(0.15)+/^/*{j_2}?(0.5)+/^/*{j'_2}?(0.89)+/^/*{j_2}}; 
	{\ar@{->} "pt25";"pt25"+(-0.7,0)*{p^{\prime*}}}; 
	{\ar@{->} "pt75";"pt75"+(0.7,0)*+{q'}}; 
	{\arr^{j_8} a(150)+a(150)+"u";a(150)+a(150)+a(210)+"u"};
	{\arr_{j_{10}} a(210)+a(210)+a(150);a(210)+a(210)};
	{a(150)+a(150)+a(210)+"u";a(210)+a(210)+a(150)**@{-}?(0.25)="pt25"?(0.75)="pt75"
		?(0.1)*@{<}?(0.5)*@{<}?(0.85)*@{<}
		?(0.15)+/_/*{j_9}?(0.5)+/_/*{j'_9}?(0.89)+/_/*{j_9}}; 
	{\ar@{->} "pt25";"pt25"+(0.7,0)*+{q}}; 
	{\ar@{->} "pt75";"pt75"+(-0.7,0)*{p^*}}; 
	a(150)+a(150)+a(90)+"u";a(150)+a(150)+"u"**@{-}?(0.4)="pt"?(0.25)*@{>}?(0.2)+/_/*{l_1}?(0.7)*@{>}+/_/*{l_1}; 
	{\ar@{.>} "pt";"pt"+(0.7,0)*+{0}}; 
	a(210)+a(210)+a(270);a(210)+a(210)**@{-}?(0.4)="pt"?(0.25)*@{>}?(0.2)+/^/*{l_2}?(0.7)*@{>}+/^/*{l_2}; 
	{\ar@{.>} "pt";"pt"+(0.7,0)*+{0}}; 
	{\arl^<<{l_8} a(150)+a(210)+a(150)+a(150)+"u";a(150)+a(210)+a(150)+"u"};
	{\arl_<<{l_9} a(210)+a(210)+a(210)+a(150);a(210)+a(210)+a(150)};
	\endxy
}

\newcommand{\GroundStateTailedThreePlaquette}{
	\xy
	0;<7mm,0mm>:
	<0mm,8mm>::
	(0,0.4)="u";
	(0,-0.4)="d";
	{\arl^{j_7} a(150)+a(150)+"u";a(150)+"u"};
	a(90)+a(90)+"u";a(90)+"u"**@{-}?(0.4)="pt"?(0.25)*@{>}?(0.2)+/_/*{l_6}?(0.7)*@{>}+/_/*{l_6}; 
	{\ar@{.>} "pt";"pt"+(0.7,0)*+{0}}; 
	{\arl_<<{l_5} a(30)+a(30)+"u";a(30)+"u"};
	{\arr^<<{l_4} a(330)+a(330);a(330)};
	a(270)+a(270);a(270)**@{-}?(0.4)="pt"?(0.25)*@{>}?(0.2)+/^/*{l_3}?(0.7)*@{>}+/^/*{l_3}; 
	{\ar@{.>} "pt";"pt"+(0.7,0)*+{0}}; 
	{\arl_{j_{11}} a(210)+a(210);a(210)};
	{\arr^{j_1} a(150)+"u";a(90)+"u"};
	{\arr^{j_6} a(90)+"u";a(30)+"u"};
	{a(30)+"u";a(330)**@{-}?(0.25)="pt25"?(0.75)="pt75"
		?(0.1)*@{<}?(0.5)*@{<}?(0.85)*@{<}
		?(0.15)+/_/*{j_5}?(0.5)+/_/*{j_5}?(0.89)+/_/*{j_5}}; 
	{\ar@{.>} "pt25";"pt25"+(0.7,0)*+{0}}; 
	{\ar@{.>} "pt75";"pt75"+(0.7,0)*+{0}}; 
	{\arr^{j_4} a(330);a(270)};
	{\arr^{j_3} a(270);a(210)};
	{a(210);a(150)+"u"**@{-}?(0.25)="pt25"?(0.75)="pt75"
		?(0.1)*@{>}?(0.5)*@{>}?(0.85)*@{>}
		?(0.15)+/^/*{j_2}?(0.5)+/^/*{j_2}?(0.89)+/^/*{j_2}}; 
	{\ar@{.>} "pt25";"pt25"+(0.7,0)*+{0}}; 
	{\ar@{.>} "pt75";"pt75"+(0.7,0)*+{0}}; 
	{\arr^{j_8} a(150)+a(150)+"u";a(150)+a(150)+a(210)+"u"};
	{\arr_{j_{10}} a(210)+a(210)+a(150);a(210)+a(210)};
	{a(150)+a(150)+a(210)+"u";a(210)+a(210)+a(150)**@{-}?(0.25)="pt25"?(0.75)="pt75"
		?(0.1)*@{<}?(0.5)*@{<}?(0.85)*@{<}
		?(0.15)+/_/*{j_9}?(0.5)+/_/*{j_9}?(0.89)+/_/*{j_9}}; 
	{\ar@{.>} "pt25";"pt25"+(0.7,0)*+{0}}; 
	{\ar@{.>} "pt75";"pt75"+(0.7,0)*+{0}}; 
	a(150)+a(150)+a(90)+"u";a(150)+a(150)+"u"**@{-}?(0.4)="pt"?(0.25)*@{>}?(0.2)+/_/*{l_1}?(0.7)*@{>}+/_/*{l_1}; 
	{\ar@{.>} "pt";"pt"+(0.7,0)*+{0}}; 
	a(210)+a(210)+a(270);a(210)+a(210)**@{-}?(0.4)="pt"?(0.25)*@{>}?(0.2)+/^/*{l_2}?(0.7)*@{>}+/^/*{l_2}; 
	{\ar@{.>} "pt";"pt"+(0.7,0)*+{0}}; 
	{\arl^<<{l_8} a(150)+a(210)+a(150)+a(150)+"u";a(150)+a(210)+a(150)+"u"};
	{\arl_<<{l_9} a(210)+a(210)+a(210)+a(150);a(210)+a(210)+a(150)};
	{\arr^{} a(150)+a(150)+a(150)+a(210)+"u";a(150)+a(150)+a(210)+a(150)+a(210)+"u"};
	{\arr_{} a(210)+a(210)+a(150)+a(150)+a(210);a(210)+a(210)+a(150)+a(210)};
	{a(150)+a(150)+a(210)+a(150)+a(210)+"u";a(210)+a(210)+a(150)+a(150)+a(210)**@{-}?(0.25)="pt25"?(0.75)="pt75"
		?(0.1)*@{<}?(0.5)*@{<}?(0.85)*@{<}
		?(0.15)+/_/*{}?(0.5)+/_/*{}?(0.89)+/_/*{}}; 
	{\ar@{.>} "pt25";"pt25"+(0.7,0)*+{}}; 
	{\ar@{.>} "pt75";"pt75"+(0.7,0)*+{}}; 
	a(150)+a(150)+a(90)+a(150)+a(210)+"u";a(150)+a(150)+a(150)+a(210)+"u"**@{-}?(0.4)="pt"?(0.25)*@{>}?(0.2)+/_/*{}?(0.7)*@{>}+/_/*{}; 
	{\ar@{.>} "pt";"pt"+(0.7,0)*+{}}; 
	a(210)+a(210)+a(270)+a(150)+a(210);a(210)+a(210)+a(150)+a(210)**@{-}?(0.4)="pt"?(0.25)*@{>}?(0.2)+/^/*{}?(0.7)*@{>}+/^/*{}; 
	{\ar@{.>} "pt";"pt"+(0.7,0)*+{}}; 
	{\arl^<<{} a(150)+a(210)+a(150)+a(150)+a(150)+a(210)+"u";a(150)+a(210)+a(150)+a(150)+a(210)+"u"};
	{\arl_<<{} a(210)+a(210)+a(210)+a(150)+a(150)+a(210);a(210)+a(210)+a(150)+a(150)+a(210)};
	\endxy
}

\newcommand{\TwoDyonThreePlaquette}{
	\xy
	0;<7mm,0mm>:
	<0mm,8mm>::
	(0,0.4)="u";
	(0,-0.4)="d";
	{\arl^{j_7} a(150)+a(150)+"u";a(150)+"u"};
	a(90)+a(90)+"u";a(90)+"u"**@{-}?(0.4)="pt"?(0.25)*@{>}?(0.2)+/_/*{l_6}?(0.7)*@{>}+/_/*{l_6}; 
	{\ar@{.>} "pt";"pt"+(0.7,0)*+{0}}; 
	{\arl_<<{l_5} a(30)+a(30)+"u";a(30)+"u"};
	{\arr^<<{l_4} a(330)+a(330);a(330)};
	a(270)+a(270);a(270)**@{-}?(0.4)="pt"?(0.25)*@{>}?(0.2)+/^/*{l_3}?(0.7)*@{>}+/^/*{l_3}; 
	{\ar@{.>} "pt";"pt"+(0.7,0)*+{0}}; 
	{\arl_{j_{11}} a(210)+a(210);a(210)};
	{\arr^{j_1} a(150)+"u";a(90)+"u"};
	{\arr^{j_6} a(90)+"u";a(30)+"u"};
	{a(30)+"u";a(330)**@{-}?(0.25)="pt25"?(0.75)="pt75"
		?(0.1)*@{<}?(0.5)*@{<}?(0.85)*@{<}
		?(0.15)+/_/*{j_5}?(0.5)+/_/*{j_5}?(0.89)+/_/*{j_5}}; 
	{\ar@{.>} "pt25";"pt25"+(0.7,0)*+{0}}; 
	{\ar@{.>} "pt75";"pt75"+(0.7,0)*+{0}}; 
	{\arr^{j_4} a(330);a(270)};
	{\arr^{j_3} a(270);a(210)};
	{a(210);a(150)+"u"**@{-}?(0.25)="pt25"?(0.75)="pt75"
		?(0.1)*@{>}?(0.5)*@{>}?(0.85)*@{>}
		?(0.15)+/^/*{j_2}?(0.5)+/^/*{j'_2}?(0.89)+/^/*{j_2}}; 
	{\ar@{->} "pt25";"pt25"+(-0.7,0)*{q^{\prime}}}; 
	{\ar@{->} "pt75";"pt75"+(0.7,0)*+{q^*}}; 
	{\arr^{j_8} a(150)+a(150)+"u";a(150)+a(150)+a(210)+"u"};
	{\arr_{j_{10}} a(210)+a(210)+a(150);a(210)+a(210)};
	{a(150)+a(150)+a(210)+"u";a(210)+a(210)+a(150)**@{-}?(0.25)="pt25"?(0.75)="pt75"
		?(0.1)*@{<}?(0.5)*@{<}?(0.85)*@{<}
		?(0.15)+/_/*{j_9}?(0.5)+/_/*{j'_9}?(0.89)+/_/*{j_9}}; 
	{\ar@{->} "pt25";"pt25"+(0.7,0)*+{q^{\prime*}}}; 
	{\ar@{->} "pt75";"pt75"+(-0.7,0)*{p}}; 
	a(150)+a(150)+a(90)+"u";a(150)+a(150)+"u"**@{-}?(0.4)="pt"?(0.25)*@{>}?(0.2)+/_/*{l_1}?(0.7)*@{>}+/_/*{l_1}; 
	{\ar@{.>} "pt";"pt"+(0.7,0)*+{0}}; 
	a(210)+a(210)+a(270);a(210)+a(210)**@{-}?(0.4)="pt"?(0.25)*@{>}?(0.2)+/^/*{l_2}?(0.7)*@{>}+/^/*{l_2}; 
	{\ar@{.>} "pt";"pt"+(0.7,0)*+{0}}; 
	{\arl^<<{l_8} a(150)+a(210)+a(150)+a(150)+"u";a(150)+a(210)+a(150)+"u"};
	{\arl_<<{l_9} a(210)+a(210)+a(210)+a(150);a(210)+a(210)+a(150)};
	{\arr^{} a(150)+a(150)+a(210)+a(150)+"u";a(150)+a(150)+a(210)+a(210)+a(150)+"u"};
	{\arr_{} a(210)+a(210)+a(150)+a(210)+a(150);a(210)+a(210)+a(210)+a(150)};
	{a(150)+a(150)+a(210)+a(210)+a(150)+"u";a(210)+a(210)+a(150)+a(210)+a(150)**@{-}?(0.25)="pt25"?(0.75)="pt75"
		?(0.1)*@{<}?(0.5)*@{<}?(0.85)*@{<}
		?(0.15)+/_/*{}?(0.5)+/_/*{}?(0.89)+/_/*{}}; 
	{\ar@{.>} "pt25";"pt25"+(0.7,0)*+{0}}; 
	{\ar@{.>} "pt75";"pt75"+(0.7,0)*+{0}}; 
	a(150)+a(150)+a(90)+a(210)+a(150)+"u";a(150)+a(150)+a(210)+a(150)+"u"**@{-}?(0.4)="pt"?(0.25)*@{>}?(0.2)+/_/*{}?(0.7)*@{>}+/_/*{}; 
	{\ar@{.>} "pt";"pt"+(0.7,0)*+{}}; 
	a(210)+a(210)+a(270)+a(210)+a(150);a(210)+a(210)+a(210)+a(150)**@{-}?(0.4)="pt"?(0.25)*@{>}?(0.2)+/^/*{}?(0.7)*@{>}+/^/*{}; 
	{\ar@{.>} "pt";"pt"+(0.7,0)*+{}}; 
	{\arl^<<{} a(150)+a(210)+a(150)+a(150)+a(210)+a(150)+"u";a(150)+a(210)+a(150)+a(210)+a(150)+"u"};
	{\arl_<<{} a(210)+a(210)+a(210)+a(150)+a(210)+a(150);a(210)+a(210)+a(150)+a(210)+a(150)};
	\endxy
}

\newcommand{\TwoPlaquettes}{
\xy
{
0;/r0.07pc/:;
(8,0)*{}="r";
(0,0)*{}="v1";
(0,20)*{}="v2";
(-13,30)*{}="v3";
(-26,20)*{}="v4";
(-26,0)*{}="v5";
(-13,-10)*{}="v6";
(13,-10)*{}="v7";
(26,0)*{}="v8";
(26,20)*{}="v9";
(13,30)*{}="v10";
"v1"; "v2"** \dir{-} ?(0.6) *\dir{>} +(-7,-1)*{j_e}?(0.3)*{}="pt3"?(0.7)*{}="pt7";
{\ar@{.>} "pt3";"pt3"+"r"};
{\ar@{.>} "pt7";"pt7"+"r"};
"v2"; "v3"** \dir{-} ?(0.6) *\dir{>};
"v4"; "v3"** \dir{-} ?(0.6) *\dir{>};
"v5"; "v4"** \dir{-} ?(0.6) *\dir{>}?(0.3)*{}="pt3"?(0.7)*{}="pt7";
{\ar@{.>} "pt3";"pt3"+"r"};
{\ar@{.>} "pt7";"pt7"+"r"};
"v6"; "v5"** \dir{-} ?(0.6) *\dir{>};
"v6"; "v1"** \dir{-} ?(0.6) *\dir{>};
"v7"; "v1"** \dir{-} ?(0.6) *\dir{>};
"v7"; "v8"** \dir{-} ?(0.6) *\dir{>};
"v8"; "v9"** \dir{-} ?(0.6) *\dir{>}?(0.3)*{}="pt3"?(0.7)*{}="pt7";
{\ar@{.>} "pt3";"pt3"+"r"};
{\ar@{.>} "pt7";"pt7"+"r"};
"v9"; "v10"** \dir{-} ?(0.6) *\dir{>};
"v2"; "v10"** \dir{-} ?(0.6) *\dir{>};}
\endxy
}

\newcommand{\latticeBare}
{
\renewcommand\latticebody
{\drop{{\ar@{-}|@{>} c+<-0.866025pc,-0.5pc>;c}
{\ar@{-}|@{>} c+<0.866025pc,-0.5pc>;c}{\ar@{-}|@{>} c;c+<0pc,1.2pc>}}}
\xy
0;<0.866025pc,0pc>:<0.433013pc,0.85pc>::;
(0,10)*{};
\croplattice{-6}5{-5}{6}%
            {-2}{2}{1}{4}
\endxy
}

\newcommand{\latticeWithTail}
{\renewcommand\latticebody
{\drop{{\ar@{-}|@{>} c+<-0.866025pc,-0.5pc>;c}
{\ar@{-}|@{>} c+<0.866025pc,-0.5pc>;c}
{\ar@{-}|@{>}|(0.16)@{>}|(0.9)@{>} c;c+<0pc,1.2pc>}
{\ar c+<0pc,0.3pc>;c+<0.4pc,0.3pc>}
{\ar c+<0pc,0.7pc>;c+<0.4pc,0.7pc>}}}
\xy
0;<0.866025pc,0pc>:<0.433013pc,0.85pc>::;
(0,10)*{};
\croplattice{-5}5{-5}{6}%
            {-2}{2}{1}{4}
\endxy
}

\newcommand{\VertexA}
{
\def\objectstyle{\scriptscriptstyle}
\def\labelstyle{\scriptscriptstyle}
\xy
0;<8mm,0mm>:
<0mm,8mm>::
0="o";
a(150)="l";
a(30)="r";
a(270)="b";
(0,-0.5)="i";
(0.55,-0.5)*+{q}="f";
\xygraph{
"l"-"o"|@{>}^{i}
"r"-"o"|@{>}_{j}
"b"-"o"|(0.25)@{>}|(0.75)@{>}^(0.25){l}^(0.75){u}
"i":"f"
}
\endxy}

\newcommand{\VertexB}
{
\def\objectstyle{\scriptscriptstyle}
\def\labelstyle{\scriptscriptstyle}
\xy
0;<8mm,0mm>:
<0mm,8mm>::
0="o";
a(150)="l";
a(30)="r";
a(270)="b";
(0.433013,0.25)="i";
(0.433013,0.25)+(0.3,-0.51962)*+{q}="f";
\xygraph{
"l"-"o"|@{>}_{i}
"b"-"o"|@{>}^{l}
"r"-"o"|(0.25)@{>}|(0.75)@{>}_(0.25){j}_(0.75){v}
"i":"f"
}
\endxy}

\newcommand{\VertexC}
{
\def\objectstyle{\scriptscriptstyle}
\def\labelstyle{\scriptscriptstyle}
\xy
0;<8mm,0mm>:
<0mm,8mm>::
0="o";
a(150)="l";
a(30)="r";
a(270)="b";
(0.433013,0.25)="i";
(0.433013,0.25)-(0.3,-0.51962)*+{q}="f";
\xygraph{
"l"-"o"|@{>}_{i}
"b"-"o"|@{>}^{l}
"r"-"o"|(0.25)@{>}|(0.75)@{>}^(0.25){j}^(0.75){v'}
"i":"f"
}
\endxy}

\newcommand{\VertexD}
{
\def\objectstyle{\scriptscriptstyle}
\def\labelstyle{\scriptscriptstyle}
\xy
0;<8mm,0mm>:
<0mm,8mm>::
0="o";
a(150)="l";
a(30)="r";
a(270)="b";
(-0.433013,0.25)="i";
(-0.433013,0.25)+(0.3,0.51962)*+{q}="f";
\xygraph{
"b"-"o"|@{>}_{l}
"r"-"o"|@{>}^{j}
"l"-"o"|(0.25)@{>}|(0.75)@{>}_(0.25){i}_(0.75){w}
"i":"f"
}
\endxy}

\newcommand{\VertexE}
{
\def\objectstyle{\scriptscriptstyle}
\def\labelstyle{\scriptscriptstyle}
\xy
0;<8mm,0mm>:
<0mm,8mm>::
0="o";
a(150)="l";
a(30)="r";
a(270)="b";
(-0.433013,0.25)="i";
(-0.433013,0.25)-(0.3,0.51962)*+{q}="f";
\xygraph{
"b"-"o"|@{>}_{l}
"r"-"o"|@{>}^{j}
"l"-"o"|(0.25)@{>}|(0.75)@{>}^(0.25){i}^(0.75){w'}
"i":"f"
}
\endxy}

\newcommand{\VertexF}
{
\def\objectstyle{\scriptscriptstyle}
\def\labelstyle{\scriptscriptstyle}
\xy
0;<8mm,0mm>:
<0mm,8mm>::
0="o";
a(150)="l";
a(30)="r";
a(270)="b";
(0,-0.5)="i";
(-0.55,-0.5)*+{q}="f";
 \xygraph{
 "l"-"o"|@{>}^{i}
 "r"-"o"|@{>}_{j}
 "b"-"o"|(0.25)@{>}|(0.75)@{>}_(0.25){l}_(0.75){u'}
 "i":"f"
 }
\endxy}

\newcommand{\BareCircle}{
\xy
0;<8mm,0mm>:
<0mm,8mm>::
(-0.3,0)*{}="l";
{\ar@{-}@`{a(0)+a(90),a(90),a(90)+a(180)} a(0)*@{*};a(180)*@{*}};
{\ar@{-}@`{a(180)+a(-90),a(-90),a(-90)+a(0)} a(180);a(0)};
a(0)+"l"*{v_1};
a(180)+"l"*{v_2};
(0,-0.3)*{p_1};
(-2,-0.3)*{p_2};
\endxy
}

\newcommand{\EqTailedCircle}{
\xy
0;<5mm,0mm>:
<0mm,5mm>::
(-0.4,0)*{}="l";
{\arl^{x}@`{a(0)+a(90),a(90),a(90)+a(180)} a(0);a(180)};
{\arl^{y}@`{a(180)+a(-90),a(-90),a(-90)+a(0)} a(180);a(0)};
{\ar_>{q} a(180);a(180)+"l"};
{\ar a(0);a(0)+"l"+"l"*+{p^*}};
\endxy
}

\newcommand{\EqTailedCirclePrimeC}{
	\xy
	0;<5mm,0mm>:
	<0mm,5mm>::
	(-0.4,0)*{}="l";
	{\arl^{x}@`{a(0)+a(90),a(90),a(90)+a(180)} a(0);a(180)};
	{\arl^{y}@`{a(180)+a(-90),a(-90),a(-90)+a(0)} a(180);a(0)};
	{\ar_>{q'} a(180);a(180)+"l"};
	{\ar a(0);a(0)+"l"+"l"*+{p^{\prime*}}};
	\endxy
}

\newcommand{\GroundStateTailedCircle}{
	\xy
	0;<5mm,0mm>:
	<0mm,5mm>::
	(-0.4,0)*{}="l";
	{\arl^{x}@`{a(0)+a(90),a(90),a(90)+a(180)} a(0);a(180)};
	{\arl^{x}@`{a(180)+a(-90),a(-90),a(-90)+a(0)} a(180);a(0)};
	{\ar@{.>}_>{0} a(180);a(180)+"l"};
	{\ar@{.>} a(0);a(0)+"l"+"l"*+{0}};
	\endxy
}

\newcommand{\FourTailsCircleZero}{
	\xy
	0;<7mm,0mm>:
	<0mm,5mm>::
	(-0.7,0)*{}="l";
	(0.7,0)*{}="r";
	a(90)*{};a(270)*{} **\crv{a(90)&a(90)+a(180)&a(270)+a(180)&a(270)} ?(0.1)*\dir{>}+/_/*{x} ?(0.4)*{}="lu" ?(0.51)*\dir{>}+/^/*{w} ?(0.6)*{}="ld";
	{\ar@{.>} "ld";"ld"+"r"*+{0}};
	{\ar "lu";"lu"+"l"*+{p^*_1}};
	a(90)*{};a(270)*{} **\crv{a(90)&a(90)+a(0)&a(270)+a(0)&a(270)} ?(0.4)*{}="ru" ?(0.49)*\dir{<}+/^/*{w} ?(0.6)*{}="rd" ?(0.9)*\dir{<}+/^/*{w};
	{\ar@{.>} "rd";"rd"+"l"*+{0}};
	{\ar "ru";"ru"+"r"*+{q_2}};
	\endxy
}

\newcommand{\EqTailedCirclePrime}{
\xy
0;<5mm,0mm>:
<0mm,5mm>::
(-0.4,0)*{}="l";
{\arl^{x'}@`{a(0)+a(90),a(90),a(90)+a(180)} a(0);a(180)};
{\arl^{y'}@`{a(180)+a(-90),a(-90),a(-90)+a(0)} a(180);a(0)};
{\ar_>{q} a(180);a(180)+"l"};
{\ar a(0);a(0)+"l"+"l"*+{p^*}};
\endxy
}

\newcommand{\EqTailedCircleTwistPrime}{
	\xy
	0;<5mm,0mm>:
	<0mm,5mm>::
	(-0.4,0)*{}="l";
	{\arl^{y}@`{a(0)+a(90),a(90),a(90)+a(180)} a(0);a(180)};
	{\arl^{y'}@`{a(180)+a(-90),a(-90),a(-90)+a(0)} a(180);a(0)};
	{\ar_>{q} a(180);a(180)+"l"};
	{\ar a(0);a(0)+"l"+"l"*+{p^*}};
	\endxy
}

\newcommand{\tetrahedron}[6]{
	\xy
		0;<8mm,0mm>:
		<0mm,8mm>::
		(0,0)="a";
		(0.8,-0.75)="b";
		(2.5,0)="c";
		(1.25,2)="d";
		{\arl_{#1} "a";"b"};
		{\arl_{#2} "b";"c"};
		{\arl^{#3}|(0.64)\hole "c";"a"};
		{\arl_{#4} "c";"d"};
		{\arl_{#5} "d";"a"};
		{\arl_{#6} "b";"d"};
	\endxy
}

\newcommand{\ChargeAnnihilateA}{
	\xy
	0;/r0.25pc/:;
	(12,16)*{}="12v16";
	(12,34)*{}="12v34";
	(14,21)*{}="14v21";
	(14,26)*{}="14v26";
	(14,30)*{}="14v30";
	(16,19)*{}="16v19";
	(16,23)*{}="16v23";
	(16,28)*{}="16v28";
	(16,31)*{}="16v31";
	(20,16)*{}="20v16";
	(19,23)*{}="19v23";
	(19,28)*{}="19v28";
	(20,34)*{}="20v34";
	(21,23)*{}="21v23";
	(21,28)*{}="21v28";
	"14v26"*{\scriptstyle j};
	"14v30"*{\scriptstyle k_1};
	"14v21"*{\scriptstyle k_2};
	"21v28"*{\scriptstyle q};
	"21v23"*{\scriptstyle q^*};
	{\ar@{>} "16v23";"19v23"};
	{\ar@{>} "16v28";"19v28"};
	{\ar@{-}|@{>} "12v16";"16v19"};
	{\ar@{-}|@{>} "12v34";"16v31"};
	{\ar@{-}|@{>} "16v23";"16v19"};
	{\ar@{-}|@{>} "16v28";"16v23"};
	{\ar@{-}|@{>} "16v31";"16v28"};
	{\ar@{-}|@{>} "20v16";"16v19"};
	{\ar@{-}|@{>} "20v34";"16v31"};
	\endxy
}

\newcommand{\ChargeAnnihilateB}{
	\xy
	0;/r0.25pc/:;
	(12,16)*{}="12v16";
	(12,34)*{}="12v34";
	(14,21)*{}="14v21";
	(14,26)*{}="14v26";
	(14,30)*{}="14v30";
	(16,19)*{}="16v19";
	(16,23)*{}="16v23";
	(16,28)*{}="16v28";
	(16,31)*{}="16v31";
	(20,16)*{}="20v16";
	(19,23)*{}="19v23";
	(19,28)*{}="19v28";
	(20,34)*{}="20v34";
	(21,23)*{}="21v23";
	(21,28)*{}="21v28";
	"14v26"*{\scriptstyle k_1};
	"14v30"*{\scriptstyle k_1};
	"14v21"*{\scriptstyle k_1};
	{\ar@{.>} "16v23";"19v23"};
	{\ar@{.>} "16v28";"19v28"};
	{\ar@{-}|@{>} "12v16";"16v19"};
	{\ar@{-}|@{>} "12v34";"16v31"};
	{\ar@{-}|@{>} "16v23";"16v19"};
	{\ar@{-}|@{>} "16v28";"16v23"};
	{\ar@{-}|@{>} "16v31";"16v28"};
	{\ar@{-}|@{>} "20v16";"16v19"};
	{\ar@{-}|@{>} "20v34";"16v31"};
	\endxy
}

\newcommand{\WPhi}[1]{
\xy
0;<3pt,0pt>:
<0pt,4pt>::
(8,22)*{}="08v22";
(8,30)*{}="08v30";
(11,24)*{}="11v24";
(11,28)*{}="11v28";
(16,18)*{}="16v18";
(16,21)*{}="16v21";
(16,27)*{}="16v27";
(16,31)*{}="16v31";
(16,34)*{}="16v34";
(18,26)*{}="18v26";
(19,23)*{}="19v23";
(21,24)*{}="21v24";
(21,28)*{}="21v28";
(24,27)*{}="24v27";
(25,25)*{}="25v25";
(26,18)*{}="26v18";
(26,21)*{}="26v21";
(26,27)*{}="26v27";
(26,31)*{}="26v31";
(26,34)*{}="26v34";
(31,24)*{}="31v24";
(31,28)*{}="31v28";
(34,22)*{}="34v22";
(34,30)*{}="34v30";
"18v26"*@{*};
"25v25"*@{*};
"26v27"*{\scriptstyle 0};
"16v27"*{\scriptstyle 0};
{\ar@{-} "11v24";"08v22"};
{\ar@{-} "11v24";"16v21"};
{\ar@{-} "11v28";"08v30"};
{\ar@{-} "11v28";"11v24"};
{\ar@{-} "16v21";"16v18"};
{\ar@{-} "16v21";"21v24"};
{\ar@{-} "16v31";"11v28"};
{\ar@{-} "16v31";"16v34"};
{\ar@{-} "16v31";"21v28"};
{\ar@{-} "21v24";"26v21"};
{\ar@{-} "21v28";"26v31"};
{\ar@{-} "26v21";"26v18"};
{\ar@{-} "26v31";"26v34"};
{\ar@{-} "26v31";"31v28"};
{\ar@{-} "31v24";"26v21"};
{\ar@{-} "31v24";"34v22"};
{\ar@{-} "31v28";"31v24"};
{\ar@{-} "31v28";"34v30"};
{\ar@{-}|(0.3){\hole}|(0.7)@{#1} "21v24";"21v28"};
{\ar@{=}@`{"19v23", "24v27"}|(0.7){\SelectTips{cm}{12}\object@{>>}} "18v26";"25v25"};
(23.5,23.5)*{\scriptstyle A};
\endxy
}

\newcommand{\WWA}{
	\xy
	0;<4pt,0pt>:
	<0pt,4pt>::
	(17,23)*{}="17v23";
	(17,28)*{}="17v28";
	(17,30)*{}="17v30";
	(21,22)*{}="21v22";
	(21,26)*{}="21v26";
	(21,30)*{}="21v30";
	(22,24)*{}="22v24";
	(25,23)*{}="25v23";
	(25,28)*{}="25v28";
	(25,30)*{}="25v30";
	"22v24"+(0.5,0)*{\scriptstyle e_1};
	"17v28"*{\scriptstyle e_2};
	"25v28"*{\scriptstyle e_3};
	"17v23"*{\scriptstyle p_0};
	"21v30"*{\scriptstyle p_1};
	"25v23"*{\scriptstyle p_2};
	{\ar@{-}|@{>} "21v22";"21v26"};
	{\ar@{-}|@{>} "21v26";"17v30"};
	{\ar@{-}|@{>} "21v26";"25v30"};
	\endxy
	}

\newcommand{\WWB}{
\xy
0;<5pt,0pt>:
<0pt,4pt>::
(17,30)*{}="17v30";
(18,24)*{}="18v24";
(18,31)*{}="18v31";
(20,24)*{}="20v24";
(20,31)*{}="20v31";
(21,22)*{}="21v22";
(21,26)*{}="21v26";
(22,24)*{}="22v24";
(22,31)*{}="22v31";
(24,24)*{}="24v24";
(24,31)*{}="24v31";
(25,30)*{}="25v30";
"18v24"*@{*};
"20v31"*@{*};
"22v31"*@{*};
"24v24"*@{*};
"20v24"+(-2,2)*{\scriptstyle A};
"22v24"+(2,2)*{\scriptstyle A};
"18v31"*{\scriptstyle 0};
"22v24"*{\scriptstyle 0};
"20v24"*{\scriptstyle 0};
"24v31"*{\scriptstyle 0};
{\ar@{-}|@{>} "21v22";"21v26"};
{\ar@{-}|(0.5){\hole}|(0.8)@{>} "21v26";"17v30"};
{\ar@{-}|(0.5){\hole}|(0.8)@{>} "21v26";"25v30"};
{\ar@{=}@`{"20v24", "18v31"}|(0.35){\SelectTips{cm}{12}\object@{>>}} "18v24";"20v31"};
{\ar@{=}@`{"24v31", "22v24"}|(0.75){\SelectTips{cm}{12}\object@{>>}} "22v31";"24v24"};
\endxy
}

\newcommand{\WWC}{
	\xy
	0;<5pt,0pt>:
	<0pt,4pt>::
	(17,30)*{}="17v30";
	(18,24)*{}="18v24";
	(18,31)*{}="18v31";
	(20,24)*{}="20v24";
	(20,31)*{}="20v31";
	(21,22)*{}="21v22";
	(21,26)*{}="21v26";
	(21,31)*{}="21v31";
	(22,24)*{}="22v24";
	(22,31)*{}="22v31";
	(24,24)*{}="24v24";
	(24,31)*{}="24v31";
	(25,30)*{}="25v30";
	"18v24"*@{*};
	"20v31"*@{*};
	"22v31"*@{*};
	"24v24"*@{*};
	"20v24"+(-2,2)*{\scriptstyle A};
	"22v24"+(2,2)*{\scriptstyle A};
	"18v31"*{\scriptstyle 0};
	"22v24"*{\scriptstyle 0};
	"20v24"*{\scriptstyle 0};
	"24v31"*{\scriptstyle 0};
	"21v31"*{\xycircle(4,2.2){.}};
	{\ar@{-}|@{>} "21v22";"21v26"};
	{\ar@{-}|(0.5){\hole}|(0.8)@{>} "21v26";"17v30"};
	{\ar@{-}|(0.5){\hole}|(0.8)@{>} "21v26";"25v30"};
	{\ar@{=}@`{"20v24", "18v31"}|(0.35){\SelectTips{cm}{12}\object@{>>}} "18v24";"20v31"};
	{\ar@{=}@`{"24v31", "22v24"}|(0.75){\SelectTips{cm}{12}\object@{>>}} "22v31";"24v24"};
	\endxy
}

\newcommand{\WWD}{
\xy
0;<5pt,0pt>:
<0pt,4pt>::
(17,30)*{}="17v30";
(18,24)*{}="18v24";
(18,31)*{}="18v31";
(20,24)*{}="20v24";
(20,31)*{}="20v31";
(21,22)*{}="21v22";
(21,26)*{}="21v26";
(21,31)*{}="21v31";
(22,24)*{}="22v24";
(22,31)*{}="22v31";
(24,24)*{}="24v24";
(24,31)*{}="24v31";
(25,30)*{}="25v30";
(28,26)*{}="28v26";
(32,30)*{}="32v30";
(33,24)*{}="33v24";
(34,23)*{}="34v23";
(34,27)*{}="34v27";
(36,22)*{}="36v22";
(36,26)*{}="36v26";
(38,22)*{}="38v22";
(38,25)*{}="38v25";
(39,24)*{}="39v24";
(40,30)*{}="40v30";
"18v24"*@{*};
"24v24"*@{*};
"33v24"*@{*};
"39v24"*@{*};
"38v25"+(-1,-3)*{\scriptstyle A};
(21,33)*{\scriptstyle A};
"28v26"*{\scriptstyle =};
"22v24"*{\scriptstyle 0};
"38v25"*{\scriptstyle 0};
"20v24"*{\scriptstyle 0};
"34v23"*{\scriptstyle 0};
{\ar@{-}|@{>} "21v22";"21v26"};
{\ar@{-}|(0.45){\hole}|(0.8)@{>} "21v26";"17v30"};
{\ar@{-}|(0.45){\hole}|(0.8)@{>} "21v26";"25v30"};
{\ar@{-}|(0.35)@{>}|(0.65){\hole} "36v22";"36v26"};
{\ar@{-}|@{>} "36v26";"32v30"};
{\ar@{-}|@{>} "36v26";"40v30"};
{\ar@{=}@`{"34v27", "38v22"}|(0.75){\SelectTips{cm}{12}\object@{>>}} "33v24";"39v24"};
{\ar@{=}@`{"20v24", "18v31",  "20v31", "22v31", "24v31", "22v24"}|(0.51){\SelectTips{cm}{12}\object@{>>}} "18v24";"24v24"};
\endxy
}

\newcommand{\QDlinePair}{
\xy
0;/r0.3pc/:;
(2,24)*{}="02v24";
(2,35)*{}="02v35";
(6,27)*{}="06v27";
(6,32)*{}="06v32";
(11,19)*{}="11v19";
(11,23)*{}="11v23";
(11,36)*{}="11v36";
(11,40)*{}="11v40";
(13,28)*{}="13v28";
(13,30)*{}="13v30";
(15,27)*{}="15v27";
(16,27)*{}="16v27";
(16,32)*{}="16v32";
(17,32)*{}="17v32";
(19,31)*{}="19v31";
(21,19)*{}="21v19";
(21,23)*{}="21v23";
(21,36)*{}="21v36";
(21,40)*{}="21v40";
(26,27)*{}="26v27";
(26,32)*{}="26v32";
(31,19)*{}="31v19";
(31,23)*{}="31v23";
(31,36)*{}="31v36";
(31,40)*{}="31v40";
(36,27)*{}="36v27";
(36,32)*{}="36v32";
(40,24)*{}="40v24";
(40,35)*{}="40v35";
"13v28"*@{*};
"19v31"*@{*};
"13v30"*{\scriptstyle J};
{\ar@{-} "06v27";"02v24"};
{\ar@{-} "06v27";"11v23"};
{\ar@{-} "06v32";"02v35"};
{\ar@{-} "06v32";"06v27"};
{\ar@{-} "11v23";"11v19"};
{\ar@{-} "11v23";"16v27"};
{\ar@{-} "11v36";"06v32"};
{\ar@{-} "11v36";"11v40"};
{\ar@{-} "11v36";"16v32"};
{\ar@{-}|(0.5){\hole} "16v27";"16v32"};
(17,28)*{\scriptstyle e};
{\ar@{-} "16v27";"21v23"};
{\ar@{-} "16v32";"21v36"};
{\ar@{-} "21v23";"21v19"};
{\ar@{-} "21v36";"21v40"};
{\ar@{-} "21v36";"26v32"};
{\ar@{-} "26v27";"21v23"};
{\ar@{-} "26v27";"31v23"};
{\ar@{-} "26v32";"26v27"};
{\ar@{-} "26v32";"31v36"};
{\ar@{-} "31v23";"31v19"};
{\ar@{-} "31v23";"36v27"};
{\ar@{-} "31v36";"31v40"};
{\ar@{-} "36v27";"36v32"};
{\ar@{-} "36v27";"40v24"};
{\ar@{-} "36v32";"31v36"};
{\ar@{-} "36v32";"40v35"};
{\ar@2{-}@`{"15v27", "17v32"}|(0.7){\SelectTips{cm}{12}\object@{>>}} "13v28";"19v31"};
(12,28)*{\scriptstyle p};
(20.5,31)*{\scriptstyle q^*};
\endxy
}

\newcommand{\QDlineTwoPairs}{
\xy
0;/r0.23pc/:;
(2,24)*{}="02v24";
(2,35)*{}="02v35";
(6,27)*{}="06v27";
(6,32)*{}="06v32";
(11,19)*{}="11v19";
(11,23)*{}="11v23";
(11,36)*{}="11v36";
(11,40)*{}="11v40";
(13,28)*{}="13v28";
(13,30)*{}="13v30";
(15,27)*{}="15v27";
(16,27)*{}="16v27";
(16,32)*{}="16v32";
(17,32)*{}="17v32";
(19,31)*{}="19v31";
(21,19)*{}="21v19";
(21,23)*{}="21v23";
(21,36)*{}="21v36";
(21,40)*{}="21v40";
(23,28)*{}="23v28";
(25,27)*{}="25v27";
(26,27)*{}="26v27";
(26,32)*{}="26v32";
(27,32)*{}="27v32";
(29,29)*{}="29v29";
(29,31)*{}="29v31";
(31,19)*{}="31v19";
(31,23)*{}="31v23";
(31,36)*{}="31v36";
(31,40)*{}="31v40";
(36,27)*{}="36v27";
(36,32)*{}="36v32";
(40,24)*{}="40v24";
(40,35)*{}="40v35";
"13v28"*@{*};
"19v31"*@{*};
"23v28"*@{*};
"29v31"*@{*};
"13v30"*{\scriptstyle J};
"29v29"*{\scriptstyle J};
{\ar@{-} "06v27";"02v24"};
{\ar@{-} "06v27";"11v23"};
{\ar@{-} "06v32";"02v35"};
{\ar@{-} "06v32";"06v27"};
{\ar@{-} "11v23";"11v19"};
{\ar@{-} "11v23";"16v27"};
{\ar@{-} "11v36";"06v32"};
{\ar@{-} "11v36";"11v40"};
{\ar@{-} "11v36";"16v32"};
{\ar@{-}|(0.5){\hole} "16v27";"16v32"};
{\ar@{-} "16v27";"21v23"};
{\ar@{-} "16v32";"21v36"};
{\ar@{-} "21v23";"21v19"};
{\ar@{-} "21v36";"21v40"};
{\ar@{-} "21v36";"26v32"};
{\ar@{-} "26v27";"21v23"};
{\ar@{-} "26v27";"31v23"};
{\ar@{-}|(0.5){\hole} "26v32";"26v27"};
{\ar@{-} "26v32";"31v36"};
{\ar@{-} "31v23";"31v19"};
{\ar@{-} "31v23";"36v27"};
{\ar@{-} "31v36";"31v40"};
{\ar@{-} "36v27";"36v32"};
{\ar@{-} "36v27";"40v24"};
{\ar@{-} "36v32";"31v36"};
{\ar@{-} "36v32";"40v35"};
{\ar@{.} "19v31";"23v28"};
{\ar@{=}@`{"15v27", "17v32"}|(0.75){\SelectTips{cm}{12}\object@{>>}} "13v28";"19v31"};
{\ar@{=}@`{"25v27", "27v32"}|(0.75){\SelectTips{cm}{12}\object@{>>}} "23v28";"29v31"};
(43,29)*{=};
(11.5,28)*{\scriptstyle p};
(31,31)*{\scriptstyle q^*};
\endxy
}

\newcommand{\QDlineString}{
\xy
0;/r0.23pc/:;
(2,24)*{}="02v24";
(2,35)*{}="02v35";
(6,27)*{}="06v27";
(6,32)*{}="06v32";
(11,19)*{}="11v19";
(11,23)*{}="11v23";
(11,36)*{}="11v36";
(11,40)*{}="11v40";
(13,28)*{}="13v28";
(13,30)*{}="13v30";
(15,27)*{}="15v27";
(16,27)*{}="16v27";
(16,32)*{}="16v32";
(17,32)*{}="17v32";
(19,31)*{}="19v31";
(21,19)*{}="21v19";
(21,23)*{}="21v23";
(21,36)*{}="21v36";
(21,40)*{}="21v40";
(23,28)*{}="23v28";
(25,27)*{}="25v27";
(26,27)*{}="26v27";
(26,32)*{}="26v32";
(27,32)*{}="27v32";
(29,31)*{}="29v31";
(31,19)*{}="31v19";
(31,23)*{}="31v23";
(31,36)*{}="31v36";
(31,40)*{}="31v40";
(36,27)*{}="36v27";
(36,32)*{}="36v32";
(40,24)*{}="40v24";
(40,35)*{}="40v35";
"13v28"*@{*};
"29v31"*@{*};
(23,32)*{\scriptstyle J};
{\ar@{-} "06v27";"02v24"};
{\ar@{-} "06v27";"11v23"};
{\ar@{-} "06v32";"02v35"};
{\ar@{-} "06v32";"06v27"};
{\ar@{-} "11v23";"11v19"};
{\ar@{-} "11v23";"16v27"};
{\ar@{-} "11v36";"06v32"};
{\ar@{-} "11v36";"11v40"};
{\ar@{-} "11v36";"16v32"};
{\ar@{-}|(0.5){\hole} "16v27";"16v32"};
{\ar@{-} "16v27";"21v23"};
{\ar@{-} "16v32";"21v36"};
{\ar@{-} "21v23";"21v19"};
{\ar@{-} "21v36";"21v40"};
{\ar@{-} "21v36";"26v32"};
{\ar@{-} "26v27";"21v23"};
{\ar@{-} "26v27";"31v23"};
{\ar@{-}|(0.5){\hole} "26v32";"26v27"};
{\ar@{-} "26v32";"31v36"};
{\ar@{-} "31v23";"31v19"};
{\ar@{-} "31v23";"36v27"};
{\ar@{-} "31v36";"31v40"};
{\ar@{-} "36v27";"36v32"};
{\ar@{-} "36v27";"40v24"};
{\ar@{-} "36v32";"31v36"};
{\ar@{-} "36v32";"40v35"};
{\ar@{=}@`{"15v27", "17v32", "19v31", "23v28", "25v27", "27v32"}|(0.6){\SelectTips{cm}{12}\object@{>>}} "13v28";"29v31"};
(11.5,28)*{\scriptstyle p};
(31,31)*{\scriptstyle q^*};
\endxy
}

\newcommand{\QDlineStringLong}{
\xy
0;/r0.3pc/:;
(2,24)*{}="02v24";
(2,35)*{}="02v35";
(6,27)*{}="06v27";
(6,32)*{}="06v32";
(11,19)*{}="11v19";
(11,23)*{}="11v23";
(11,26)*{}="11v26";
(11,36)*{}="11v36";
(11,40)*{}="11v40";
(12,25)*{}="12v25";
(13,28)*{}="13v28";
(15,25)*{}="15v25";
(16,27)*{}="16v27";
(16,32)*{}="16v32";
(17,25)*{}="17v25";
(19,21)*{}="19v21";
(20,19)*{}="20v19";
(21,19)*{}="21v19";
(21,23)*{}="21v23";
(21,36)*{}="21v36";
(21,40)*{}="21v40";
(22,23)*{}="22v23";
(26,27)*{}="26v27";
(26,32)*{}="26v32";
(29,21)*{}="29v21";
(29,31)*{}="29v31";
(29,33)*{}="29v33";
(30,19)*{}="30v19";
(31,19)*{}="31v19";
(31,23)*{}="31v23";
(31,36)*{}="31v36";
(31,40)*{}="31v40";
(32,23)*{}="32v23";
(32,34)*{}="32v34";
(35,34)*{}="35v34";
(36,25)*{}="36v25";
(36,27)*{}="36v27";
(36,32)*{}="36v32";
(37,28)*{}="37v28";
(38,32)*{}="38v32";
(39,23)*{}="39v23";
(40,24)*{}="40v24";
(40,34)*{}="40v34";
(40,35)*{}="40v35";
"13v28"*@{*};
"29v31"*@{*};
"11v26"*{\scriptstyle J};
{\ar@{-} "06v27";"02v24"};
{\ar@{-} "06v27";"11v23"};
{\ar@{-} "06v32";"02v35"};
{\ar@{-} "06v32";"06v27"};
{\ar@{-} "11v23";"11v19"};
{\ar@{-}|(0.5){\hole} "11v23";"16v27"};
{\ar@{-} "11v36";"06v32"};
{\ar@{-} "11v36";"11v40"};
{\ar@{-} "11v36";"16v32"};
{\ar@{-} "16v27";"16v32"};
{\ar@{-} "16v27";"21v23"};
{\ar@{-} "16v32";"21v36"};
{\ar@{-}|(0.5){\hole} "21v23";"21v19"};
{\ar@{-} "21v36";"21v40"};
{\ar@{-} "21v36";"26v32"};
{\ar@{-} "26v27";"21v23"};
{\ar@{-} "26v27";"31v23"};
{\ar@{-} "26v32";"26v27"};
{\ar@{-} "26v32";"31v36"};
{\ar@{-}|(0.5){\hole} "31v23";"31v19"};
{\ar@{-} "31v23";"36v27"};
{\ar@{-} "31v36";"31v40"};
{\ar@{-} "36v27";"36v32"};
{\ar@{-}|(0.5){\hole} "36v27";"40v24"};
{\ar@{-}|(0.5){\hole} "36v32";"31v36"};
{\ar@{-}|(0.5){\hole} "36v32";"40v35"};
{\ar@{=}@`{"12v25", "15v25", "17v25", "19v21", "20v19", "22v23", "29v21", "30v19", "32v23", "36v25", "39v23", "37v28", "38v32", "40v34", "35v34", "32v34", "29v33"}|(0.7){\SelectTips{cm}{12}\object@{>>}} "13v28";"29v31"};
(12,28)*{\scriptstyle p};
(30.5,31)*{\scriptstyle q^*};
\endxy
}

\newcommand{\TwistStringL}{
	\xy
	0;/r0.23pc/:;
	(2,24)*{}="02v24";
	(2,35)*{}="02v35";
	(6,27)*{}="06v27";
	(6,32)*{}="06v32";
	(11,19)*{}="11v19";
	(11,23)*{}="11v23";
	(11,36)*{}="11v36";
	(11,40)*{}="11v40";
	(13,28)*{}="13v28";
	(15,27)*{}="15v27";
	(15,28)*{}="15v28";
	(16,27)*{}="16v27";
	(16,32)*{}="16v32";
	(17,28)*{}="17v28";
	(17,32)*{}="17v32";
	(17,33)*{}="17v33";
	(18,31)*{}="18v31";
	(19,29)*{}="19v29";
	(19,31)*{}="19v31";
	(19,33)*{}="19v33";
	(20,25)*{}="20v25";
	(21,19)*{}="21v19";
	(21,23)*{}="21v23";
	(21,34)*{}="21v34";
	(21,36)*{}="21v36";
	(21,40)*{}="21v40";
	(22,25)*{}="22v25";
	(22,30)*{}="22v30";
	(23,28)*{}="23v28";
	(23,33)*{}="23v33";
	(24,26)*{}="24v26";
	(25,27)*{}="25v27";
	(25,31)*{}="25v31";
	(26,27)*{}="26v27";
	(26,32)*{}="26v32";
	(27,32)*{}="27v32";
	(29,31)*{}="29v31";
	(31,19)*{}="31v19";
	(31,23)*{}="31v23";
	(31,36)*{}="31v36";
	(31,40)*{}="31v40";
	(36,27)*{}="36v27";
	(36,32)*{}="36v32";
	(40,24)*{}="40v24";
	(40,35)*{}="40v35";
	"13v28"*@{*};
	"29v31"*@{*};
	"22v30"*{\scriptstyle J};
	{\ar@{-} "06v27";"02v24"};
	{\ar@{-} "06v27";"11v23"};
	{\ar@{-} "06v32";"02v35"};
	{\ar@{-} "06v32";"06v27"};
	{\ar@{-} "11v23";"11v19"};
	{\ar@{-} "11v23";"16v27"};
	{\ar@{-} "11v36";"06v32"};
	{\ar@{-} "11v36";"11v40"};
	{\ar@{-} "11v36";"16v32"};
	{\ar@{-}|(0.65){\hole} "16v27";"16v32"};
	{\ar@{-} "16v27";"21v23"};
	{\ar@{-} "16v32";"21v36"};
	{\ar@{-} "21v23";"21v19"};
	{\ar@{-} "21v36";"21v40"};
	{\ar@{-} "21v36";"26v32"};
	{\ar@{-} "26v27";"21v23"};
	{\ar@{-} "26v27";"31v23"};
	{\ar@{-}|(0.4){\hole} "26v32";"26v27"};
	{\ar@{-} "26v32";"31v36"};
	{\ar@{-} "31v23";"31v19"};
	{\ar@{-} "31v23";"36v27"};
	{\ar@{-} "31v36";"31v40"};
	{\ar@{-} "36v27";"36v32"};
	{\ar@{-} "36v27";"40v24"};
	{\ar@{-} "36v32";"31v36"};
	{\ar@{-} "36v32";"40v35"};
	{\ar@{=}@`{"15v28", "17v33", "17v28", "20v25", "22v25", "25v27", "25v31", "23v33", "21v34", "19v33", "18v31", "19v29", "24v26", "27v32"}|(0.7){\SelectTips{eu}{12}\object@{>>}}|(0.94){\hole}  "13v28";"29v31"};
	"13v28"+(-2,0)*{\scriptstyle p};
	"29v31"+(2.5,0)*{\scriptstyle q^*};
	\endxy
}

\newcommand{\TwistStringR}{
\xy
	0;/r0.23pc/:;
	(2,24)*{}="02v24";
	(2,35)*{}="02v35";
	(6,27)*{}="06v27";
	(6,32)*{}="06v32";
	(11,19)*{}="11v19";
	(11,23)*{}="11v23";
	(11,36)*{}="11v36";
	(11,40)*{}="11v40";
	(13,28)*{}="13v28";
	(15,27)*{}="15v27";
	(16,27)*{}="16v27";
	(16,32)*{}="16v32";
	(17,28)*{}="17v28";
	(17,31)*{}="17v31";
	(18,30)*{}="18v30";
	(19,26)*{}="19v26";
	(19,31)*{}="19v31";
	(19,32)*{}="19v32";
	(19,34)*{}="19v34";
	(21,19)*{}="21v19";
	(21,23)*{}="21v23";
	(21,25)*{}="21v25";
	(21,32)*{}="21v32";
	(21,36)*{}="21v36";
	(21,40)*{}="21v40";
	(22,30)*{}="22v30";
	(22,34)*{}="22v34";
	(23,26)*{}="23v26";
	(23,28)*{}="23v28";
	(24,30)*{}="24v30";
	(24,32)*{}="24v32";
	(25,27)*{}="25v27";
	(25,28)*{}="25v28";
	(25,30)*{}="25v30";
	(26,27)*{}="26v27";
	(26,32)*{}="26v32";
	(27,30)*{}="27v30";
	(29,31)*{}="29v31";
	(31,19)*{}="31v19";
	(31,23)*{}="31v23";
	(31,36)*{}="31v36";
	(31,40)*{}="31v40";
	(36,27)*{}="36v27";
	(36,32)*{}="36v32";
	(40,24)*{}="40v24";
	(40,35)*{}="40v35";
	"13v28"*@{*};
	"29v31"*@{*};
	(19,29)*{\scriptstyle J};
	{\ar@{-} "06v27";"02v24"};
	{\ar@{-} "06v27";"11v23"};
	{\ar@{-} "06v32";"02v35"};
	{\ar@{-} "06v32";"06v27"};
	{\ar@{-} "11v23";"11v19"};
	{\ar@{-} "11v23";"16v27"};
	{\ar@{-} "11v36";"06v32"};
	{\ar@{-} "11v36";"11v40"};
	{\ar@{-} "11v36";"16v32"};
	{\ar@{-}|(0.5){\hole} "16v27";"16v32"};
	{\ar@{-} "16v27";"21v23"};
	{\ar@{-} "16v32";"21v36"};
	{\ar@{-} "21v23";"21v19"};
	{\ar@{-} "21v36";"21v40"};
	{\ar@{-} "21v36";"26v32"};
	{\ar@{-} "26v27";"21v23"};
	{\ar@{-} "26v27";"31v23"};
	{\ar@{-}|(0.4){\hole} "26v32";"26v27"};
	{\ar@{-} "26v32";"31v36"};
	{\ar@{-} "31v23";"31v19"};
	{\ar@{-} "31v23";"36v27"};
	{\ar@{-} "31v36";"31v40"};
	{\ar@{-} "36v27";"36v32"};
	{\ar@{-} "36v27";"40v24"};
	{\ar@{-} "36v32";"31v36"};
	{\ar@{-} "36v32";"40v35"};
	{\ar@{=}@`{"15v27", "17v31", "19v34", "22v34", "24v32", "25v30", "25v28", "23v26", "21v25", "19v26", "17v28", "18v30", "19v32", "21v32", "24v30", "27v30"}|(0.65){\SelectTips{eu}{12}\object@{>>}}|(0.94){\hole} "13v28";"29v31"};
	"13v28"+(-2,0)*{\scriptstyle p};
	"29v31"+(2.5,0)*{\scriptstyle q^*};
	\endxy
}

\newcommand{\RibbonId}{
\xy
0;/r0.3pc/:;
(16,16)*{}="16v16";
(16,26)*{}="16v26";
(18,22)*{}="18v22";
"16v16"*@{*};
"16v26"*@{*};
"18v22"*{\scriptstyle J};
{\ar@{=}|(0.5){\SelectTips{eu}{12}\object@{>>}} "16v16";"16v26"};
\endxy
}

\newcommand{\RibbonTwistL}{
\xy
0;/r0.3pc/:;
(18,23)*{}="18v23";
(16,16)*{}="16v16";
(16,19)*{}="16v19";
(16,24)*{}="16v24";
(16,26)*{}="16v26";
(15,19)*{}="15v19";
(15,24)*{}="15v24";
(12,20)*{}="12v20";
(12,23)*{}="12v23";
"16v16"*@{*};
"16v26"*@{*};
"18v23"*{\scriptstyle J};
{\ar@{=}@`{"16v19", "15v24", "12v23", "12v20", "15v19", "16v24"}|(0.15){\hole}|(0.93){\SelectTips{eu}{12}\object@{>>}} "16v16";"16v26"};
\endxy
}

\newcommand{\RibbonTwistR}{
\xy
0;/r0.3pc/:;
(14,23)*{}="14v23";
(16,16)*{}="16v16";
(16,19)*{}="16v19";
(16,24)*{}="16v24";
(16,26)*{}="16v26";
(17,19)*{}="17v19";
(17,24)*{}="17v24";
(20,20)*{}="20v20";
(20,23)*{}="20v23";
"16v16"*@{*};
"16v26"*@{*};
"14v23"*{\scriptstyle J};
{\ar@{=}@`{"16v19", "17v24", "20v23", "20v20", "17v19", "16v24"}|(0.85){\hole}|(0.93){\SelectTips{eu}{12}\object@{>>}} "16v16";"16v26"};
\endxy
}

\newcommand{\ContractString}{
	\xy
	0;/r0.2pc/:;
	(14,21)*{}="14v21";
	(14,26)*{}="14v26";
	(18,18)*{}="18v18";
	(18,29)*{}="18v29";
	(21,21)*{}="21v21";
	(21,26)*{}="21v26";
	"21v21"*@{*};
	"21v26"*@{*};
	{\ar@{.} "21v26";"21v21"};
	{\ar@{=}@`{"18v18", "14v21", "14v26", "18v29"}|(0.5){\SelectTips{eu}{12}\object@{>>}} "21v21";"21v26"};
	(12,23)*{\scriptstyle J};
	\endxy
}

\newcommand{\LoopString}{\xy
	0;/r0.2pc/:;
	(17,23)*{}="17v23";
	(17,28)*{}="17v28";
	(21,21)*{}="21v21";
	(21,31)*{}="21v31";
	(25,23)*{}="25v23";
	(25,25)*{}="25v25";
	(25,26)*{}="25v26";
	(25,27)*{}="25v27";
	(24,29)*{}="24v29";
	{\ar@{=}@`{"25v25", "25v23", "21v21", "17v23", "17v28", "21v31", "24v29", "25v27"}|(0.5){\SelectTips{eu}{12}\object@{>>}} "25v26";"25v26"};
	(15,25)*{\scriptstyle J};
	\endxy
}

\newcommand{\SmatrixString}{
\xy
0;/r0.25pc/:;
(15,26)*{}="15v26";
(23,22)*{}="23v22";
(23,29)*{}="23v29";
(27,21)*{}="27v21";
(27,30)*{}="27v30";
(17,24)*{}="17v24";
(17,26)*{}="17v26";
(17,27)*{}="17v27";
(18,22)*{}="18v22";
(18,29)*{}="18v29";
(30,22)*{}="30v22";
(30,29)*{}="30v29";
(31,24)*{}="31v24";
(31,26)*{}="31v26";
(31,27)*{}="31v27";
(21,21)*{}="21v21";
(21,30)*{}="21v30";
(26,22)*{}="26v22";
(26,29)*{}="26v29";
(33,26)*{}="33v26";
"15v26"*{\scriptstyle J};
"33v26"*{\scriptstyle K};
{\ar@{=}@`{"17v24", "18v22", "21v21", "26v22", "26v29", "21v30", "18v29", "17v27"}|(0.2){\SelectTips{eu}{12}\object@{<<}}|(0.61){\hole} "17v26";"17v26"};
{\ar@{=}@`{"31v24", "30v22", "27v21", "23v22", "23v29", "27v30", "30v29", "31v27"}|(0.2){\SelectTips{eu}{12}\object@{>>}}|(0.39){\hole} "31v26";"31v26"};
\endxy
}

\newcommand{\SmatrixStringB}{
	\xy
	0;/r0.25pc/:;
	(15,26)*{}="15v26";
	(23,22)*{}="23v22";
	(23,29)*{}="23v29";
	(27,21)*{}="27v21";
	(27,30)*{}="27v30";
	(17,24)*{}="17v24";
	(17,26)*{}="17v26";
	(17,27)*{}="17v27";
	(18,22)*{}="18v22";
	(18,29)*{}="18v29";
	(30,22)*{}="30v22";
	(30,29)*{}="30v29";
	(31,24)*{}="31v24";
	(31,26)*{}="31v26";
	(31,27)*{}="31v27";
	(21,21)*{}="21v21";
	(21,30)*{}="21v30";
	(26,22)*{}="26v22";
	(26,29)*{}="26v29";
	(33,26)*{}="33v26";
	"15v26"*{\scriptstyle J};
	"33v26"*{\scriptstyle K};
	{\ar@{=}@`{"17v24", "18v22", "21v21", "26v22", "26v29", "21v30", "18v29", "17v27"}|(0.2){\SelectTips{eu}{12}\object@{<<}}|(0.39){\hole} "17v26";"17v26"};
	{\ar@{=}@`{"31v24", "30v22", "27v21", "23v22", "23v29", "27v30", "30v29", "31v27"}|(0.2){\SelectTips{eu}{12}\object@{>>}}|(0.61){\hole} "31v26";"31v26"};
	\endxy
}

\newcommand{\trfL}{
\xy
0;<3pt,0pt>:
<0pt,3pt>::
(11,11)*{}="11v11";
(11,41)*{}="11v41";
(16,16)*{}="16v16";
(16,21)*{}="16v21";
(16,46)*{}="16v46";
(17,12)*{}="17v12";
(17,42)*{}="17v42";
(18,21)*{}="18v21";
(18,24)*{}="18v24";
(18,25)*{}="18v25";
(18,26)*{}="18v26";
(18,29)*{}="18v29";
(18,34)*{}="18v34";
(18,36)*{}="18v36";
(21,20)*{}="21v20";
(21,30)*{}="21v30";
(23,18)*{}="23v18";
(24,21)*{}="24v21";
(24,24)*{}="24v24";
(24,26)*{}="24v26";
(24,29)*{}="24v29";
(25,39)*{}="25v39";
(26,14)*{}="26v14";
(26,25)*{}="26v25";
(26,44)*{}="26v44";
(27,40)*{}="27v40";
(28,31)*{}="28v31";
(30,35)*{}="30v35";
(31,16)*{}="31v16";
(35,24)*{}="35v24";
(36,11)*{}="36v11";
(36,41)*{}="36v41";
(37,27)*{}="37v27";
(39,33)*{}="39v33";
(40,27)*{}="40v27";
(41,16)*{}="41v16";
(41,46)*{}="41v46";
"17v12"*@{*};
"17v42"*@{*};
"26v14"*@{*};
"26v44"*@{*};
{\ar@{-} "11v11";"36v11"};
{\ar@{-} "11v41";"16v46"};
{\ar@{-} "16v16";"11v11"};
{\ar@{-}|(0.05){\hole}|(0.35){\hole} "16v16";"41v16"};
{\ar@{-} "16v46";"41v46"};
{\ar@{-} "36v11";"41v16"};
{\ar@{-} "36v41";"11v41"};
{\ar@{-} "41v46";"36v41"};
{\ar@{=}@`{"27v40", "18v34", "39v33", "40v27", "35v24", "26v25", "23v18"}|(0.04){\hole} |(0.11){\hole}|(0.23){\hole}|(0.4){\SelectTips{eu}{12}\object@{<<}}|(0.66){\hole}   "26v44";"26v14"};
{\ar@{=}@`{"18v24", "18v21", "24v21", "24v24", "24v26", "24v29", "18v29", "18v26"}|(0.15){\hole}|(0.26){\hole}|(0.74){\hole}|(0.88){\SelectTips{eu}{12}\object@{>>}} "18v25";"18v25"};
{\ar@{=}@`{"18v36", "25v39", "30v35", "21v30", "21v20", "31v16", "37v27", "28v31", "16v21"}|(0.012){\hole}|(0.05){\SelectTips{eu}{12}\object@{<<}}|(0.444){\hole}|(0.536){\hole}|(0.896){\hole} "17v42";"17v12"};
\endxy
}

\newcommand{\trfLcontraction}{
\xy
0;<3pt,0pt>:
<0pt,3pt>::
(11,11)*{}="11v11";
(11,41)*{}="11v41";
(16,16)*{}="16v16";
(16,21)*{}="16v21";
(16,46)*{}="16v46";
(17,12)*{}="17v12";
(17,42)*{}="17v42";
(18,8)*{}="18v08";
(18,9)*{}="18v09";
(18,21)*{}="18v21";
(18,24)*{}="18v24";
(18,25)*{}="18v25";
(18,26)*{}="18v26";
(18,29)*{}="18v29";
(18,34)*{}="18v34";
(18,36)*{}="18v36";
(18,49)*{}="18v49";
(18,50)*{}="18v50";
(21,20)*{}="21v20";
(21,30)*{}="21v30";
(23,18)*{}="23v18";
(24,21)*{}="24v21";
(24,24)*{}="24v24";
(24,26)*{}="24v26";
(24,29)*{}="24v29";
(25,39)*{}="25v39";
(26,14)*{}="26v14";
(26,25)*{}="26v25";
(26,44)*{}="26v44";
(27,10)*{}="27v10";
(27,11)*{}="27v11";
(27,40)*{}="27v40";
(26,47)*{}="26v47";
(26,48)*{}="26v48";
(28,31)*{}="28v31";
(30,35)*{}="30v35";
(31,16)*{}="31v16";
(35,24)*{}="35v24";
(36,11)*{}="36v11";
(36,41)*{}="36v41";
(37,27)*{}="37v27";
(39,33)*{}="39v33";
(40,27)*{}="40v27";
(41,16)*{}="41v16";
(41,46)*{}="41v46";
(44,10)*{}="44v10";
(44,48)*{}="44v48";
(45,12)*{}="45v12";
(45,46)*{}="45v46";
(46,8)*{}="46v08";
(46,50)*{}="46v50";
(47,9)*{}="47v09";
(47,49)*{}="47v49";
"17v12"*@{*};
"17v42"*@{*};
"26v14"*@{*};
"26v44"*@{*};
{\ar@{-} "11v11";"36v11"};
{\ar@{-} "11v41";"16v46"};
{\ar@{-} "16v16";"11v11"};
{\ar@{-}|(0.05){\hole}|(0.35){\hole} "16v16";"41v16"};
{\ar@{-} "16v46";"41v46"};
{\ar@{-} "36v11";"41v16"};
{\ar@{-} "36v41";"11v41"};
{\ar@{-} "41v46";"36v41"};
{\ar@{=}@`{"27v40", "18v34", "39v33", "40v27", "35v24", "26v25", "23v18"}|(0.04){\hole} |(0.11){\hole}|(0.23){\hole}|(0.4){\SelectTips{eu}{12}\object@{<<}}|(0.66){\hole}   "26v44";"26v14"};
{\ar@{=}@`{"18v24", "18v21", "24v21", "24v24", "24v26", "24v29", "18v29", "18v26"}|(0.15){\hole}|(0.26){\hole}|(0.74){\hole}|(0.88){\SelectTips{eu}{12}\object@{<<}} "18v25";"18v25"};
{\ar@{=}@`{"18v36", "25v39", "30v35", "21v30", "21v20", "31v16", "37v27", "28v31", "16v21"}|(0.012){\hole}|(0.05){\SelectTips{eu}{12}\object@{<<}}|(0.444){\hole}|(0.536){\hole}|(0.896){\hole} "17v42";"17v12"};
{\ar@{.}@`{"18v49", "18v50", "46v50", "47v49", "47v09", "46v08", "18v08", "18v09"} "17v42";"17v12"};
{\ar@{.}@`{"26v47", "26v48", "44v48", "45v46", "45v12", "44v10", "27v10", "27v11"} "26v44";"26v14"};
\endxy
}

\newcommand{\trfLink}{
	\xy
	0;<3pt,0pt>:
	<0pt,3pt>::
	(11,11)*{}="11v11";
	(11,41)*{}="11v41";
	(16,16)*{}="16v16";
	(16,21)*{}="16v21";
	(16,46)*{}="16v46";
	(17,12)*{}="17v12";
	(17,42)*{}="17v42";
	(18,8)*{}="18v08";
	(18,9)*{}="18v09";
	(18,21)*{}="18v21";
	(18,24)*{}="18v24";
	(18,25)*{}="18v25";
	(18,26)*{}="18v26";
	(18,29)*{}="18v29";
	(18,34)*{}="18v34";
	(18,36)*{}="18v36";
	(18,49)*{}="18v49";
	(18,50)*{}="18v50";
	(21,20)*{}="21v20";
	(21,30)*{}="21v30";
	(23,18)*{}="23v18";
	(24,21)*{}="24v21";
	(24,24)*{}="24v24";
	(24,26)*{}="24v26";
	(24,29)*{}="24v29";
	(25,39)*{}="25v39";
	(26,14)*{}="26v14";
	(26,25)*{}="26v25";
	(26,44)*{}="26v44";
	(27,10)*{}="27v10";
	(27,11)*{}="27v11";
	(27,40)*{}="27v40";
	(26,47)*{}="26v47";
	(26,48)*{}="26v48";
	(28,31)*{}="28v31";
	(30,35)*{}="30v35";
	(31,16)*{}="31v16";
	(35,24)*{}="35v24";
	(36,11)*{}="36v11";
	(36,41)*{}="36v41";
	(37,27)*{}="37v27";
	(39,33)*{}="39v33";
	(40,27)*{}="40v27";
	(41,16)*{}="41v16";
	(41,46)*{}="41v46";
	(44,10)*{}="44v10";
	(44,48)*{}="44v48";
	(45,12)*{}="45v12";
	(45,46)*{}="45v46";
	(46,8)*{}="46v08";
	(46,50)*{}="46v50";
	(47,9)*{}="47v09";
	(47,49)*{}="47v49";
	{\ar@{=}@`{"27v40", "18v34", "39v33", "40v27", "35v24", "26v25", "23v18"} |(0.11){\hole}|(0.23){\hole}|(0.4){\SelectTips{eu}{12}\object@{<<}}|(0.66){\hole}   "26v44";"26v14"};
	{\ar@{=}@`{"18v24", "18v21", "24v21", "24v24", "24v26", "24v29", "18v29", "18v26"}|(0.15){\hole}|(0.26){\hole}|(0.74){\hole}|(0.88){\SelectTips{eu}{12}\object@{<<}} "18v25";"18v25"};
	{\ar@{=}@`{"18v36", "25v39", "30v35", "21v30", "21v20", "31v16", "37v27", "28v31", "16v21"}|(0.05){\SelectTips{eu}{12}\object@{<<}}|(0.444){\hole}|(0.536){\hole}|(0.896){\hole} "17v42";"17v12"};
	{\ar@{=}@`{"18v49", "18v50", "46v50", "47v49", "47v09", "46v08", "18v08", "18v09"} "17v42";"17v12"};
	{\ar@{=}@`{"26v47", "26v48", "44v48", "45v46", "45v12", "44v10", "27v10", "27v11"} "26v44";"26v14"};
	\endxy
}

\newcommand{\SmatrixProcedureA}{
	\xy
	0;<4pt,0pt>:
	<0pt,4pt>::
	(5,19)*{}="05v19";
	(5,26)*{}="05v26";
	(5,29)*{}="05v29";
	(5,36)*{}="05v36";
	(8,21)*{}="08v21";
	(8,24)*{}="08v24";
	(8,31)*{}="08v31";
	(8,34)*{}="08v34";
	(11,17)*{}="11v17";
	(11,19)*{}="11v19";
	(11,26)*{}="11v26";
	(11,29)*{}="11v29";
	(11,36)*{}="11v36";
	(11,38)*{}="11v38";
	(14,21)*{}="14v21";
	(14,24)*{}="14v24";
	(14,27)*{}="14v27";
	(14,31)*{}="14v31";
	(14,34)*{}="14v34";
	(15,29)*{}="15v29";
	(17,17)*{}="17v17";
	(17,19)*{}="17v19";
	(17,26)*{}="17v26";
	(17,29)*{}="17v29";
	(17,36)*{}="17v36";
	(17,38)*{}="17v38";
	(19,26)*{}="19v26";
	(20,21)*{}="20v21";
	(20,24)*{}="20v24";
	(20,31)*{}="20v31";
	(20,34)*{}="20v34";
	(21,29)*{}="21v29";
	(23,17)*{}="23v17";
	(23,19)*{}="23v19";
	(23,26)*{}="23v26";
	(23,29)*{}="23v29";
	(23,36)*{}="23v36";
	(23,38)*{}="23v38";
	(25,26)*{}="25v26";
	(26,21)*{}="26v21";
	(26,24)*{}="26v24";
	(26,28)*{}="26v28";
	(26,31)*{}="26v31";
	(26,34)*{}="26v34";
	(29,19)*{}="29v19";
	(29,26)*{}="29v26";
	(29,29)*{}="29v29";
	(29,36)*{}="29v36";
	"14v27"*@{*};
	"26v28"*@{*};
	"21v29"*{\scriptstyle K};
	{\ar@{-} "05v26";"05v29"};
	{\ar@{-} "05v29";"08v31"};
	{\ar@{-} "08v21";"05v19"};
	{\ar@{-} "08v21";"11v19"};
	{\ar@{-} "08v24";"05v26"};
	{\ar@{-} "08v24";"08v21"};
	{\ar@{-} "08v31";"11v29"};
	{\ar@{-} "08v34";"05v36"};
	{\ar@{-} "08v34";"08v31"};
	{\ar@{-} "11v19";"11v17"};
	{\ar@{-} "11v19";"14v21"};
	{\ar@{-} "11v26";"08v24"};
	{\ar@{-} "11v29";"11v26"};
	{\ar@{-} "11v29";"14v31"};
	{\ar@{-} "11v36";"08v34"};
	{\ar@{-} "11v36";"11v38"};
	{\ar@{-} "11v36";"14v34"};
	{\ar@{-} "14v21";"17v19"};
	{\ar@{-} "14v24";"11v26"};
	{\ar@{-} "14v24";"14v21"};
	{\ar@{-} "14v31";"14v34"};
	{\ar@{-} "14v31";"17v29"};
	{\ar@{-} "14v34";"17v36"};
	{\ar@{-} "17v19";"17v17"};
	{\ar@{-} "17v19";"20v21"};
	{\ar@{-} "17v26";"14v24"};
	{\ar@{-} "17v26";"20v24"};
	{\ar@{-}|(0.5){\hole} "17v29";"17v26"};
	{\ar@{-} "17v36";"17v38"};
	{\ar@{-} "17v36";"20v34"};
	{\ar@{-} "20v21";"23v19"};
	{\ar@{-} "20v24";"20v21"};
	{\ar@{-} "20v24";"23v26"};
	{\ar@{-} "20v31";"17v29"};
	{\ar@{-} "20v31";"23v29"};
	{\ar@{-} "20v34";"20v31"};
	{\ar@{-} "20v34";"23v36"};
	{\ar@{-} "23v19";"23v17"};
	{\ar@{-} "23v19";"26v21"};
	{\ar@{-} "23v26";"26v24"};
	{\ar@{-}|(0.5){\hole} "23v29";"23v26"};
	{\ar@{-} "23v36";"23v38"};
	{\ar@{-} "23v36";"26v34"};
	{\ar@{-} "26v21";"26v24"};
	{\ar@{-} "26v21";"29v19"};
	{\ar@{-} "26v24";"29v26"};
	{\ar@{-} "26v31";"23v29"};
	{\ar@{-} "26v31";"29v29"};
	{\ar@{-} "26v34";"26v31"};
	{\ar@{-} "26v34";"29v36"};
	{\ar@{-} "29v26";"29v29"};
	{\ar@{=}@`{"15v29", "19v26", "21v29", "25v26"}|(0.63){\SelectTips{eu}{12}\object@{<<}} "14v27";"26v28"};
	\endxy
}

\newcommand{\SmatrixProcedureB}{
\xy
0;<4pt,0pt>:
<0pt,4pt>::
(5,19)*{}="05v19";
(5,26)*{}="05v26";
(5,29)*{}="05v29";
(5,36)*{}="05v36";
(8,21)*{}="08v21";
(8,24)*{}="08v24";
(8,26)*{}="08v26";
(8,27)*{}="08v27";
(8,29)*{}="08v29";
(8,31)*{}="08v31";
(8,34)*{}="08v34";
(11,17)*{}="11v17";
(11,19)*{}="11v19";
(11,22)*{}="11v22";
(11,26)*{}="11v26";
(11,29)*{}="11v29";
(11,33)*{}="11v33";
(11,36)*{}="11v36";
(11,38)*{}="11v38";
(14,21)*{}="14v21";
(14,24)*{}="14v24";
(14,27)*{}="14v27";
(14,31)*{}="14v31";
(14,34)*{}="14v34";
(15,29)*{}="15v29";
(17,17)*{}="17v17";
(17,19)*{}="17v19";
(17,22)*{}="17v22";
(17,26)*{}="17v26";
(17,29)*{}="17v29";
(17,33)*{}="17v33";
(17,36)*{}="17v36";
(17,38)*{}="17v38";
(19,26)*{}="19v26";
(20,21)*{}="20v21";
(20,24)*{}="20v24";
(20,27)*{}="20v27";
(20,31)*{}="20v31";
(20,34)*{}="20v34";
(21,29)*{}="21v29";
(23,17)*{}="23v17";
(23,19)*{}="23v19";
(23,26)*{}="23v26";
(23,29)*{}="23v29";
(23,36)*{}="23v36";
(23,38)*{}="23v38";
(25,26)*{}="25v26";
(26,21)*{}="26v21";
(26,24)*{}="26v24";
(26,28)*{}="26v28";
(26,31)*{}="26v31";
(26,34)*{}="26v34";
(29,19)*{}="29v19";
(29,26)*{}="29v26";
(29,29)*{}="29v29";
(29,36)*{}="29v36";
"14v27"*@{*};
"26v28"*@{*};
"21v29"*{\scriptstyle K};
"11v33"*{\scriptstyle J};
{\ar@{-} "05v26";"05v29"};
{\ar@{-} "05v29";"08v31"};
{\ar@{-} "08v21";"05v19"};
{\ar@{-} "08v21";"11v19"};
{\ar@{-} "08v24";"05v26"};
{\ar@{-} "08v24";"08v21"};
{\ar@{-}|(0.5){\hole} "08v31";"11v29"};
{\ar@{-} "08v34";"05v36"};
{\ar@{-} "08v34";"08v31"};
{\ar@{-} "11v19";"11v17"};
{\ar@{-} "11v19";"14v21"};
{\ar@{-}|(0.5){\hole} "11v26";"08v24"};
{\ar@{-} "11v29";"11v26"};
{\ar@{-} "11v29";"14v31"};
{\ar@{-} "11v36";"08v34"};
{\ar@{-} "11v36";"11v38"};
{\ar@{-} "11v36";"14v34"};
{\ar@{-} "14v21";"17v19"};
{\ar@{-} "14v24";"11v26"};
{\ar@{-}|(0.6){\hole} "14v24";"14v21"};
{\ar@{-}|(0.5){\hole} "14v31";"14v34"};
{\ar@{-} "14v31";"17v29"};
{\ar@{-} "14v34";"17v36"};
{\ar@{-} "17v19";"17v17"};
{\ar@{-} "17v19";"20v21"};
{\ar@{-} "17v26";"14v24"};
{\ar@{-}|(0.5){\hole} "17v26";"20v24"};
{\ar@{-}|(0.5){\hole} "17v29";"17v26"};
{\ar@{-} "17v36";"17v38"};
{\ar@{-} "17v36";"20v34"};
{\ar@{-} "20v21";"23v19"};
{\ar@{-} "20v24";"20v21"};
{\ar@{-} "20v24";"23v26"};
{\ar@{-}|(0.5){\hole} "20v31";"17v29"};
{\ar@{-} "20v31";"23v29"};
{\ar@{-} "20v34";"20v31"};
{\ar@{-} "20v34";"23v36"};
{\ar@{-} "23v19";"23v17"};
{\ar@{-} "23v19";"26v21"};
{\ar@{-} "23v26";"26v24"};
{\ar@{-}|(0.5){\hole} "23v29";"23v26"};
{\ar@{-} "23v36";"23v38"};
{\ar@{-} "23v36";"26v34"};
{\ar@{-} "26v21";"26v24"};
{\ar@{-} "26v21";"29v19"};
{\ar@{-} "26v24";"29v26"};
{\ar@{-} "26v31";"23v29"};
{\ar@{-} "26v31";"29v29"};
{\ar@{-} "26v34";"26v31"};
{\ar@{-} "26v34";"29v36"};
{\ar@{-} "29v26";"29v29"};
{\ar@{=}@`{"15v29", "19v26", "21v29", "25v26"}|(0.41){\hole}|(0.63){\SelectTips{eu}{12}\object@{<<}} "14v27";"26v28"};
{\ar@{=}@`{"08v26", "11v22", "17v22", "20v27", "17v33", "11v33", "08v29"}|(0.88){\SelectTips{eu}{12}\object@{<<}} "08v27";"08v27"};
\endxy
}

\newcommand{\SmatrixProcedureC}{
\xy
0;<4pt,0pt>:
<0pt,4pt>::
(5,19)*{}="05v19";
(5,26)*{}="05v26";
(5,29)*{}="05v29";
(5,36)*{}="05v36";
(8,21)*{}="08v21";
(8,24)*{}="08v24";
(8,26)*{}="08v26";
(8,27)*{}="08v27";
(8,29)*{}="08v29";
(8,31)*{}="08v31";
(8,34)*{}="08v34";
(11,17)*{}="11v17";
(11,19)*{}="11v19";
(11,22)*{}="11v22";
(11,26)*{}="11v26";
(11,29)*{}="11v29";
(11,33)*{}="11v33";
(11,36)*{}="11v36";
(11,38)*{}="11v38";
(14,21)*{}="14v21";
(14,24)*{}="14v24";
(14,26)*{}="14v26";
(14,31)*{}="14v31";
(14,34)*{}="14v34";
(15,28)*{}="15v28";
(15,29)*{}="15v29";
(16,24)*{}="16v24";
(17,17)*{}="17v17";
(17,19)*{}="17v19";
(17,22)*{}="17v22";
(17,26)*{}="17v26";
(17,29)*{}="17v29";
(17,33)*{}="17v33";
(17,36)*{}="17v36";
(17,38)*{}="17v38";
(18,23)*{}="18v23";
(19,26)*{}="19v26";
(20,21)*{}="20v21";
(20,24)*{}="20v24";
(20,27)*{}="20v27";
(20,31)*{}="20v31";
(20,34)*{}="20v34";
(21,22)*{}="21v22";
(21,29)*{}="21v29";
(23,17)*{}="23v17";
(23,19)*{}="23v19";
(23,23)*{}="23v23";
(23,26)*{}="23v26";
(23,29)*{}="23v29";
(23,36)*{}="23v36";
(23,38)*{}="23v38";
(24,25)*{}="24v25";
(25,26)*{}="25v26";
(26,21)*{}="26v21";
(26,24)*{}="26v24";
(26,29)*{}="26v29";
(26,31)*{}="26v31";
(26,34)*{}="26v34";
(27,26)*{}="27v26";
(29,19)*{}="29v19";
(29,26)*{}="29v26";
(29,29)*{}="29v29";
(29,36)*{}="29v36";
"14v26"*@{*};
"15v28"*@{*};
"26v29"*@{*};
"27v26"*@{*};
"21v29"*{\scriptstyle K};
"11v33"*{\scriptstyle J};
(23,21)*{\scriptstyle K};
{\ar@{-} "05v26";"05v29"};
{\ar@{-} "05v29";"08v31"};
{\ar@{-} "08v21";"05v19"};
{\ar@{-} "08v21";"11v19"};
{\ar@{-} "08v24";"05v26"};
{\ar@{-} "08v24";"08v21"};
{\ar@{-}|(0.5){\hole} "08v31";"11v29"};
{\ar@{-} "08v34";"05v36"};
{\ar@{-} "08v34";"08v31"};
{\ar@{-} "11v19";"11v17"};
{\ar@{-} "11v19";"14v21"};
{\ar@{-}|(0.5){\hole} "11v26";"08v24"};
{\ar@{-} "11v29";"11v26"};
{\ar@{-} "11v29";"14v31"};
{\ar@{-} "11v36";"08v34"};
{\ar@{-} "11v36";"11v38"};
{\ar@{-} "11v36";"14v34"};
{\ar@{-} "14v21";"17v19"};
{\ar@{-} "14v24";"11v26"};
{\ar@{-}|(0.5){\hole} "14v24";"14v21"};
{\ar@{-}|(0.5){\hole} "14v31";"14v34"};
{\ar@{-} "14v31";"17v29"};
{\ar@{-} "14v34";"17v36"};
{\ar@{-} "17v19";"17v17"};
{\ar@{-} "17v19";"20v21"};
{\ar@{-}|(0.55){\hole} "17v26";"14v24"};
{\ar@{-}|(0.5){\hole} "17v26";"20v24"};
{\ar@{-}|(0.5){\hole} "17v29";"17v26"};
{\ar@{-} "17v36";"17v38"};
{\ar@{-} "17v36";"20v34"};
{\ar@{-} "20v21";"23v19"};
{\ar@{-}|(0.5){\hole} "20v24";"20v21"};
{\ar@{-} "20v24";"23v26"};
{\ar@{-}|(0.5){\hole} "20v31";"17v29"};
{\ar@{-} "20v31";"23v29"};
{\ar@{-} "20v34";"20v31"};
{\ar@{-} "20v34";"23v36"};
{\ar@{-} "23v19";"23v17"};
{\ar@{-} "23v19";"26v21"};
{\ar@{-}|(0.5){\hole} "23v26";"26v24"};
{\ar@{-}|(0.5){\hole} "23v29";"23v26"};
{\ar@{-} "23v36";"23v38"};
{\ar@{-} "23v36";"26v34"};
{\ar@{-} "26v21";"26v24"};
{\ar@{-} "26v21";"29v19"};
{\ar@{-} "26v24";"29v26"};
{\ar@{-} "26v31";"23v29"};
{\ar@{-} "26v31";"29v29"};
{\ar@{-} "26v34";"26v31"};
{\ar@{-} "26v34";"29v36"};
{\ar@{-} "29v26";"29v29"};
{\ar@{.} "14v26";"15v28"};
{\ar@{.} "27v26";"26v29"};
{\ar@{=}@`{"15v29", "19v26", "21v29", "25v26"}|(0.41){\hole}|(0.63){\SelectTips{eu}{12}\object@{<<}} "15v28"+(0.1,-0.2);"26v29"};
{\ar@{=}@`{"08v26", "11v22", "17v22", "20v27", "17v33", "11v33", "08v29"}|(0.36){\hole}|(0.88){\SelectTips{eu}{12}\object@{<<}} "08v27";"08v27"};
{\ar@{=}@`{"16v24", "18v23", "21v22", "23v23", "24v25"}|(0.6){\SelectTips{eu}{12}\object@{>>}} "14v26";"27v26"};
\endxy
}

\newcommand{\SliceCreation}{
\xy
0;<4pt,0pt>:
<0pt,4pt>::
(11,31)*{}="11v31";
(11,41)*{}="11v41";
(16,36)*{}="16v36";
(16,46)*{}="16v46";
(17,43)*{}="17v43";
(18,37)*{}="18v37";
(24,37)*{}="24v37";
(24,44)*{}="24v44";
(26,31)*{}="26v31";
(26,41)*{}="26v41";
(31,36)*{}="31v36";
(31,46)*{}="31v46";
"17v43"*@{*};
"24v44"*@{*};
{\ar@{-} "16v36";"11v31"};
{\ar@{-} "16v36";"31v36"};
{\ar@{-} "16v46";"11v41"};
{\ar@{-} "16v46";"31v46"};
{\ar@{-} "26v31";"11v31"};
{\ar@{-} "26v41";"11v41"};
{\ar@{-} "31v36";"26v31"};
{\ar@{-} "31v46";"26v41"};
{\ar@{=}@`{"18v37", "24v37"}|(0.15){\hole}|(0.6){\SelectTips{eu}{12}\object@{>>}}|(0.82){\hole} "17v43";"24v44"};
(24,38)*{\scriptstyle J};
\endxy
}

\newcommand{\Graphqm}{
\xy
0;<4pt,0pt>:
<0pt,4pt>::
(9,44)*{}="09v44";
(9,54)*{}="09v54";
(11,45)*{}="11v45";
(11,47)*{}="11v47";
(11,51)*{}="11v51";
(11,53)*{}="11v53";
(14,33)*{}="14v33";
(14,47)*{}="14v47";
(14,51)*{}="14v51";
(15,46)*{}="15v46";
(15,52)*{}="15v52";
(16,34)*{}="16v34";
(16,36)*{}="16v36";
(16,40)*{}="16v40";
(16,42)*{}="16v42";
(16,56)*{}="16v56";
(16,58)*{}="16v58";
(19,36)*{}="19v36";
(19,40)*{}="19v40";
(20,35)*{}="20v35";
(20,41)*{}="20v41";
(21,29)*{}="21v29";
(21,31)*{}="21v31";
(21,45)*{}="21v45";
(21,47)*{}="21v47";
(21,51)*{}="21v51";
(21,53)*{}="21v53";
(24,47)*{}="24v47";
(24,51)*{}="24v51";
(25,46)*{}="25v46";
(25,52)*{}="25v52";
(26,34)*{}="26v34";
(26,36)*{}="26v36";
(26,40)*{}="26v40";
(26,42)*{}="26v42";
(26,56)*{}="26v56";
(26,58)*{}="26v58";
(29,36)*{}="29v36";
(29,40)*{}="29v40";
(30,35)*{}="30v35";
(30,41)*{}="30v41";
(31,29)*{}="31v29";
(31,31)*{}="31v31";
(31,45)*{}="31v45";
(31,47)*{}="31v47";
(31,51)*{}="31v51";
(31,53)*{}="31v53";
(34,47)*{}="34v47";
(34,51)*{}="34v51";
(35,46)*{}="35v46";
(35,52)*{}="35v52";
(36,34)*{}="36v34";
(36,36)*{}="36v36";
(36,40)*{}="36v40";
(36,42)*{}="36v42";
(36,56)*{}="36v56";
(36,58)*{}="36v58";
(39,36)*{}="39v36";
(39,40)*{}="39v40";
(40,35)*{}="40v35";
(40,41)*{}="40v41";
(41,29)*{}="41v29";
(41,31)*{}="41v31";
(41,45)*{}="41v45";
(41,47)*{}="41v47";
(41,51)*{}="41v51";
(41,53)*{}="41v53";
(43,54)*{}="43v54";
(44,47)*{}="44v47";
(44,51)*{}="44v51";
(46,34)*{}="46v34";
(46,42)*{}="46v42";
(48,33)*{}="48v33";
(48,43)*{}="48v43";
"30v35"*{\scriptstyle q_{10}m_{10}};
"40v41"*{\scriptstyle q_{11}m_{11}};
"40v35"*{\scriptstyle q_{12}m_{12}};
"15v52"*{\scriptstyle q_ 1m_ 1};
"15v46"*{\scriptstyle q_ 2m_ 2};
"25v52"*{\scriptstyle q_ 3m_ 3};
"25v46"*{\scriptstyle q_ 4m_ 4};
"35v52"*{\scriptstyle q_ 5m_ 5};
"35v46"*{\scriptstyle q_ 6m_ 6};
"20v41"*{\scriptstyle q_{7}m_{7}};
"20v35"*{\scriptstyle q_{8}m_{8}};
"30v41"*{\scriptstyle q_{9}m_{9}};
{\ar@{-} "09v44";"11v45"};
{\ar@{-} "11v45";"11v53"};
{\ar@{-} "11v51";"14v51"};
{\ar@{-} "11v53";"09v54"};
{\ar@{-} "11v53";"11v45"};
{\ar@{-} "11v53";"16v56"};
{\ar@{-} "14v47";"11v47"};
{\ar@{-} "16v34";"14v33"};
{\ar@{-} "16v34";"21v31"};
{\ar@{-} "16v36";"19v36"};
{\ar@{-} "16v40";"19v40"};
{\ar@{-} "16v42";"11v45"};
{\ar@{-} "16v42";"16v34"};
{\ar@{-} "16v56";"16v58"};
{\ar@{-} "16v56";"21v53"};
{\ar@{-} "21v29";"21v31"};
{\ar@{-} "21v31";"26v34"};
{\ar@{-} "21v45";"16v42"};
{\ar@{-} "21v47";"24v47"};
{\ar@{-} "21v53";"21v45"};
{\ar@{-} "21v53";"26v56"};
{\ar@{-} "24v51";"21v51"};
{\ar@{-} "26v34";"26v42"};
{\ar@{-} "26v34";"31v31"};
{\ar@{-} "26v36";"29v36"};
{\ar@{-} "26v40";"29v40"};
{\ar@{-} "26v42";"21v45"};
{\ar@{-} "26v56";"26v58"};
{\ar@{-} "26v56";"31v53"};
{\ar@{-} "31v31";"31v29"};
{\ar@{-} "31v31";"36v34"};
{\ar@{-} "31v45";"26v42"};
{\ar@{-} "31v47";"34v47"};
{\ar@{-} "31v51";"34v51"};
{\ar@{-} "31v53";"31v45"};
{\ar@{-} "31v53";"36v56"};
{\ar@{-} "36v34";"36v42"};
{\ar@{-} "36v34";"41v31"};
{\ar@{-} "36v36";"39v36"};
{\ar@{-} "36v40";"39v40"};
{\ar@{-} "36v42";"31v45"};
{\ar@{-} "36v56";"36v58"};
{\ar@{-} "36v56";"41v53"};
{\ar@{-} "41v31";"41v29"};
{\ar@{-} "41v31";"46v34"};
{\ar@{-} "41v45";"36v42"};
{\ar@{-} "41v47";"44v47"};
{\ar@{-} "41v51";"44v51"};
{\ar@{-} "41v53";"41v45"};
{\ar@{-} "41v53";"43v54"};
{\ar@{-} "46v34";"46v42"};
{\ar@{-} "46v34";"48v33"};
{\ar@{-} "46v42";"41v45"};
{\ar@{-} "46v42";"48v43"};
\endxy
}

\newcommand{\TriangulationGraph}{
\xy
0;<4pt,0pt>:
<0pt,4pt>::
(6,49)*{}="06v49";
(9,44)*{}="09v44";
(9,54)*{}="09v54";
(11,38)*{}="11v38";
(11,45)*{}="11v45";
(11,47)*{}="11v47";
(11,51)*{}="11v51";
(11,53)*{}="11v53";
(11,60)*{}="11v60";
(14,33)*{}="14v33";
(14,47)*{}="14v47";
(14,51)*{}="14v51";
(15,46)*{}="15v46";
(15,52)*{}="15v52";
(16,27)*{}="16v27";
(16,34)*{}="16v34";
(16,36)*{}="16v36";
(16,40)*{}="16v40";
(16,42)*{}="16v42";
(16,49)*{}="16v49";
(16,56)*{}="16v56";
(16,58)*{}="16v58";
(19,36)*{}="19v36";
(19,40)*{}="19v40";
(20,35)*{}="20v35";
(20,41)*{}="20v41";
(21,29)*{}="21v29";
(21,31)*{}="21v31";
(21,38)*{}="21v38";
(21,45)*{}="21v45";
(21,47)*{}="21v47";
(21,51)*{}="21v51";
(21,53)*{}="21v53";
(21,60)*{}="21v60";
(24,47)*{}="24v47";
(24,51)*{}="24v51";
(25,46)*{}="25v46";
(25,52)*{}="25v52";
(26,27)*{}="26v27";
(26,34)*{}="26v34";
(26,36)*{}="26v36";
(26,40)*{}="26v40";
(26,42)*{}="26v42";
(26,49)*{}="26v49";
(26,56)*{}="26v56";
(26,58)*{}="26v58";
(29,36)*{}="29v36";
(29,40)*{}="29v40";
(30,35)*{}="30v35";
(30,41)*{}="30v41";
(31,29)*{}="31v29";
(31,31)*{}="31v31";
(31,38)*{}="31v38";
(31,45)*{}="31v45";
(31,47)*{}="31v47";
(31,51)*{}="31v51";
(31,53)*{}="31v53";
(31,60)*{}="31v60";
(34,47)*{}="34v47";
(34,51)*{}="34v51";
(35,46)*{}="35v46";
(35,52)*{}="35v52";
(36,27)*{}="36v27";
(36,34)*{}="36v34";
(36,36)*{}="36v36";
(36,40)*{}="36v40";
(36,42)*{}="36v42";
(36,49)*{}="36v49";
(36,56)*{}="36v56";
(36,58)*{}="36v58";
(39,36)*{}="39v36";
(39,40)*{}="39v40";
(40,35)*{}="40v35";
(40,41)*{}="40v41";
(41,29)*{}="41v29";
(41,31)*{}="41v31";
(41,38)*{}="41v38";
(41,45)*{}="41v45";
(41,47)*{}="41v47";
(41,51)*{}="41v51";
(41,53)*{}="41v53";
(41,60)*{}="41v60";
(43,54)*{}="43v54";
(44,27)*{}="44v27";
(44,47)*{}="44v47";
(44,51)*{}="44v51";
(46,34)*{}="46v34";
(46,42)*{}="46v42";
(46,49)*{}="46v49";
(48,33)*{}="48v33";
(48,43)*{}="48v43";
(51,38)*{}="51v38";
{\ar@{.} "09v44";"11v45"};
{\ar@{.} "11v45";"11v53"};
{\ar@{.} "11v51";"14v51"};
{\ar@{.} "11v53";"09v54"};
{\ar@{.} "11v53";"11v45"};
{\ar@{.} "11v53";"16v56"};
{\ar@{.} "14v47";"11v47"};
{\ar@{.} "16v34";"14v33"};
{\ar@{.} "16v34";"21v31"};
{\ar@{.} "16v36";"19v36"};
{\ar@{.} "16v40";"19v40"};
{\ar@{.} "16v42";"11v45"};
{\ar@{.} "16v42";"16v34"};
{\ar@{.} "16v56";"16v58"};
{\ar@{.} "16v56";"21v53"};
{\ar@{.} "21v29";"21v31"};
{\ar@{.} "21v31";"26v34"};
{\ar@{.} "21v45";"16v42"};
{\ar@{.} "21v47";"24v47"};
{\ar@{.} "21v53";"21v45"};
{\ar@{.} "21v53";"26v56"};
{\ar@{.} "24v51";"21v51"};
{\ar@{.} "26v34";"26v42"};
{\ar@{.} "26v34";"31v31"};
{\ar@{.} "26v36";"29v36"};
{\ar@{.} "26v40";"29v40"};
{\ar@{.} "26v42";"21v45"};
{\ar@{.} "26v56";"26v58"};
{\ar@{.} "26v56";"31v53"};
{\ar@{.} "31v31";"31v29"};
{\ar@{.} "31v31";"36v34"};
{\ar@{.} "31v45";"26v42"};
{\ar@{.} "31v47";"34v47"};
{\ar@{.} "31v51";"34v51"};
{\ar@{.} "31v53";"31v45"};
{\ar@{.} "31v53";"36v56"};
{\ar@{.} "36v34";"36v42"};
{\ar@{.} "36v34";"41v31"};
{\ar@{.} "36v36";"39v36"};
{\ar@{.} "36v40";"39v40"};
{\ar@{.} "36v42";"31v45"};
{\ar@{.} "36v56";"36v58"};
{\ar@{.} "36v56";"41v53"};
{\ar@{.} "41v31";"41v29"};
{\ar@{.} "41v31";"46v34"};
{\ar@{.} "41v45";"36v42"};
{\ar@{.} "41v47";"44v47"};
{\ar@{.} "41v51";"44v51"};
{\ar@{.} "41v53";"41v45"};
{\ar@{.} "41v53";"43v54"};
{\ar@{.} "46v34";"46v42"};
{\ar@{.} "46v34";"48v33"};
{\ar@{.} "46v42";"41v45"};
{\ar@{.} "46v42";"48v43"};
{\ar@{-} "06v49";"11v60"};
{\ar@{-} "11v38";"06v49"};
{\ar@{-} "11v38";"16v27"};
{\ar@{-} "11v60";"16v49"};
{\ar@{-} "16v27";"21v38"};
{\ar@{-} "16v49";"06v49"};
{\ar@{-} "16v49";"11v38"};
{\ar@{-} "16v49";"21v38"};
{\ar@{-} "16v49";"21v60"};
{\ar@{-} "16v49";"26v49"};
{\ar@{-} "21v38";"11v38"};
{\ar@{-} "21v38";"26v27"};
{\ar@{-} "21v38";"26v49"};
{\ar@{-} "21v60";"26v49"};
{\ar@{-} "26v27";"31v38"};
{\ar@{-} "26v49";"31v38"};
{\ar@{-} "26v49";"31v60"};
{\ar@{-} "26v49";"36v49"};
{\ar@{-} "31v38";"21v38"};
{\ar@{-} "31v38";"36v27"};
{\ar@{-} "31v38";"36v49"};
{\ar@{-} "31v60";"36v49"};
{\ar@{-} "36v27";"41v38"};
{\ar@{-} "36v49";"41v38"};
{\ar@{-} "36v49";"41v60"};
{\ar@{-} "36v49";"46v49"};
{\ar@{-} "41v38";"31v38"};
{\ar@{-} "41v38";"44v27"};
{\ar@{-} "41v60";"46v49"};
{\ar@{-} "44v27";"51v38"};
{\ar@{-} "46v49";"41v38"};
{\ar@{-} "51v38";"46v49"};
\endxy
}

\newcommand{\VertexTail}{
\xy
0;<4pt,0pt>:
<0pt,4pt>::
(6,17)*{}="06v17";
(9,17)*{}="09v17";
(9,23)*{}="09v23";
(9,27)*{}="09v27";
(11,21)*{}="11v21";
(11,24)*{}="11v24";
(11,28)*{}="11v28";
(13,17)*{}="13v17";
(14,24)*{}="14v24";
(15,23)*{}="15v23";
(16,17)*{}="16v17";
"09v27"*{\scriptstyle i};
"09v17"*{\scriptstyle j};
"13v17"*{\scriptstyle k};
"09v23"*{\scriptstyle l};
"15v23"*{\scriptstyle qm};
{\ar@{>} "11v24";"14v24"};
{\ar@{-}|@{>} "06v17";"11v21"};
{\ar@{-}|@{>} "11v24";"11v21"};
{\ar@{-}|@{>} "11v28";"11v24"};
{\ar@{-}|@{>} "16v17";"11v21"};
\endxy
}

\newcommand{\TailedLattice}{
\xy
0;<2pt,0pt>:
<0pt,3pt>::
(4,26)*{}="04v26";
(4,38)*{}="04v38";
(6,30)*{}="06v30";
(8,28)*{}="08v28";
(8,30)*{}="08v30";
(8,34)*{}="08v34";
(8,36)*{}="08v36";
(12,14)*{}="12v14";
(12,34)*{}="12v34";
(12,42)*{}="12v42";
(12,50)*{}="12v50";
(14,18)*{}="14v18";
(16,16)*{}="16v16";
(16,18)*{}="16v18";
(16,22)*{}="16v22";
(16,24)*{}="16v24";
(16,40)*{}="16v40";
(16,42)*{}="16v42";
(16,46)*{}="16v46";
(16,48)*{}="16v48";
(20,22)*{}="20v22";
(20,30)*{}="20v30";
(20,46)*{}="20v46";
(24,8)*{}="24v08";
(24,12)*{}="24v12";
(24,28)*{}="24v28";
(24,30)*{}="24v30";
(24,34)*{}="24v34";
(24,36)*{}="24v36";
(24,52)*{}="24v52";
(24,56)*{}="24v56";
(28,18)*{}="28v18";
(28,34)*{}="28v34";
(28,42)*{}="28v42";
(32,16)*{}="32v16";
(32,18)*{}="32v18";
(32,22)*{}="32v22";
(32,24)*{}="32v24";
(32,40)*{}="32v40";
(32,42)*{}="32v42";
(32,46)*{}="32v46";
(32,48)*{}="32v48";
(36,22)*{}="36v22";
(36,30)*{}="36v30";
(36,46)*{}="36v46";
(40,8)*{}="40v08";
(40,12)*{}="40v12";
(40,28)*{}="40v28";
(40,30)*{}="40v30";
(40,34)*{}="40v34";
(40,36)*{}="40v36";
(40,52)*{}="40v52";
(40,56)*{}="40v56";
(44,18)*{}="44v18";
(44,34)*{}="44v34";
(44,42)*{}="44v42";
(48,16)*{}="48v16";
(48,18)*{}="48v18";
(48,22)*{}="48v22";
(48,24)*{}="48v24";
(48,40)*{}="48v40";
(48,42)*{}="48v42";
(48,46)*{}="48v46";
(48,48)*{}="48v48";
(52,22)*{}="52v22";
(52,30)*{}="52v30";
(52,46)*{}="52v46";
(56,8)*{}="56v08";
(56,12)*{}="56v12";
(56,28)*{}="56v28";
(56,30)*{}="56v30";
(56,34)*{}="56v34";
(56,36)*{}="56v36";
(56,52)*{}="56v52";
(56,56)*{}="56v56";
(60,18)*{}="60v18";
(60,34)*{}="60v34";
(60,42)*{}="60v42";
(64,16)*{}="64v16";
(64,18)*{}="64v18";
(64,22)*{}="64v22";
(64,24)*{}="64v24";
(64,40)*{}="64v40";
(64,42)*{}="64v42";
(64,46)*{}="64v46";
(64,48)*{}="64v48";
(66,46)*{}="66v46";
(68,14)*{}="68v14";
(68,22)*{}="68v22";
(68,30)*{}="68v30";
(68,50)*{}="68v50";
(72,28)*{}="72v28";
(72,30)*{}="72v30";
(72,34)*{}="72v34";
(72,36)*{}="72v36";
(74,34)*{}="74v34";
(76,26)*{}="76v26";
(76,38)*{}="76v38";
{\ar@{-} "04v38";"08v36"};
{\ar@{-} "08v28";"04v26"};
{\ar@{-} "08v28";"16v24"};
{\ar@{-} "08v30";"06v30"};
{\ar@{-} "08v34";"12v34"};
{\ar@{-} "08v36";"08v28"};
{\ar@{-} "16v16";"12v14"};
{\ar@{-} "16v16";"16v24"};
{\ar@{-} "16v18";"14v18"};
{\ar@{-} "16v22";"20v22"};
{\ar@{-} "16v24";"24v28"};
{\ar@{-} "16v40";"08v36"};
{\ar@{-} "16v40";"16v48"};
{\ar@{-} "16v42";"12v42"};
{\ar@{-} "16v46";"20v46"};
{\ar@{-} "16v48";"12v50"};
{\ar@{-} "16v48";"24v52"};
{\ar@{-} "24v08";"24v12"};
{\ar@{-} "24v12";"16v16"};
{\ar@{-} "24v28";"32v24"};
{\ar@{-} "24v30";"20v30"};
{\ar@{-} "24v34";"28v34"};
{\ar@{-} "24v36";"16v40"};
{\ar@{-} "24v36";"24v28"};
{\ar@{-} "24v52";"24v56"};
{\ar@{-} "24v52";"32v48"};
{\ar@{-} "32v16";"24v12"};
{\ar@{-} "32v16";"40v12"};
{\ar@{-} "32v18";"28v18"};
{\ar@{-} "32v22";"36v22"};
{\ar@{-} "32v24";"32v16"};
{\ar@{-} "32v24";"40v28"};
{\ar@{-} "32v40";"24v36"};
{\ar@{-} "32v42";"28v42"};
{\ar@{-} "32v46";"36v46"};
{\ar@{-} "32v48";"32v40"};
{\ar@{-} "40v12";"40v08"};
{\ar@{-} "40v12";"48v16"};
{\ar@{-} "40v28";"40v36"};
{\ar@{-} "40v30";"36v30"};
{\ar@{-} "40v34";"44v34"};
{\ar@{-} "40v36";"32v40"};
{\ar@{-} "40v36";"48v40"};
{\ar@{-} "40v52";"32v48"};
{\ar@{-} "40v52";"40v56"};
{\ar@{-} "48v16";"48v24"};
{\ar@{-} "48v16";"56v12"};
{\ar@{-} "48v18";"44v18"};
{\ar@{-} "48v22";"52v22"};
{\ar@{-} "48v24";"40v28"};
{\ar@{-} "48v40";"56v36"};
{\ar@{-} "48v42";"44v42"};
{\ar@{-} "48v46";"52v46"};
{\ar@{-} "48v48";"40v52"};
{\ar@{-} "48v48";"48v40"};
{\ar@{-} "56v12";"56v08"};
{\ar@{-} "56v12";"64v16"};
{\ar@{-} "56v28";"48v24"};
{\ar@{-} "56v28";"64v24"};
{\ar@{-} "56v30";"52v30"};
{\ar@{-} "56v34";"60v34"};
{\ar@{-} "56v36";"56v28"};
{\ar@{-} "56v36";"64v40"};
{\ar@{-} "56v52";"48v48"};
{\ar@{-} "56v52";"56v56"};
{\ar@{-} "64v16";"64v24"};
{\ar@{-} "64v16";"68v14"};
{\ar@{-} "64v18";"60v18"};
{\ar@{-} "64v22";"68v22"};
{\ar@{-} "64v24";"72v28"};
{\ar@{-} "64v40";"64v48"};
{\ar@{-} "64v42";"60v42"};
{\ar@{-} "64v46";"66v46"};
{\ar@{-} "64v48";"56v52"};
{\ar@{-} "64v48";"68v50"};
{\ar@{-} "72v28";"72v36"};
{\ar@{-} "72v28";"76v26"};
{\ar@{-} "72v30";"68v30"};
{\ar@{-} "72v34";"74v34"};
{\ar@{-} "72v36";"64v40"};
{\ar@{-} "72v36";"76v38"};
\endxy
}

\newcommand{\fusionTwoDyonPair}{
\xy
0;<3pt,0pt>:
<0pt,3pt>::
(24,28)*{}="24v28";
(24,48)*{}="24v48";
(26,38)*{}="26v38";
(32,30)*{}="32v30";
(32,32)*{}="32v32";
(32,44)*{}="32v44";
(32,46)*{}="32v46";
(38,38)*{}="38v38";
(40,28)*{}="40v28";
(40,48)*{}="40v48";
(46,38)*{}="46v38";
(60,28)*{}="60v28";
(60,48)*{}="60v48";
(62,38)*{}="62v38";
(66,40)*{}="66v40";
(68,32)*{}="68v32";
(68,44)*{}="68v44";
(70,36)*{}="70v36";
(74,38)*{}="74v38";
(76,28)*{}="76v28";
(76,48)*{}="76v48";
"26v38"*@{*};
"38v38"*@{*};
"62v38"*@{*};
"74v38"*@{*};
"46v38"*{\scriptstyle =\,\sum_L};
{\ar@{-} "32v32";"24v28"};
{\ar@{-}|(0.2)\hole|(0.8)\hole "32v32";"32v44"};
{\ar@{-} "32v32";"40v28"};
{\ar@{-} "32v44";"24v48"};
{\ar@{-} "32v44";"40v48"};
{\ar@{-} "60v48";"68v44"};
{\ar@{-} "68v32";"60v28"};
{\ar@{-} "68v32";"76v28"};
{\ar@{-}|(0.5)\hole "68v44";"68v32"};
{\ar@{-} "68v44";"76v48"};
{\ar@{=}@`{"32v30"}|(0.65){\SelectTips{eu}{12}\object@{>>}} "26v38";"38v38"};
{\ar@{=}@`{"32v46"}|(0.65){\SelectTips{eu}{12}\object@{>>}} "26v38";"38v38"};
{\ar@{=}@`{"66v40", "70v36"}|(0.65){\SelectTips{eu}{12}\object@{>>}} "62v38";"74v38"};
(37,42)*{\scriptstyle J};
(37,32)*{\scriptstyle K};
(72,40)*{\scriptstyle L};
\endxy
}

\newcommand{\FusionThreeDyon}{
\xy
0;<3pt,0pt>:
<0pt,3pt>::
(4,26)*{}="04v26";
(4,38)*{}="04v38";
(8,28)*{}="08v28";
(8,36)*{}="08v36";
(12,14)*{}="12v14";
(12,50)*{}="12v50";
(16,16)*{}="16v16";
(16,24)*{}="16v24";
(16,32)*{}="16v32";
(16,40)*{}="16v40";
(16,48)*{}="16v48";
(20,34)*{}="20v34";
(22,26)*{}="22v26";
(24,8)*{}="24v08";
(24,12)*{}="24v12";
(24,28)*{}="24v28";
(24,34)*{}="24v34";
(24,36)*{}="24v36";
(24,42)*{}="24v42";
(24,52)*{}="24v52";
(24,56)*{}="24v56";
(26,40)*{}="26v40";
(26,46)*{}="26v46";
(28,30)*{}="28v30";
(30,42)*{}="30v42";
(32,16)*{}="32v16";
(32,24)*{}="32v24";
(32,32)*{}="32v32";
(32,36)*{}="32v36";
(32,40)*{}="32v40";
(32,48)*{}="32v48";
(36,32)*{}="36v32";
(38,34)*{}="38v34";
(40,8)*{}="40v08";
(40,12)*{}="40v12";
(40,28)*{}="40v28";
(40,36)*{}="40v36";
(40,52)*{}="40v52";
(40,56)*{}="40v56";
(42,30)*{}="42v30";
(46,28)*{}="46v28";
(48,16)*{}="48v16";
(48,24)*{}="48v24";
(48,32)*{}="48v32";
(48,40)*{}="48v40";
(48,48)*{}="48v48";
(56,8)*{}="56v08";
(56,12)*{}="56v12";
(56,28)*{}="56v28";
(56,36)*{}="56v36";
(56,52)*{}="56v52";
(56,56)*{}="56v56";
(64,16)*{}="64v16";
(64,24)*{}="64v24";
(64,40)*{}="64v40";
(64,48)*{}="64v48";
(68,14)*{}="68v14";
(68,50)*{}="68v50";
(72,28)*{}="72v28";
(72,36)*{}="72v36";
(76,26)*{}="76v26";
(76,38)*{}="76v38";
"16v32"*@{*};
"26v46"*@{*};
"28v30"*@{*};
"32v36"*@{*};
"36v32"*@{*};
"48v32"*@{*};
"20v34"*{\scriptstyle J};
"24v42"*{\scriptstyle K};
"46v28"*{\scriptstyle L};
"32v32"*{\xycircle(6,6){--}};
{\ar@{-} "04v38";"08v36"};
{\ar@{-} "08v28";"04v26"};
{\ar@{-} "08v28";"16v24"};
{\ar@{-} "08v36";"08v28"};
{\ar@{-} "16v16";"12v14"};
{\ar@{-} "16v16";"16v24"};
{\ar@{-} "16v24";"24v28"};
{\ar@{-} "16v40";"08v36"};
{\ar@{-} "16v40";"16v48"};
{\ar@{-} "16v48";"12v50"};
{\ar@{-} "16v48";"24v52"};
{\ar@{-} "24v08";"24v12"};
{\ar@{-} "24v12";"16v16"};
{\ar@{-} "24v28";"32v24"};
{\ar@{-} "24v36";"16v40"};
{\ar@{-}|(0.65)\hole "24v36";"24v28"};
{\ar@{-} "24v52";"24v56"};
{\ar@{-} "24v52";"32v48"};
{\ar@{-} "32v16";"24v12"};
{\ar@{-} "32v16";"40v12"};
{\ar@{-} "32v24";"32v16"};
{\ar@{-} "32v24";"40v28"};
{\ar@{-}|(0.2)\hole "32v40";"24v36"};
{\ar@{-} "32v48";"32v40"};
{\ar@{-} "40v12";"40v08"};
{\ar@{-} "40v12";"48v16"};
{\ar@{-}|(0.5)\hole "40v28";"40v36"};
{\ar@{-} "40v36";"32v40"};
{\ar@{-} "40v36";"48v40"};
{\ar@{-} "40v52";"32v48"};
{\ar@{-} "40v52";"40v56"};
{\ar@{-} "48v16";"48v24"};
{\ar@{-} "48v16";"56v12"};
{\ar@{-} "48v24";"40v28"};
{\ar@{-} "48v40";"56v36"};
{\ar@{-} "48v48";"40v52"};
{\ar@{-} "48v48";"48v40"};
{\ar@{-} "56v12";"56v08"};
{\ar@{-} "56v12";"64v16"};
{\ar@{-} "56v28";"48v24"};
{\ar@{-} "56v28";"64v24"};
{\ar@{-} "56v36";"56v28"};
{\ar@{-} "56v36";"64v40"};
{\ar@{-} "56v52";"48v48"};
{\ar@{-} "56v52";"56v56"};
{\ar@{-} "64v16";"64v24"};
{\ar@{-} "64v16";"68v14"};
{\ar@{-} "64v24";"72v28"};
{\ar@{-} "64v40";"64v48"};
{\ar@{-} "64v48";"56v52"};
{\ar@{-} "64v48";"68v50"};
{\ar@{-} "72v28";"72v36"};
{\ar@{-} "72v28";"76v26"};
{\ar@{-} "72v36";"64v40"};
{\ar@{-} "72v36";"76v38"};
{\ar@{.} "28v30";"32v32"};
{\ar@{.} "32v32";"32v36"};
{\ar@{.} "32v32";"36v32"};
{\ar@{=}@`{"22v26", "24v34"}|(0.35){\SelectTips{eu}{12}\object@{>>}} "16v32";"28v30"};
{\ar@{=}@`{"26v40", "30v42"}|(0.35){\SelectTips{eu}{12}\object@{>>}} "26v46";"32v36"};
{\ar@{=}@`{"42v30", "38v34"}|(0.35){\SelectTips{eu}{12}\object@{>>}} "48v32";"36v32"};
"16v32"+(-2,0)*{\scriptstyle j};
"26v46"+(-2,2)*{\scriptstyle k};
"48v32"+(2,2)*{\scriptstyle l};
\endxy
}

\newcommand{\ThreeDyonState}{
\xy
0;<3pt,0pt>:
<0pt,3pt>::
(4,26)*{}="04v26";
(4,38)*{}="04v38";
(8,28)*{}="08v28";
(8,36)*{}="08v36";
(12,14)*{}="12v14";
(12,50)*{}="12v50";
(16,16)*{}="16v16";
(16,24)*{}="16v24";
(16,32)*{}="16v32";
(16,40)*{}="16v40";
(16,48)*{}="16v48";
(20,34)*{}="20v34";
(22,26)*{}="22v26";
(24,8)*{}="24v08";
(24,12)*{}="24v12";
(24,28)*{}="24v28";
(24,34)*{}="24v34";
(24,36)*{}="24v36";
(24,42)*{}="24v42";
(24,52)*{}="24v52";
(24,56)*{}="24v56";
(26,40)*{}="26v40";
(26,46)*{}="26v46";
(32,32)*{}="32v32";
(30,42)*{}="30v42";
(32,16)*{}="32v16";
(32,24)*{}="32v24";
(32,32)*{}="32v32";
(32,32)*{}="32v32";
(32,40)*{}="32v40";
(32,48)*{}="32v48";
(32,32)*{}="32v32";
(38,34)*{}="38v34";
(40,8)*{}="40v08";
(40,12)*{}="40v12";
(40,28)*{}="40v28";
(40,36)*{}="40v36";
(40,52)*{}="40v52";
(40,56)*{}="40v56";
(42,30)*{}="42v30";
(46,28)*{}="46v28";
(48,16)*{}="48v16";
(48,24)*{}="48v24";
(48,32)*{}="48v32";
(48,40)*{}="48v40";
(48,48)*{}="48v48";
(56,8)*{}="56v08";
(56,12)*{}="56v12";
(56,28)*{}="56v28";
(56,36)*{}="56v36";
(56,52)*{}="56v52";
(56,56)*{}="56v56";
(64,16)*{}="64v16";
(64,24)*{}="64v24";
(64,40)*{}="64v40";
(64,48)*{}="64v48";
(68,14)*{}="68v14";
(68,50)*{}="68v50";
(72,28)*{}="72v28";
(72,36)*{}="72v36";
(76,26)*{}="76v26";
(76,38)*{}="76v38";
"16v32"*@{*};
"26v46"*@{*};
"48v32"*@{*};
"20v34"*{\scriptstyle J};
"24v42"*{\scriptstyle K};
"46v28"*{\scriptstyle L};
{\ar@{-} "04v38";"08v36"};
{\ar@{-} "08v28";"04v26"};
{\ar@{-} "08v28";"16v24"};
{\ar@{-} "08v36";"08v28"};
{\ar@{-} "16v16";"12v14"};
{\ar@{-} "16v16";"16v24"};
{\ar@{-} "16v24";"24v28"};
{\ar@{-} "16v40";"08v36"};
{\ar@{-} "16v40";"16v48"};
{\ar@{-} "16v48";"12v50"};
{\ar@{-} "16v48";"24v52"};
{\ar@{-} "24v08";"24v12"};
{\ar@{-} "24v12";"16v16"};
{\ar@{-} "24v28";"32v24"};
{\ar@{-} "24v36";"16v40"};
{\ar@{-}|(0.7)\hole "24v36";"24v28"};
{\ar@{-} "24v52";"24v56"};
{\ar@{-} "24v52";"32v48"};
{\ar@{-} "32v16";"24v12"};
{\ar@{-} "32v16";"40v12"};
{\ar@{-} "32v24";"32v16"};
{\ar@{-} "32v24";"40v28"};
{\ar@{-}|(0.2)\hole "32v40";"24v36"};
{\ar@{-} "32v48";"32v40"};
{\ar@{-} "40v12";"40v08"};
{\ar@{-} "40v12";"48v16"};
{\ar@{-}|(0.5)\hole "40v28";"40v36"};
{\ar@{-} "40v36";"32v40"};
{\ar@{-} "40v36";"48v40"};
{\ar@{-} "40v52";"32v48"};
{\ar@{-} "40v52";"40v56"};
{\ar@{-} "48v16";"48v24"};
{\ar@{-} "48v16";"56v12"};
{\ar@{-} "48v24";"40v28"};
{\ar@{-} "48v40";"56v36"};
{\ar@{-} "48v48";"40v52"};
{\ar@{-} "48v48";"48v40"};
{\ar@{-} "56v12";"56v08"};
{\ar@{-} "56v12";"64v16"};
{\ar@{-} "56v28";"48v24"};
{\ar@{-} "56v28";"64v24"};
{\ar@{-} "56v36";"56v28"};
{\ar@{-} "56v36";"64v40"};
{\ar@{-} "56v52";"48v48"};
{\ar@{-} "56v52";"56v56"};
{\ar@{-} "64v16";"64v24"};
{\ar@{-} "64v16";"68v14"};
{\ar@{-} "64v24";"72v28"};
{\ar@{-} "64v40";"64v48"};
{\ar@{-} "64v48";"56v52"};
{\ar@{-} "64v48";"68v50"};
{\ar@{-} "72v28";"72v36"};
{\ar@{-} "72v28";"76v26"};
{\ar@{-} "72v36";"64v40"};
{\ar@{-} "72v36";"76v38"};
{\ar@{=}@`{"22v26", "24v34"}|(0.35){\SelectTips{eu}{12}\object@{>>}} "16v32";"32v32"};
{\ar@{=}@`{"26v40", "30v42"}|(0.45){\SelectTips{eu}{12}\object@{>>}} "26v46";"32v32"};
{\ar@{=}@`{"42v30", "38v34"}|(0.35){\SelectTips{eu}{12}\object@{>>}} "48v32";"32v32"};
"16v32"+(-2,0)*{\scriptstyle j};
"26v46"+(-2,2)*{\scriptstyle k};
"48v32"+(2,2)*{\scriptstyle l};
"32v32"*{\scriptstyle \circ};
\endxy
}

\newcommand{\TwoDyonContraction}{
\xy
0;<3pt,0pt>:
<0pt,3pt>::
(12,14)*{}="12v14";
(12,30)*{}="12v30";
(16,16)*{}="16v16";
(16,28)*{}="16v28";
(20,24)*{}="20v24";
(24,8)*{}="24v08";
(24,12)*{}="24v12";
(24,22)*{}="24v22";
(24,32)*{}="24v32";
(24,36)*{}="24v36";
(28,14)*{}="28v14";
(28,24)*{}="28v24";
(32,16)*{}="32v16";
(32,28)*{}="32v28";
(34,26)*{}="34v26";
(38,26)*{}="38v26";
(40,8)*{}="40v08";
(40,12)*{}="40v12";
(40,32)*{}="40v32";
(40,36)*{}="40v36";
(42,20)*{}="42v20";
(46,16)*{}="46v16";
(48,16)*{}="48v16";
(48,28)*{}="48v28";
(52,18)*{}="52v18";
(52,26)*{}="52v26";
(56,8)*{}="56v08";
(56,12)*{}="56v12";
(56,20)*{}="56v20";
(56,32)*{}="56v32";
(56,36)*{}="56v36";
(60,18)*{}="60v18";
(64,16)*{}="64v16";
(64,28)*{}="64v28";
(68,14)*{}="68v14";
(68,30)*{}="68v30";
"24v22"*@{*};
"38v26"*@{*};
"42v20"*@{*};
"56v20"*@{*};
"28v24"*{\scriptstyle J};
"52v18"*{\scriptstyle J};
"20v24"*{\scriptstyle p};
"60v18"*{\scriptstyle q};
{\ar@{-} "16v16";"12v14"};
{\ar@{-} "16v16";"24v12"};
{\ar@{-} "16v28";"12v30"};
{\ar@{-} "16v28";"16v16"};
{\ar@{-} "24v12";"24v08"};
{\ar@{-} "24v12";"32v16"};
{\ar@{-} "24v32";"16v28"};
{\ar@{-} "24v32";"24v36"};
{\ar@{-}|(0.5)\hole "32v16";"32v28"};
{\ar@{-} "32v16";"40v12"};
{\ar@{-} "32v28";"24v32"};
{\ar@{-} "32v28";"40v32"};
{\ar@{-} "40v12";"40v08"};
{\ar@{-} "40v12";"48v16"};
{\ar@{-} "40v32";"40v36"};
{\ar@{-}|(0.5)\hole "48v16";"48v28"};
{\ar@{-} "48v16";"56v12"};
{\ar@{-} "48v28";"40v32"};
{\ar@{-} "48v28";"56v32"};
{\ar@{-} "56v12";"56v08"};
{\ar@{-} "56v32";"56v36"};
{\ar@{-} "56v32";"64v28"};
{\ar@{-} "64v16";"56v12"};
{\ar@{-} "64v16";"68v14"};
{\ar@{-} "64v28";"64v16"};
{\ar@{-} "64v28";"68v30"};
{\ar@{.} "38v26";"42v20"};
{\ar@{=}@`{"28v14", "34v26"} "24v22";"38v26"};
{\ar@{=}@`{"52v26", "46v16"} "56v20";"42v20"};
\endxy
}

\newcommand{\TwoDyonFusion}{
\xy
0;<4pt,0pt>:
<0pt,4pt>::
(12,14)*{}="12v14";
(12,30)*{}="12v30";
(16,16)*{}="16v16";
(16,28)*{}="16v28";
(20,24)*{}="20v24";
(24,8)*{}="24v08";
(24,12)*{}="24v12";
(24,22)*{}="24v22";
(24,32)*{}="24v32";
(24,36)*{}="24v36";
(28,14)*{}="28v14";
(28,24)*{}="28v24";
(32,16)*{}="32v16";
(32,28)*{}="32v28";
(34,26)*{}="34v26";
(38,18)*{}="38v18";
(38,26)*{}="38v26";
(40,8)*{}="40v08";
(40,12)*{}="40v12";
(40,22)*{}="40v22";
(40,28)*{}="40v28";
(40,32)*{}="40v32";
(40,36)*{}="40v36";
(42,20)*{}="42v20";
(46,16)*{}="46v16";
(48,16)*{}="48v16";
(48,28)*{}="48v28";
(52,18)*{}="52v18";
(52,26)*{}="52v26";
(56,8)*{}="56v08";
(56,12)*{}="56v12";
(56,20)*{}="56v20";
(56,32)*{}="56v32";
(56,36)*{}="56v36";
(60,18)*{}="60v18";
(64,16)*{}="64v16";
(64,28)*{}="64v28";
(68,14)*{}="68v14";
(68,30)*{}="68v30";
"24v22"*@{*};
"38v26"*@{*};
"42v20"*@{*};
"56v20"*@{*};
"28v24"*{\scriptstyle J};
"52v18"*{\scriptstyle J};
"20v24"*{\scriptstyle p};
"40v28"*{\scriptstyle k};
"38v18"*{\scriptstyle k^*};
"60v18"*{\scriptstyle k};
"40v22"*{\xycircle(7,8){--}};
{\ar@{-} "16v16";"12v14"};
{\ar@{-} "16v16";"24v12"};
{\ar@{-} "16v28";"12v30"};
{\ar@{-} "16v28";"16v16"};
{\ar@{-} "24v12";"24v08"};
{\ar@{-} "24v12";"32v16"};
{\ar@{-} "24v32";"16v28"};
{\ar@{-} "24v32";"24v36"};
{\ar@{-}|(0.5)\hole "32v16";"32v28"};
{\ar@{-} "32v16";"40v12"};
{\ar@{-} "32v28";"24v32"};
{\ar@{-} "32v28";"40v32"};
{\ar@{-} "40v12";"40v08"};
{\ar@{-} "40v12";"48v16"};
{\ar@{-} "40v32";"40v36"};
{\ar@{-}|(0.5)\hole "48v16";"48v28"};
{\ar@{-} "48v16";"56v12"};
{\ar@{-} "48v28";"40v32"};
{\ar@{-} "48v28";"56v32"};
{\ar@{-} "56v12";"56v08"};
{\ar@{-} "56v32";"56v36"};
{\ar@{-} "56v32";"64v28"};
{\ar@{-} "64v16";"56v12"};
{\ar@{-} "64v16";"68v14"};
{\ar@{-} "64v28";"64v16"};
{\ar@{-} "64v28";"68v30"};
{\ar@{=}@`{"28v14", "34v26"}|(0.35){\SelectTips{eu}{12}\object@{>>}} "24v22";"38v26"};
{\ar@{=}@`{"52v26", "46v16"}|(0.35){\SelectTips{eu}{12}\object@{>>}} "56v20";"42v20"};
\endxy
}

\newcommand{\TwoDyonFusionEq}{
\xy
0;<2.5pt,0pt>:
<0pt,2.5pt>::
(16,32)*{}="16v32";
(20,36)*{}="20v36";
(22,36)*{}="22v36";
(26,28)*{}="26v28";
(30,30)*{}="30v30";
(30,34)*{}="30v34";
(32,28)*{}="32v28";
(32,32)*{}="32v32";
(34,34)*{}="34v34";
(38,36)*{}="38v36";
(42,28)*{}="42v28";
(42,36)*{}="42v36";
(46,32)*{}="46v32";
"16v32"*@{*};
"30v30"*@{*};
"34v34"*@{*};
"46v32"*@{*};
"20v36"*{\scriptstyle J};
"42v36"*{\scriptstyle J};
"32v28"*{\scriptstyle k};
"30v34"+(2,0)*{\scriptstyle k^*};
"32v32"*{\xycircle(6,6){--}};
{\ar@{=}@`{"22v36", "26v28"}|(0.35){\SelectTips{eu}{12}\object@{>>}} "16v32";"30v30"};
{\ar@{=}@`{"38v36", "42v28"}|(0.65){\SelectTips{eu}{12}\object@{>>}} "34v34";"46v32"};
\endxy
}

\newcommand{\TwoDyonContractionEq}{
\xy
0;<2.5pt,0pt>:
<0pt,2.5pt>::
(16,32)*{}="16v32";
(20,36)*{}="20v36";
(22,36)*{}="22v36";
(26,28)*{}="26v28";
(30,30)*{}="30v30";
(34,34)*{}="34v34";
(38,36)*{}="38v36";
(42,28)*{}="42v28";
(42,36)*{}="42v36";
(46,32)*{}="46v32";
"16v32"*@{*};
"30v30"*@{*};
"34v34"*@{*};
"46v32"*@{*};
"20v36"*{\scriptstyle J};
"42v36"*{\scriptstyle J};
{\ar@{.} "30v30";"34v34"};
{\ar@{=}@`{"22v36", "26v28"}|(0.35){\SelectTips{eu}{12}\object@{>>}} "16v32";"30v30"};
{\ar@{=}@`{"38v36", "42v28"}|(0.35){\SelectTips{eu}{12}\object@{>>}} "34v34";"46v32"};
\endxy
}

\newcommand{\StringOperatorJ}{
\xy
0;<2pt,0pt>:
<0pt,2pt>::
(16,16)*{}="16v16";
(22,12)*{}="22v12";
(26,20)*{}="26v20";
(28,14)*{}="28v14";
(32,16)*{}="32v16";
"16v16"*@{*};
"32v16"*@{*};
"28v14"*{\scriptstyle J};
{\ar@{=}@`{"22v12", "26v20"}|(0.45){\SelectTips{eu}{12}\object@{>>}} "16v16";"32v16"};
\endxy
}

\newcommand{\UnitCellGraph}{
\xy
0;<2pt,0pt>:
<0pt,2pt>::
(4,26)*{}="04v26";
(4,38)*{}="04v38";
(8,28)*{}="08v28";
(8,30)*{}="08v30";
(8,34)*{}="08v34";
(8,36)*{}="08v36";
(12,14)*{}="12v14";
(12,30)*{}="12v30";
(12,34)*{}="12v34";
(12,50)*{}="12v50";
(16,16)*{}="16v16";
(16,18)*{}="16v18";
(16,22)*{}="16v22";
(16,24)*{}="16v24";
(16,40)*{}="16v40";
(16,42)*{}="16v42";
(16,46)*{}="16v46";
(16,48)*{}="16v48";
(20,18)*{}="20v18";
(20,22)*{}="20v22";
(20,42)*{}="20v42";
(20,46)*{}="20v46";
(24,8)*{}="24v08";
(24,12)*{}="24v12";
(24,28)*{}="24v28";
(24,30)*{}="24v30";
(24,34)*{}="24v34";
(24,36)*{}="24v36";
(24,52)*{}="24v52";
(24,56)*{}="24v56";
(28,30)*{}="28v30";
(28,34)*{}="28v34";
(32,16)*{}="32v16";
(32,18)*{}="32v18";
(32,22)*{}="32v22";
(32,24)*{}="32v24";
(32,40)*{}="32v40";
(32,42)*{}="32v42";
(32,46)*{}="32v46";
(32,48)*{}="32v48";
(36,18)*{}="36v18";
(36,22)*{}="36v22";
(36,42)*{}="36v42";
(36,46)*{}="36v46";
(40,8)*{}="40v08";
(40,12)*{}="40v12";
(40,28)*{}="40v28";
(40,30)*{}="40v30";
(40,34)*{}="40v34";
(40,36)*{}="40v36";
(40,52)*{}="40v52";
(40,56)*{}="40v56";
(44,30)*{}="44v30";
(44,34)*{}="44v34";
(48,16)*{}="48v16";
(48,18)*{}="48v18";
(48,22)*{}="48v22";
(48,24)*{}="48v24";
(48,40)*{}="48v40";
(48,42)*{}="48v42";
(48,46)*{}="48v46";
(48,48)*{}="48v48";
(52,18)*{}="52v18";
(52,22)*{}="52v22";
(52,42)*{}="52v42";
(52,46)*{}="52v46";
(56,8)*{}="56v08";
(56,12)*{}="56v12";
(56,28)*{}="56v28";
(56,30)*{}="56v30";
(56,34)*{}="56v34";
(56,36)*{}="56v36";
(56,52)*{}="56v52";
(56,56)*{}="56v56";
(60,30)*{}="60v30";
(60,34)*{}="60v34";
(64,16)*{}="64v16";
(64,18)*{}="64v18";
(64,22)*{}="64v22";
(64,24)*{}="64v24";
(64,40)*{}="64v40";
(64,42)*{}="64v42";
(64,46)*{}="64v46";
(64,48)*{}="64v48";
(66,18)*{}="66v18";
(66,22)*{}="66v22";
(66,42)*{}="66v42";
(66,46)*{}="66v46";
(68,14)*{}="68v14";
(68,50)*{}="68v50";
(72,28)*{}="72v28";
(72,30)*{}="72v30";
(72,34)*{}="72v34";
(72,36)*{}="72v36";
(74,30)*{}="74v30";
(74,34)*{}="74v34";
(76,26)*{}="76v26";
(76,38)*{}="76v38";
{\ar@{-} "04v38";"08v36"};
{\ar@{-} "08v28";"04v26"};
{\ar@{-} "08v28";"16v24"};
{\ar@{-} "08v34";"12v34"};
{\ar@{-} "08v36";"08v28"};
{\ar@{-} "16v16";"12v14"};
{\ar@{-} "16v16";"16v24"};
{\ar@{-} "16v22";"20v22"};
{\ar@{-} "16v24";"24v28"};
{\ar@{-} "16v40";"08v36"};
{\ar@{-} "16v40";"16v48"};
{\ar@{-} "16v46";"20v46"};
{\ar@{-} "16v48";"12v50"};
{\ar@{-} "16v48";"24v52"};
{\ar@{-} "24v08";"24v12"};
{\ar@{-} "24v12";"16v16"};
{\ar@{-} "24v28";"32v24"};
{\ar@{-} "24v34";"28v34"};
{\ar@{-} "24v36";"16v40"};
{\ar@{-} "24v36";"24v28"};
{\ar@{-} "24v52";"24v56"};
{\ar@{-} "24v52";"32v48"};
{\ar@{-} "32v16";"24v12"};
{\ar@{-} "32v16";"40v12"};
{\ar@{-} "32v22";"36v22"};
{\ar@{-} "32v24";"32v16"};
{\ar@{-} "32v24";"40v28"};
{\ar@{-} "32v40";"24v36"};
{\ar@{-} "32v46";"36v46"};
{\ar@{-} "32v48";"32v40"};
{\ar@{-} "40v12";"40v08"};
{\ar@{-} "40v12";"48v16"};
{\ar@{-} "40v28";"40v36"};
{\ar@{-} "40v34";"44v34"};
{\ar@{-} "40v36";"32v40"};
{\ar@{-} "40v36";"48v40"};
{\ar@{-} "40v52";"32v48"};
{\ar@{-} "40v52";"40v56"};
{\ar@{-} "48v16";"48v24"};
{\ar@{-} "48v16";"56v12"};
{\ar@{-} "48v22";"52v22"};
{\ar@{-} "48v24";"40v28"};
{\ar@{-} "48v40";"56v36"};
{\ar@{-} "48v46";"52v46"};
{\ar@{-} "48v48";"40v52"};
{\ar@{-} "48v48";"48v40"};
{\ar@{-} "56v12";"56v08"};
{\ar@{-} "56v12";"64v16"};
{\ar@{-} "56v28";"48v24"};
{\ar@{-} "56v28";"64v24"};
{\ar@{-} "56v34";"60v34"};
{\ar@{-} "56v36";"56v28"};
{\ar@{-} "56v36";"64v40"};
{\ar@{-} "56v52";"48v48"};
{\ar@{-} "56v52";"56v56"};
{\ar@{-} "64v16";"64v24"};
{\ar@{-} "64v16";"68v14"};
{\ar@{-} "64v22";"66v22"};
{\ar@{-} "64v24";"72v28"};
{\ar@{-} "64v40";"64v48"};
{\ar@{-} "64v46";"66v46"};
{\ar@{-} "64v48";"56v52"};
{\ar@{-} "64v48";"68v50"};
{\ar@{-} "72v28";"72v36"};
{\ar@{-} "72v28";"76v26"};
{\ar@{-} "72v34";"74v34"};
{\ar@{-} "72v36";"64v40"};
{\ar@{-} "72v36";"76v38"};
{\ar@{.} "08v30";"12v30"};
{\ar@{.} "16v18";"20v18"};
{\ar@{.} "16v42";"20v42"};
{\ar@{.} "24v30";"28v30"};
{\ar@{.} "32v18";"36v18"};
{\ar@{.} "32v42";"36v42"};
{\ar@{.} "40v30";"44v30"};
{\ar@{.} "48v18";"52v18"};
{\ar@{.} "48v42";"52v42"};
{\ar@{.} "56v30";"60v30"};
{\ar@{.} "64v18";"66v18"};
{\ar@{.} "64v42";"66v42"};
{\ar@{.} "72v30";"74v30"};
\endxy
}

\newcommand{\TreeStructure}{
\xy
0;<2.5pt,0pt>:
<0pt,2pt>::
(12,48)*{}="12v48";
(14,48)*{}="14v48";
(16,42)*{}="16v42";
(16,46)*{}="16v46";
(16,48)*{}="16v48";
(24,34)*{}="24v34";
(28,48)*{}="28v48";
(30,48)*{}="30v48";
(32,26)*{}="32v26";
(32,42)*{}="32v42";
(32,46)*{}="32v46";
(32,48)*{}="32v48";
(44,14)*{}="44v14";
(44,48)*{}="44v48";
(46,48)*{}="46v48";
(48,10)*{}="48v10";
(48,42)*{}="48v42";
(48,46)*{}="48v46";
(48,48)*{}="48v48";
(60,48)*{}="60v48";
(68,48)*{}="68v48";
(70,48)*{}="70v48";
(72,42)*{}="72v42";
(72,46)*{}="72v46";
(72,48)*{}="72v48";
"60v48"*{\scriptstyle \dots};
"16v46"*{\scriptstyle p_1};
"32v46"*{\scriptstyle p_2};
"48v46"*{\scriptstyle p_3};
"72v46"*{\scriptstyle p_{P-1}};
"16v48"*{\xycircle(4,6){}};
"32v48"*{\xycircle(4,6){}};
"48v48"*{\xycircle(4,6){}};
"72v48"*{\xycircle(4,6){}};
{\ar@{-} "12v48";"14v48"};
{\ar@{-} "16v42";"48v10"};
{\ar@{-} "28v48";"30v48"};
{\ar@{-} "32v42";"24v34"};
{\ar@{-} "44v48";"46v48"};
{\ar@{-} "48v42";"32v26"};
{\ar@{-} "68v48";"70v48"};
{\ar@{-} "72v42";"44v14"};
\endxy
}

\newcommand{\DyonBasis}{
\xy
0;<2.5pt,0pt>:
<0pt,2.5pt>::
(16,42)*{}="16v42";
(16,46)*{}="16v46";
(24,34)*{}="24v34";
(26,28)*{}="26v28";
(32,26)*{}="32v26";
(32,42)*{}="32v42";
(32,46)*{}="32v46";
(34,18)*{}="34v18";
(44,8)*{}="44v08";
(44,14)*{}="44v14";
(48,42)*{}="48v42";
(48,46)*{}="48v46";
(52,6)*{}="52v06";
(56,2)*{}="56v02";
(58,42)*{}="58v42";
(62,2)*{}="62v02";
(72,42)*{}="72v42";
(72,46)*{}="72v46";
(88,42)*{}="88v42";
(88,46)*{}="88v46";
"16v42"*@{*};
"32v42"*@{*};
"48v42"*@{*};
"56v02"*@{*};
"72v42"*@{*};
"88v42"*@{*};
"24v34"*{\scriptstyle \circ};
"32v26"*{\scriptstyle \circ};
"44v14"*{\scriptstyle \circ};
"52v06"*{\scriptstyle \circ};
"34v18"*{\scriptstyle \dots};
"58v42"*{\scriptstyle \dots};
"16v46"*{\scriptstyle J_1q_1};
"32v46"*{\scriptstyle J_2q_2};
"48v46"*{\scriptstyle J_3q_3};
"88v46"*{\scriptstyle J_{P-1}q_{P-1}};
"72v46"*{\scriptstyle J_{P-2}q_{P-2}};
"62v02"*{\scriptstyle J_Pq_P};
"26v28"*{\scriptstyle K_1};
"44v08"*{\scriptstyle K_{P-3}};
{\ar@{=} "16v42";"56v02"};
{\ar@{=} "32v42";"24v34"};
{\ar@{=} "44v14";"72v42"};
{\ar@{=} "48v42";"32v26"};
{\ar@{=} "88v42";"52v06"};
\endxy
}

\newcommand{\FourFluxonBasis}{
\xy
0;<3pt,0pt>:
<0pt,3pt>::
(18,40)*{}="18v40";
(18,44)*{}="18v44";
(20,32)*{}="20v32";
(24,34)*{}="24v34";
(28,24)*{}="28v24";
(28,32)*{}="28v32";
(30,40)*{}="30v40";
(30,44)*{}="30v44";
(36,24)*{}="36v24";
(40,40)*{}="40v40";
(40,44)*{}="40v44";
(42,24)*{}="42v24";
(42,32)*{}="42v32";
(46,34)*{}="46v34";
(50,32)*{}="50v32";
(52,40)*{}="52v40";
(52,44)*{}="52v44";
"18v40"*@{*};
"30v40"*@{*};
"40v40"*@{*};
"52v40"*@{*};
"24v34"*{\scriptstyle \circ};
"46v34"*{\scriptstyle \circ};
"36v24"*{\scriptstyle K};
"18v44"*{\scriptstyle \tau\overline{\tau}};
"30v44"*{\scriptstyle \tau\overline{\tau}};
"40v44"*{\scriptstyle \tau\overline{\tau}};
"52v44"*{\scriptstyle \tau\overline{\tau}};
{\ar@{=}@`{"20v32", "28v32"} "18v40";"30v40"};
{\ar@{=}@`{"28v24", "42v24"} "24v34";"46v34"};
{\ar@{=}@`{"42v32", "50v32"} "40v40";"52v40"};
\endxy
}

\begin{document}

\title{Full Dyon Excitation Spectrum in Extended Levin-Wen Models}

\author{Yuting Hu} \email{yuting@physics.utah.edu} \affiliation{Department of Physics and
	Astronomy, University of Utah, Salt Lake City, UT 84112, USA}

\author{Nathan Geer}  \email{nathan.geer@gmail.com}  \affiliation{Department of
	Mathematics and Statistics, Utah State University, Logan, UT 84322, USA} 

\author{Yong-Shi Wu} \email{wu@physics.utah.edu} \affiliation{Key State Laboratory of Surface
	Physics and Department of Physics, Fudan University, Shanghai 200433, P.R. China} 
	\affiliation{Center for Field Theory and Particle Physics, Fudan University, Shanghai 200433, China} 
	\affiliation{Collaborative Innovation Center of Advanced Microstructures, Fudan University, Shanghai 200433, China} \affiliation{Department of Physics and Astronomy, University of Utah, Salt Lake City, UT 84112, USA}

\date{\today}

\begin{abstract}
	In Levin-Wen (LW) models, a wide class of exactly solvable discrete models, for two 
	dimensional topological phases, it is relatively easy to describe only single fluxon 
	excitations, but not the charge and dyonic as well as many-fluxon excitations. To 
	incorporate charged and dyonic excitations in (doubled) topological phases, an extension 
	of the LW models is proposed in this paper. We first enlarge the Hilbert space with adding 
	a tail on one of the edges of each trivalent vertex, to describe the internal charge degrees 
	of freedom at the vertex. Then we study the full dyon spectrum of the extended LW models, 
	including both quantum numbers and wave functions for dyonic quasiparticle excitations. The 
	local operators associated with the dyonic excitations are shown to form the so-called tube 
	algebra, whose representations (modules) form the quantum double (categoric center) of the 
	input data (unitary fusion category). In physically relevant cases, the input data is from a 
	finite or quantum group (with braiding $R$-matrices), we find that the elementary excitations 
	(or dyon species), as well as any localized/isolated excited states, are characterized by three 
	quantum numbers: charge, fluxon type, and twist. They provide a ``complete basis'' for many-body 
	states in the enlarged Hilbert space. Concrete examples are presented and the relevance of our 
	results to the electric-magnetic duality existing in the models is addressed.   
	
\end{abstract} 


\maketitle
\tableofcontents

\section{Introduction} 

In recent years two-dimensional topological phases have received increasing attention from the 
science community. These phases represent a novel class of quantum matter at zero temperature 
\cite{Wen95}, whose bulk properties are robust against weak interactions and disorders. They may 
be divided into two families: \textit{doubled} (with time-reversal symmetry, or TRS, preserved), 
and \textit{chiral} (with TRS broken). Chiral phases were first discovered in integer and 
fractional quantum Hall (IQH and FQH) liquids. Mathematically, their effective low-energy 
description is given by Chern-Simons gauge theory or, more generally, topological quantum field 
theory (TQFT) \cite{Witten89,WZ91}. Doubled topological phases include topological insulators and 
some states in quantum spin liquids. Either chiral or doubled phases may be exploited to do 
fault-tolerant (or topological) quantum computing \cite{Kitaev,FKLW,NSSFD,Wang}.  

The (chiral) Chern-Simons theories are formulated in the continuum and have no lattice counterpart. On the
other hand, doubled topological phases do admit a discrete description. The first such formulation in the 
physics literature was the Kitaev's toric code model \cite{Kitaev}. (In the mathematical literature, a 
discrete version of TQFT had been constructed a bit earlier by Turaev and Viro \cite{TV}, which by now is 
known to describe certain doubled phases.) About ten years ago, Levin and Wen (LW) \cite{LW} constructed 
a wide class of discrete models on a trivalent lattice/graph, with an exactly solvable Hamiltonian, for two 
dimensional doubled topological phases. The model is now believed to be a discretized version of 
\textit{doubled} Chern-Simons theory \cite{FNSWW}, which is mathematically the same as the Turaev-Viro TQFT 
\cite{Turaev94,KMR,Wang}. The original motivation of the LW model was to generate ground states that 
exhibit the phenomenon of string-net condensation \cite{Wen03} as a physical mechanism for topological 
phases. The ground states in this model can be viewed as the fixed-point states of some renormalization 
group flow  \cite{CGW}, which look the same at all length scales and thus have no local degrees of 
freedom. Like Kitaev's toric code model \cite{Kitaev}, we expect that the subspace of degenerate ground 
states in the LW model can be used as a fault-tolerant code for quantum computation.

Two of us have studied, in a previous joint paper with another author \cite{GSD}, the ground state 
degeneracy (GSD) of the LW model on a (discretized) closed oriented surface $M$. Usually in TQFT the GSD is 
examined as a topological invariant of the 3-manifold $S^1\times M$ \cite{Turaev94,KMR,Simon}. In the LW 
Hamiltonian approach, our computation of the GSD became accessible to physicists. In this paper, we attack 
the problem of solving the full spectrum of quasiparticle excitations in the LW models with the input data 
being a unitary fusion category (see below for details). This problem is of significance for further 
interdisciplinary study of the models in physics, mathematics and quantum computation codes. It is 
generally believed that the quasiparticle excitation species are related to the quantum double that  
classifies the degenerate ground states. Several proposals about excitation spectrum in the LW models have 
been made in the literature \cite{LW,KK11,BCKA13,LW14}. In this paper, we will present a new approach to 
understanding the full elementary excitation spectrum of the LW models, that addresses both the quantum 
numbers and corresponding states (or wave functions) explicitly for all quasiparticle species. In 
particular we want to accommodate the needs in physics and quantum computation codes for concrete and 
explicit expressions to play with.    

Several new developments feature our analysis. Usually for a single (pure) fluxon, it is easy to 
characterize/specify their quantum numbers. See, for example, \cite{HSW12}. However, the fusion of 
two or more (pure) fluxons, generally leads to the appearance of charge quantum numbers. (Some examples 
are shown later in Sec.\ VII.) Namely the set of single fluxon species are not closed under fusion! 
So how to represent all dyon (charge-fluxon composite) species in the LW model presented a challenge. 
Our way to solve this problem is to enlarge the original Hilbert space of the LW model. We explicitly 
introduce internal charge degrees of freedom (d.o.f.)\ at each trivalent vertex by adding a tail on one of 
its edges. This has greatly facilitated the treatment of fusion outcomes. In this way, the LW model is 
actually extended, with the underlying graph(s) involving univalent vertices. The second important 
development is that we have identified the operator algebra for the local operators, that can be used to 
generate all quasiparticle excitations, to be the Tube algebra constructed by Ocneanu \cite{Ocneanu, EK95, 
Izumi00, Mueger}. Using the Tube algebra, the relationship of quasiparticle species to the irreducible 
representations (simple modules) of the quantum double becomes relatively easy to establish, and the above 
mentioned complicated situations for fusion of non-abelian anyons become easier to handle. It has been 
shown \cite{Mueger} that a half-braiding in the quantum double (or center) category corresponds to an  
irreducible representation (simple module) of the Tube algebra. This not only enables us to define the 
string operators, but also to account for charged as well as dyonic excitations. Our present analysis has 
clarified and emphasized the importance of supplementing the twist, as quantum number in addition to the 
usual charge and fluxon-type, to the characterization of quasiparticle species in extended LW models. 
Indeed, generally there may exist quasiparticle excitations which have the same charge and fluxon-type but 
have different twists and, therefore, should be counted as different species. 
 
A similar operator algebra approach for quasiparticles in topological phases has been proposed by Lan and 
Wen \cite{LW14}. The Q-algebra they independently invented is a presentation of the Tube algebra. 
In their approach, they added an extra (charge) index at each vertex for its internal charge d.o.f., while 
we want to add a (charge) tail at one of the links attached to a vertex. This difference makes the two 
formulations have complimentary technical advantages and disadvantages. 

Because of the interdisciplinary interests in the LW models, we have tried to adapt our presentation in 
this paper to audience with different backgrounds. Of course, the basic audience in our mind is physicists, 
and we have tried hard to make the presentation accessible to physicists. However, whenever a reference of 
the terminology or of the idea can be made to the mathematical literature, we will do it to help readers of 
mathematical background.  (Readers with physics background can safely skip these mathematical remarks 
without harming their further reading.) The last section is also devoted to the relationship between our 
approach and TQFTs. 

We will use some terminology in category theory language for convenience of physicists, because this language 
could be used widely in future physics, just like group theory has become the language of contemporary physics. 
Condensed matter physicists do not need to be worried. Whoever has learned angular momentum or crystal 
group theory in quantum mechanics is familiar with at least one fusion category, which is nothing but the 
category formed by {\it all finite-dimensional unitary representations} of the rotation group or its 
discrete subgroups in three-dimensional Euclidean space! The decomposition of the (tensor) product of two 
irreducible representations into a (direct) sum of irreducible representations just gives to the fusion 
algebra, with the non-negative integral coefficients in the direct-sum decomposation as the fusion (rule) 
coefficients. The $6j$-symbols are well-known in group theory. So the fusion category is a straightforward 
generalization of the representation theory of groups (or group algebras) to more complicated algebras 
(more precisely, weak Hopf algebras). Up to now only fusion categories associated with a finite group or a 
quantum group appear in the literature of condensed matter physics. 
                       
The paper is organized as follows. In Section II we briefly review the LW models and set up our notations.  
In Section III we review the topological symmetry of the ground states, i.e. the invariance under 
Pachner mutations of the (spatial) graph. Then we begin our study of excited states by introducing an 
extension of the Hilbert space of the LW models, as well as the extended Hamiltonian in Sec.\ IV. We devote 
Sec.\ V to the central issue of the paper, i.e.\ the study of elementary (quasiparticle) excitations, using 
local operators preserving topological symmetry, which is shown to form the tube algebra. Minimal 
projection operators and simple modules (irreducible representations) of the tube algebra are introduced. A 
dyon species is identified with an irreducible representation (simple module) of the Tube algebra, and 
fusion of all dyon species gives rise to the quantum double (or the categoric center) of the input fusion 
category. String operators are generalized to dyon-pair creation, hopping operators etc, and their 
properties that are related to important observables, such as twist and $S$-matrix, are studied using 
graphic calculus. Next two sections, Sec. \ VI and Sec/ \ VII, are devoted to studying excitation spectrum 
and emergent braiding statistics from the above-established set-up. In Sec.\ VIII, with possible a
pplications in anyon condensation, we examine the particular case -- the {\it braided} LW models -- 
with the input fusion category equipped with an $R$-matrix. Physically these models are actually a 
generalized gauge theory with gauge ``group'' being a finite or quantum group. In Sec.\ IX, we present 
several examples, including cases with input data from an abelian group, from a non-abelian group $S_3$, 
from the Kitaev's quantum double model as well as from a modular category, such as the double semion model 
and double Fibonnaci model. Sec. \ X addresses the electric-magnetic duality between two particular LW 
models with two seeming different sets of input data involving the same finite group. In Sec.\ XI we 
elaborate the relation to topological quantum field theory for the convenience of readers with mathematical 
background. The final section (Sec.\ XII) is devoted to conclusions and discussions. In addition to summarizing our main results, we present arguments that our extended models, though with enlarged Hilbert space and modified Hamiltonian, 
give rise to the same topological phase at zero temperature as the Levin-Wen models, while having different 
perspective for properties, phases and phase transitions at finite temperatures involving charged and dyonic 
excitations. We also emphasize that the local (string-like) operators we have defined in this paper and their 
algebra may be useful, even when the Hamiltonian is deformed away from our modified Hamiltonian (provided that 
the gauge symmetry,or the input category is not changed).

\section{Levin-Wen models}\label{S:LWModels}

Let us briefly review the Levin-Wen models. The input data to define the model, i.e., to specify the Hilbert space and the Hamiltonian, is the unitary fusion category $\mathcal{C}$. More specifically, we will use the tensor description of $\mathcal{C}$ in terms of $6j$-symbols.

\begin{figure}[t]
  \centering
  \subfigure[]{\includegraphics{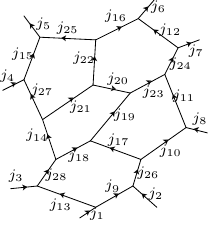}}
  \label{fig:TrivalentGraphA}\qquad
  \subfigure[]{\includegraphics{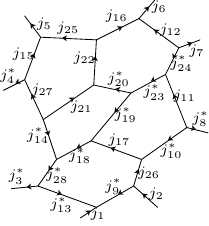}}
  \label{fig:TrivalentGraphB}
  \caption{A configuration of string types on a directed trivalent graph. The configuration (b) is treated the same as (a), with some of the directions of some edges reversed and the corresponding labels $j$ conjugated $j^*$.}
  \label{fig:TrivalentGraph}
 \end{figure}

The model is defined on a trivalent
graph embedded to a closed oriented surface. The Hilbert space is spanned by the degrees of freedom on edges. See Fig.\ref{fig:TrivalentGraph}.
For each edge, we assign a label $j$ (called string type), which
runs over a finite set of integers $L=\{j=0,1,...,N\}$.
The Hilbert space is spanned by all configurations of the labels
on edges. Each label $j$ has a ``conjugate'' $j^*$, which is also
an integer and satisfies $j^{**}=j$. If we reverse the direction
of one edge and replace the label $j$ by $j^*$ on this edge,
we require the state to be the same. See Fig.\ref{fig:TrivalentGraph}. There is unique ``trivial''
label $j=0$ satisfying $0^*=0$.

To specify the Hamiltonian, we introduce the structure on string types as follows.

A \textit{fusion rule} on $L$ is a function $N:L \times L \times L\rightarrow \ds{N}$ such that for $a,b,c,d\in L$,
\begin{align}
&N_{0a}^b=N_{a0}^b=\delta_{ab},\label{eq:N:a0b}\\
&N_{ab}^0=\delta_{ab^*},\label{eq:N:ab0}\\
&\sum_{x\in L}N_{ab}^xN_{xc}^d=\sum_{x\in L}N_{ax}^dN_{cd}^x.\label{eq:NN=NN}
\end{align}
A fusion rule is multiplicity-free if $N_{ab}^c\in \{0,1\}$ for all $a,b,c\in L$.  We 
 restrict to the multiplicity-free case throughout this paper unless specified. We define $\delta_{abc}:=N_{ab}^{c^*}$ which has 
the symmetric properties:  $\delta_{abc}=\delta_{bca}$ and $\delta_{abc}=\delta_{c^*b^*a^*}$. We say a triple $(a,b,c)$ is admissible if $\delta_{abc}=1$.

Given a fusion rule on $L$, a \textit{quantum dimension} is a map $\rmd:L\rightarrow \ds{R}$ such that $\rmd_{a^*}=\rmd_{a}$ and
\begin{equation}
\sum_{c\in L}\rmd_c\delta_{abc^*}=\rmd_a\rmd_b.
\label{eq:dimcond}
\end{equation}
In particular, $\rmd_0=1$. Let $\alpha_j=\mathrm{sgn}(\rmd_j)$ which takes values of $\pm 1$ for each label $j$, and require:
\begin{equation}
\alpha_i\alpha_j\alpha_k=1, \quad \text{if }\delta_{ijk}=1.
\end{equation}

Given a fusion rule and a quantum dimension on $L$, we may define $6j$-symbols, often denoted as $G$. A \textit{tetrahedral symmetric unitary $6j$-symbol} is a map $G:L^6\rightarrow \ds{C}$ satisfying the following conditions:
\begin{align}
\label{eq:6jcond}
\begin{array}{ll}
&G^{ijm}_{kln}=G^{mij}_{nk^{*}l^{*}}
=G^{klm^{*}}_{ijn^{*}}=\alpha_m\alpha_n\,\overline{G^{j^*i^*m^*}_{l^*k^*n}},\\
&\sum_{n}{\rmd_{n}}G^{mlq}_{kp^{*}n}G^{jip}_{mns^{*}}G^{js^{*}n}_{lkr^{*}}
=G^{jip}_{q^{*}kr^{*}}G^{riq^{*}}_{mls^{*}},\\
&\sum_{n}{\rmd_{n}}G^{mlq}_{kp^{*}n}G^{l^{*}m^{*}i^{*}}_{pk^{*}n}
=\frac{\delta_{iq}}{\rmd_{i}}\delta_{mlq}\delta_{k^{*}ip},\\
\end{array}
\end{align}

The data $\{\rmd_j,\delta_{ijk},G_{klm}^{ijm}\}$ can be derived from the representation theory of a group, 
or more generally a quantum group. (More generally, such a set of data is from a unitary fusion category.) 
For instance, we take the labels $j$ to be the irreducible representations of
a finite group. The trivial label $0$ is the trivial representation. The fusion rule
tells whether the tensor product $j_1\otimes{j_2}\otimes{j_3}$ contains
the trivial representation or not. The number $\alpha_j$ is the Frobenius-Schur indicator telling if the 
representation $j$ is real or complex, or pseudoreal, $\rmd_j=\alpha_j\mathrm{dim}(j)$ the dimension 
$\mathrm{dim}(j)$ of the corresponding representation space multiplied by the Frobenius-Schur indicator 
$\alpha_j$.  The number $G_{kln}^{ijm}$ the (symmetrized) Racah $6j$ symbol for
the group. In this example, the LW model can be mapped to the Kitaev's quantum double model. 

One important property of the $6j$-symbols is that
\begin{equation}
\label{eq:G=Gdelta}
G^{ijm}_{kln}=G^{ijm}_{kln}\delta_{ijm}\delta_{klm^*}\delta_{lin}\delta_{nk^*j^*}.
\end{equation}
To prove this, one can rewrite the orthogonality condition by
\begin{equation}
\sum_{n}\left({\rmv_{n}\rmv_{q}}G^{mlq}_{kp^{*}n}\right)\overline{\left({\rmv_{n}\rmv_{i}}G^{mlq}_{kp^{*}n}\right)}
=\delta_{iq}\delta_{mli}\delta_{k^{*}ip}.
\end{equation}
When $q=i$, the equality implies that $G^{mli}_{kp^*n}$ must varnish unless $\delta_{mlq}\delta_{k^{*}ip}=1$. By using the tetrahedral symmetry, one arrives at Eq.\ \eqref{eq:G=Gdelta}.  Here $\rmv_j=\sqrt{\rmd_j}$ is a choice of a square root of the quantum dimension. 
The number $\rmv_j$ is either real or pure imaginary, depending on the sign $\alpha_j=\mathrm{sgn}(\rmd_j)$.

Depending on how the square root is taken, $\rmv_j$ is determined up to a sign. We fix the sign as follows. From the conditions in Eq.\ \eqref{eq:6jcond}, we have
$(G^{ijk}_{0kj}\rmv_j\rmv_k)^2=\delta_{ijk}$. It is possible to fix the sign of $\rmv_j$ such that $G^{ijk}_{0kj}\rmv_j\rmv_k=\delta_{ijk}$. We define
\begin{equation}
\label{eq:v::Definition}
\rmv_j:=\frac{1}{G^{j^*j0}_{0\,0\,j}}.
\end{equation}
In particular, $\rmv_0=1$ because $\rmd_0=1$ (from Eq.\ \eqref{eq:dimcond}) and thus $G^{000}_{000}=1$ from Eq.\ \eqref{eq:6jcond}. Indeed, we can verify $\rmv_j^2=\rmd_j$ directly from the orthogonality condition in Eq.\ \eqref{eq:6jcond} together with $\rmd_0=1$. The definition in Eq.\ \eqref{eq:v::Definition} also implies
\begin{equation}
\label{eq:Gvv=delta::vConvention}
G^{ijk}_{0kj}\rmv_j\rmv_k=\delta_{ijk},
\end{equation}
which can be proved by the pentagon identity $\rmd_0G^{ijk}_{0kj}G^{j^*j0}_{0\,0\,j}G^{k^*k0}_{0\,0\,k}=G^{ijk}_{0kj}G^{j^*i^*k^*}_{k^*0j}$ and the orthogonality $\rmd_jG^{ijk}_{0kj}G^{j^*i^*k^*}_{k^*0j}=\frac{1}{\rmd_k}\delta_{ijk}$.

There are two types of local operators, ${Q}_v$ defined at
vertices $v$ and $B_p^s$ (indexed by the label $s=0,1,...,N$)
at plaquettes $p$. Let us first
define the operator ${Q}_v$.
On a trivalent graph, ${Q}_v$ acts on the labels of three
edges incoming to the vertex $v$. We define the action of ${Q}_v$
on the basis vector with $j_1,j_2,j_3$ by
\begin{align}
  {Q}_v\left|
  \Ypart{j_1}{j_3}{j_2}{>}{<}{<}
  \right\rangle
  =\delta_{j_1j_2j_3}\left|
  \Ypart{j_1}{j_3}{j_2}{>}{<}{<}
  \right\rangle
\end{align}
where the tensor $\delta_{j_1j_2j_3}$ equals either 1 or 0, which
determines whether the triple $\{j_1,j_2,j_3\}$ is ``allowed'' to
meet at the vertex. Since $\delta_{j_1j_2j_3}$ is
symmetric under permutations of the three labels:
$\delta_{j_1j_2j_3}=\delta_{j_2j_3j_1}=\delta_{j_1j_3j_2}$, the ordering in this
triple $\{j_1,j_2,j_3\}$ is not important.

The operator $B_p^s$ acts on the boundary edges of the plaquette
$p$, and has the matrix elements on a triangle plaquette,
\begin{align}
\label{eq:Bps::InLW}
&\Biggl\langle
\TriangleYpart{j_4}{j^{\prime}_1}{j^{\prime}_2}{j^{\prime}_3}{j_5}{j_6}{>}{<}{>}{<}{<}{<}
\Biggr|
B_p^s
\Biggl|\TriangleYpart{j_4}{j_1}{j_2}{j_3}{j_5}{j_6}{>}{<}{>}{<}{<}{<}\Biggr\rangle\nonumber\\
=&
\rmv_{j_1}\rmv_{j_2}\rmv_{j_3}\rmv_{j'_1}\rmv_{j'_2}\rmv_{j'_3}
G^{j_5j^*_1j_3}_{sj'_3j^{\prime*}_1}G^{j_4j^*_2j_1}_{sj'_1j^{\prime*}_2}G^{j_6j^*_3j_2}_{sj'_2j^{\prime*}_3}.
\end{align}
The same rule applies
when the plaquette $p$ is a quadrangle, a pentagon,
or a hexagon and so on. Note that the matrix is nondiagonal
only on the labels of the boundary edges (i.e., $j_1$, $j_2$, and
$j_3$ on the above graph).

The operators $B_p^s$ have the properties
\begin{align}
  &B_p^{s\dagger}=B_p^{s^*},
  \label{eq:BpsDagger}\\
  &B_p^rB_{p}^s=\sum_{t}\delta_{rst^*}B_p^t,
  \label{eq:BpsAlgebra}
\end{align}
Both can be verified by using conditions \eqref{eq:6jcond}.

The Hamiltonian of the model is (here $D=\sum_j{\rmd}_j^2$)
\begin{equation}
  \label{eq:HamiltonianLW}
  {H}=-\sum_{v}{Q}_v-\sum_{p}B_p,
  \quad
  B_p=\frac{1}{D}\sum_{s}\rmd_sB_p^{s}
\end{equation}
where the sum run over vertices $v$ and plaquettes $p$
of the trivalent graph.

The main property of ${Q}_v$ and $B_p$ is that they are mutually-commuting projection
operators: (1) $[Q_v,Q_{v^{\prime}}]=0=[B_p,B_{p^{\prime}}],[Q_v,B_p]=0$; (2) and ${Q}_v{Q}_{v}={Q}_v$
and $B_pB_{p}=B_p$.
Thus the Hamiltonian is exactly soluble.
The elementary energy eigenstates are given by common eigenvectors of all these projections.
The ground states satisfies ${Q}_v=B_p=1$
for all $v$,$p$, while the excited states violate these constraints
for some plaquettes or vertices.

In particular, if $\{\rmd,\delta,G\}$ arises from the representation theory of groups or quantum groups, we have $\delta_{rst^*}=\delta_{srt^*}$. Then the $B_p^s$'s commute with each other,
\begin{equation}
  [B_{p_1}^r,B_{p_2}^s]=0
\end{equation}
which can be verified by the conditions in \eqref{eq:6jcond} when $p_1$ and $p_2$
are the two nearest neighboring plaquettes, and by Eq.\ \eqref{eq:BpsAlgebra} together with $\delta_{rst^*}=\delta_{srt^*}$
when $p_1=p_2$.

\section{Topological symmetry for ground states}
\label{sec:TopologicalSymmetry}

To characterize the topological phases, we study the topological observables. Examples include the topological degeneracy of ground states. Behind them the topological symmetry plays an important role: topological observables are those invariant under mutations of the spatial graph. In continuum theory, they are observables invariant under the smooth deformation of the space-time manifold. In the following we analyze the mutation symmetry for the ground states.

\begin{figure}[t]
	\centering
	\includegraphics{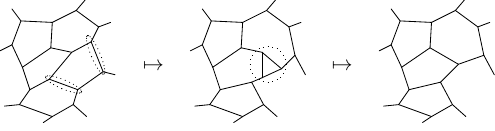}
	\caption{A mutation two graphs that discretize the same manifold. The left one is mutated to the middle one by a composition of $f_1$ moves, and the middle one is mutated to the right one by a $f_3$ move. }
	\label{fig:GraphMutation}
\end{figure}

Let us begin with \textit{any two} arbitrary trivalent graphs $\Gamma^{(1)}$ and $\Gamma^{(2)}$ discretizing the same surface (e.g., a torus). It is known that they can be mutated to each other by a composition of the Pachner moves\cite{Pachner}:
\begin{align}
	\label{eq:PachnerMoves}
	&f_1: \quad \BareHpart \rightarrow \BareRotatedHpart,\\
	&f_2: \quad\BareYpart \rightarrow \BareTriangleYpart,\\
	&f_3: \quad\BareTriangleYpart \rightarrow \BareYpart.
\end{align}
See Fig.\ref{fig:GraphMutation} for instance.

We can associate two different Hilbert spaces to $\Gamma^{(1)}$ and $\Gamma^{(2)}$, respectively, as described in the previous section. Denote by $\mathcal{H}^{(1)}$ the Hilbert space on $\Gamma^{(1)}$, and $\mathcal{H}^{(2)}$ on $\Gamma^{(2)}$.


To the elementary moves $f_1$, $f_2$, and $f_3$, we associate linear maps between the corresponding Hilbert spaces as follows:
\begin{align}
	&\hat{T}_1:\Biggl|\Hpart{j_2}{j_3}{j_5}{j_1}{j_4}{>}{>}{>}{<}{<}\Biggr\rangle\rightarrow\sum_{j'_{5}}v_{j_5}v_{j'_5}G^{j_{1}j_{2}j_{5}}_{j_{3}j_{4}j'_{5}}
	\Biggl|\RotatedHpart{j_2}{j_3}{j'_5}{j_1}{j_4}{>}{>}{>}{<}{<}\Biggr\rangle,\label{eq:T1}\\
	&\hat{T}_2:\Biggl|\Ypart{j_2}{j_1}{j_3}{>}{<}{<}\Biggr\rangle\rightarrow\sum_{j_{4}j_{5}j_{6}}\frac{v_{j_4}v_{j_5}v_{j_6}}
	{\sqrt{D}}G^{j_{2}j_{3}j_{1}}_{j^*_{6}j_{4}j^*_{5}}
	\Biggl|\TriangleYpart{j_2}{j_4}{j_5}{j_6}{j_1}{j_3}{>}{<}{>}{<}{<}{<}\Biggr\rangle,\label{eq:T2}\\
	&\hat{T}_3:\Biggl||\TriangleYpart{j_2}{j_4}{j_5}{j_6}{j_1}{j_3}{>}{<}{>}{<}{<}{<}\Biggr\rangle\rightarrow
	\frac{v_{j_4}v_{j_5}v_{j_6}}{\sqrt{D}}
	G^{j^*_{3}j^*_{2}j^*_{1}}_{j^*_{4}j_{6}j^*_{5}}
	\Biggl|\Ypart{j_2}{j_1}{j_3}{>}{<}{<}\Biggr\rangle.\label{eq:T3}
\end{align}
Note that since we can reverse any edge by conjugating the corresponding label, the above formulas do not depend on the edge directions.

Between the Hilbert spaces $\mathcal{H}^{(1)}$ and $\mathcal{H}^{(2)}$ on any two graphs, there is a mutation transformation by a composition of
these elementary maps. In particular, ${B}_p=D^{-1}\sum_{s}{\rmd}_s{B}_p^s$ is a special example.
In fact, on the particular triangle plaquette $p$ as in \eqref{eq:T3},
we can verify ${B}_{p=\triangledown}=\hat{T}_2\hat{T}_3$.


The ground states have properties:
\begin{enumerate}
	\item The mutations are unitary in the ground-state subspace.
	\item The ground states are invariant under mutations.
\end{enumerate}

By unitarity $O_1=O_2^{\dagger}$ we mean $\bra{\phi}O_1\ket{\phi'}=\overline{\bra{\phi'}O_2\ket{\phi}}$. For example, $\hat{T_2}^{\dagger}=\hat{T_3}$ can be verified by
\begin{align}
	\label{eq:T2daggerToT3}
    &\overline{\Biggl\langle\TriangleYpart{j_2}{j_4}{j_5}{j_6}{j_1}{j_3}{>}{<}{>}{<}{<}{<}\Biggr\vert
		\hat{T}_2
		\Biggl\vert\Ypart{j_2}{j_1}{j_3}{>}{<}{<}\Biggr\rangle}
	=\overline{\frac{v_{j_4}v_{j_5}v_{j_6}}
		{\sqrt{D}}G^{j_{2}j_{3}j_{1}}_{j^*_{6}j_{4}j^*_{5}}}
	\nonumber\\
	=&\frac{v_{j_4}v_{j_5}v_{j_6}}{\sqrt{D}}G^{j^*_{3}j^*_{2}j^*_{1}}_{j^*_{4}j_{6}j^*_{5}}
	=\Biggl\langle\Ypart{j_2}{j_1}{j_3}{>}{<}{<}\Biggr\vert
	\hat{T}_3
	\Biggl\vert\TriangleYpart{j_2}{j_4}{j_5}{j_6}{j_1}{j_3}{>}{<}{>}{<}{<}{<}\Biggr\rangle,
\end{align}
using condition \eqref{eq:6jcond}, $G^{j_{2}j_{3}j_{1}}_{j^*_{6}j_{4}j^*_{5}}\propto \delta_{j^*_4j_1j_6}$, and $\alpha_{j_1}=\alpha_{j_4}\alpha_{j_6}$.

\section{Extension of the Model}

To study the spectrum of Levin-Wen models, we will extend the Hilbert space. An elementary excitation $\ket{\psi}$ supports two types of quasiparticles, charge at vertex $v$ if $Q_v\ket{\psi}=0$, and fluxon at plaquette $p$, if $B_p\ket{\psi}=0$. We extend the local Hilbert space at $v$ to support distinguished charges and internal degrees of freedom 
between 
them.

\begin{figure}
\centering
\vspace{0.5cm}
\subfigure[]{\includegraphics[scale=0.6]{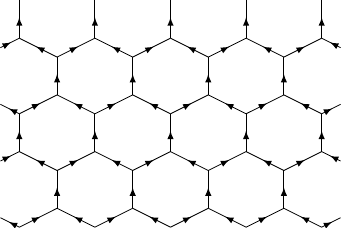}
\label{subfig:a:Lattice}}
\qquad\subfigure[]{\includegraphics[scale=0.6]{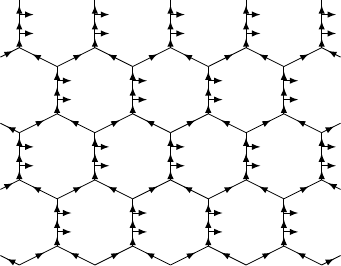}
\label{subfig:b::LatticeWithTail}}
\caption{Extension of the Hilbert space by extra tails.}
\label{fig:Lattice}
\end{figure}

Let us start with a trivalent graph, e.g., the hexagonal lattice in Fig.\ref{subfig:a:Lattice}.
There are three edges connected to each vertex. To each vertex, we associate an open edge called a tail and attach it to one of the three connected edges.   For example, in Fig.\ref{subfig:b::LatticeWithTail} each vertex carries a tail.  
To define the Hilbert space it does not matter which one of the three edges neighboring to the vertex we choose to attach the tail.  

\begin{figure}[!t]
\centering
\vspace{0.5cm}
\subfigure[]{\includegraphics[]{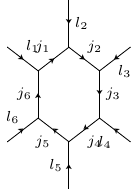}\label{subfig:PurePlaquette}}
\quad\quad
\subfigure[]{\includegraphics[]{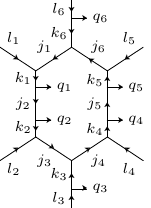}\label{subfig:TailedPlaquette}}
\caption{Around each vertex, two extra d.o.f. are needed: $q$ is assigned to the tail, and $k$ to the line connecting the vertex and the tail.}
\label{fig:Plaquette}
\end{figure}

The Hilbert space is spanned by the string types $j\in I$ on all the edges of the tailed graph. With the tails, there are two more d.o.f. around each vertex. For example, in Fig.\ref{subfig:TailedPlaquette}, these extra d.o.f. are labeled by $k$ and $q$ near each vertex. Each vertex of the spatial graph in Fig.\ref{subfig:a:Lattice} is actually presented by two vertices in Fig.\ref{subfig:b::LatticeWithTail}. Around each new vertex in Fig.\ref{subfig:b::LatticeWithTail}, we require the fusion rule $\delta_{ijk}=1$ for the three neighboring edges labeled by $i,j$, and $k$ connecting to the vertex. For example, at the left upper corner of the plaquette in Fig.\ref{subfig:TailedPlaquette}, we require $\delta_{j_1l_1k^*_1}=1$ and $\delta_{k_1j^*_2q^*_1}=1$.

The Hamiltonian has two terms:
\begin{equation}
\label{eq:HamiltonianGaugeModel}
H=-\sum_v \Q_v-\sum_p \B_p.
\end{equation}

The first term is
\begin{equation}
\label{eq:Qv::QvInGaugeModel}
\Q_v\bket{\NewVertex}=\delta_{q_1,0} \bket{\NewVertex}.
\end{equation}

The second term is
\begin{equation}
\label{eq:Bp::BpInGaugeModel}
\B_p=\frac{1}{D}\sum_s \rmd_s \B_p^s,
\end{equation}
with
\begin{widetext}
\begin{equation}
\label{eq:Bps::InGaugeModel}
\begin{aligned}
\B_p^s\bket{\bmm\NewPlaquette\emm}
=&\delta_{q_1,0}\delta_{q_2,0}\sum_{j'_1j'_2j'_3j'_4j'_5j'_6k'_1k'_3k'_4k'_6}
\rmv_{j_1}\rmv_{j_2}\rmv_{j_3}\rmv_{j_4}\rmv_{j_5}\rmv_{j_6}
\rmv_{j'_1}\rmv_{j'_2}\rmv_{j'_3}\rmv_{j'_4}\rmv_{j'_5}\rmv_{j'_6}\times\\
&\rmv_{k_4}\rmv_{k_5}\rmv_{k'_4}\rmv_{k'_5}G^{l_1k^*_1j_1}_{s^*j'_1j^{\prime*}_2}
G^{l_2j^*_3k_2}_{s^*j'_2j^{\prime*}_3}
G^{k_3j^*_4j_3}_{s^*j'_3j^{\prime*}_4}
G^{l_4k^*_4j_4}_{s^*j'_4k^{\prime*}_4}
G^{l_5j^*_6k_5}_{s^*k'_5j^{\prime*}_6}
G^{k_6j^*_1j_6}_{s^*j'_6j^{\prime*}_1}\times\\
&G^{q_4j^*_5k_4}_{s^*k'_4j^{\prime*}_5}
G^{q_5k^*_5j_5}_{s^*j'_5k^{\prime*}_5}
\bket{\BpsPrime}.
\end{aligned}
\end{equation}
\end{widetext}
The operator $\B_p^s$ is a straightforward extension of the Levin-Wen operator $B_p^s$ in Eq.\ \eqref{eq:Bps::InLW}. The two tails labeled by $q_4$ and $q_5$ are viewed as two external legs of the plaquette. With $q_1=0=q_2$, the plaquette is effectively treated as having 8 boundary vertices.  Acting on this effective plaquette by $B_p^s$ defined in Eq.\ \eqref{eq:Bps::InLW}, we arrive at 16 of $v$'s and 8 of $6j$-symbols in the Eq.\ \eqref{eq:Bps::InGaugeModel}.  

If we restrict to the $\Q=1$ subspace, we recover the traditional Levin-Wen model. Take Fig.\ref{subfig:TailedPlaquette} for an example, in the subspace with $q=0$ fixed for all tails, we have $k_1=j_2=k_2$, $k_4=j_5=k_5$, $k_3=l_3$, $k_6=l_6$, etc. Hence, $\Q=1$ subspace in Fig.\ref{subfig:TailedPlaquette} is identified with the usual $Q=1$ subspace in Fig.\ref{subfig:PurePlaquette}. In such a case, $\B_p^s$ becomes the usual $B_p^s$.

The model is exactly solvable because the local terms in the Hamiltonian \eqref{eq:HamiltonianGaugeModel} are mutually commuting projection operators.

\section{Elementary excitations}\label{sec:ElementaryExcitations}

We study elementary excitations by algebra of local operators preserving topological symmetry. Excitations support quasipartciles, which are identified with the irreducible representations of the algebra and are classified by the quantum double category.

Elementary excitations are mutual eigenvectors of all $\Q$'s and $\B$'s. In particular, the ground states are $\Q=1$ and $\B=1$ eigenvectors. Since the $\Q=1$ subspace recovers traditional $Q=1$ Levin-Wen Hilbert subspace, the ground states in the extended model are exactly the same as in the traditional Levin-Wen model.

Elementary excitations support local quasiaprticles. If an excitation $\ket{\psi}$ is a $\Q_v=0$ eigenvector, we say there is a charge quasiparticle living at $v$, which is identified by a nontrivial tail label. On the other hand, if $\B_p=0$ (with $\Q_v=1$ around the plaquette) we say there is a fluxon quasiparticle at $p$.  We call a generic quasiparticle a dyon -- a composite of charge and fluxon.

\subsection{Topological symmetry and tube algebra of observables}

As discussed above, when all tail labels are trivial, the extended Hilbert space becomes the traditional one without tails. The ground states have mutation symmetry as discussed in Sec \ref{sec:TopologicalSymmetry}. In the following we explore the topological symmetry in excitations. The corresponding topological observables under such symmetry form the ``tube algebra''. This enable us to classify the elementary excitations. In particular, the good quantum numbers can be identified by the irreducible representations of the tube algebra, which are formulated by the quantum double of the input category.

Excitations have less symmetry.  In the presence of a nontrivial quasiparticle at the triangle plaquette in Eq.\ \eqref{eq:T2} and \eqref{eq:T3}, excited states are not invariant under $\hat{T_2}$ and $\hat{T_3}$, because $\hat{T}_2\hat{T}_3={B}_{p=\triangledown}=0$.

In the generic case, the tail label --- called a charge --- within a plaquette is nontrivial. States are not invariant under $\hat{T_2}$ and $\hat{T_3}$. But they still have $\hat{T}_1$ symmetry. Another symmetry with respect to the tail is to move a tail along the plaquette boundary. Define 
\begin{align}
\hat{T}_4:\Biggl|\NewPlaquetteMoveTailA\Biggr\rangle\rightarrow\sum_{k'_{1}}v_{k_1}v_{k'_1}G^{j^*_2q^*_1k_1}_{j_1l_1k'_1}
\Biggl|\NewPlaquetteMoveTailB\Biggr\rangle,\label{eq:T4}
\end{align}
which has a similar form to $\hat{T}_1$. The tail $q_1$ can freely move along the plaquette boundary, as long as it does not cross another tail, e.g., the one labeled by $q_2$ above.

\begin{figure}[h]
	\centering
	\subfigure[]{\includegraphics[]{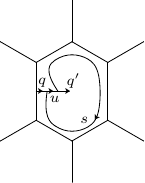}}
	\label{subfig:OneTailLoop}
	\quad
	\subfigure[]{\includegraphics[]{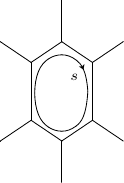}}
	\label{subfig:Loop}
	\quad
	\subfigure[]{\includegraphics[]{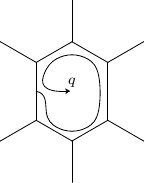}}
	\label{subfig:DyonTwist:Rotation}
	\caption{Graphical interpretation of $B_{qjq's}$ and $B_p^s$. (a) ${B}_{qsq'u}$ attaches a string and fuses it along the plaquette boundary by $\hat{T}_1$ and $\hat{T_3}$. (b). With $q=q'=0$, $B_{qjq's}$ is reduced to $B_{0s0s}=\B_p^s$. (c) With $u=0$, $\Theta$ performs a rotation of a tail along the plaquette boundary.}
	\label{fig:FuseLoop}
\end{figure}

We are interested in topological observables, i.e., local operators that preserve $\hat{T}_1$ and $\hat{T_4}$ transformations. Consider elementary excitations with at most one quasiparticle at the plaquette for simplicity. Hence we consider only one tail for simplicity as follows. Define
\begin{widetext}
	\begin{equation}
	\label{eq:LocalDyonOperator}
	\begin{aligned}
	B_{qsq'u}\bket{\bmm\NewPlaquetteOneTail\emm}
	=&\sum_{j'_1j'_2j'_3j'_4j'_5j'_6k'}
	\rmv_{j_1}\rmv_{j_2}\rmv_{j_3}\rmv_{j_4}\rmv_{j_5}\rmv_{j_6}\rmv_{k}
	\rmv_{j'_1}\rmv_{j'_2}\rmv_{j'_3}\rmv_{j'_4}\rmv_{j'_5}\rmv_{j'_6}\rmv_{k'}
	\times\\
	&\qquad
	G^{l_1k^*j_1}_{sj'_1k^{\prime*}}
	G^{l_2j^*_3j_2}_{sj'_2j^{\prime*}_3}
	G^{l_3j^*_4j_3}_{sj'_3j^{\prime*}_4}
	G^{l_4j^*_5j_4}_{sj'_4j^{\prime*}_5}
	G^{l_5j^*_6j_5}_{sj'_5j^{\prime*}_6}
	G^{l_6j^*_1j_6}_{sj'_6j^{\prime*}_1}\times\\
	&\qquad
	\left(G^{kj^*_2q^*}_{su^*j^{\prime*}_2}G^{j^{\prime*}_2u^*k}_{sk'q^{\prime*}}\right)
	\bket{\NewPlaquetteOneTailPrime}.
	\end{aligned}
	\end{equation}
\end{widetext}
This operator has a graphical presentation of fusing a string labeled by $s$ along the plaquette boundary, performed by $\hat{T}$ transformations (up to some normalization factor), see Fig.\ref{fig:FuseLoop}. 

The topological observables preserving $\hat{T}_1$ and $\hat{T}_4$ are linear combinations of $B_{pkqt}$. 
Denote such  
a generic operator by
\begin{equation}
x=\sum_{pkqt}x_{pkqt}B_{pkqt},
\label{eq:Elementx}
\end{equation}
where the summation runs over $p,k,q,t$ with $\delta_{pqt^*}=1=\delta_{kqt^*}$ (otherwise $B_{pkqt}=0$). These operators satisfy the multiplication rule $x\cdot y=z$ where $z$ is given by:
\begin{equation}
z_{pkqt}
={\rmd_k \rmd_t}\sum_{mnlrs}x_{lnqr}y_{pmls}G^{m^*sl^*}_{nr^*t}G^{m^*tr^*}_{q^*n^*k}G^{s^*pm}_{kn^*t}
\label{eq:xy=z}
\end{equation}
which can be verified in graphical presentation in Fig.\ref{fig:FuseLoop}(a).
The operators in Eq.\ \eqref{eq:Elementx} equipped such a multiplication rule form the \textit{tube algebra} $\scA$. 

The local operator $\B_p^s$ in the Hamiltonian is a special element in $\scA$:
\begin{equation}
\B_p^s=\rmd_s B_{0s0s}.
\end{equation}
Which fuses a loop to the plaquette boundary. See Fig.\ref{fig:FuseLoop}(b).

Quasiparticles in elementary excitations are identified by projection operators $\Pi\cdot \Pi=\Pi$ where $\Pi$ is minimal.  
Here minimal means that if $\Pi$ is a combination of projection operators $\Pi=\Pi_1+\Pi_2$ then $\Pi_1$ or $\Pi_2$ is zero.  
Each projection $\Pi$ projects onto states with a specific quasiparticle at $p$ called a dyon.  A ground state is a $\prod_p\B_p=1$ eigenstate; 
we say such a $p$ has a trivial dyon which we identify with the projection $\B_p$.  
Dyons identified by all other projections $\Pi\neq \B_p$ carry higher energy, because $\B_p$ is a special minimal projection and $\B_p\cdot \Pi=0$.

Consider dyons with charge $q$ at the tail fixed. Operators on such states form a subalgebra $\scA_q$ with elements $x=\sum_{kt}x_{qkqt}B_{qkqt}$. If we express a projection in $\scA_q$ by
\begin{equation}
\Pi_q=\sum_{kt}\Pi_{qkt}B_{qkqt},
\end{equation}
then $\Pi_q\cdot \Pi_q=\Pi_q$ implies
\begin{equation}
\Pi_{qkt}
={\rmd_k \rmd_t}\sum_{mnrs}\Pi_{qnr}\Pi_{qms}G^{m^*sq^*}_{nr^*t}G^{m^*tr^*}_{q^*n^*k}G^{s^*qm}_{kn^*t}.
\label{eq:PiPi=Pi}
\end{equation}

Each minimal projection $\Pi_q$ identifies a dyon at $p$ and is in one-to-one correspondence with an irreducible representation of $\scA_q$.

\subsection{Quantum double theory of dyons}

The dyons with fixed charge $q$ are identified with irreducible representations of $\scA_q$. However, the dyon species are identified with the irreducible representations of tube algebra $\scA$. A dyon with fixed charge $q$ does not form a species itself, because a generic topological observable $B_{qkq't}$ transforms the dyon with charge $q$ to other dyon(s) with charge $q'$. A dyon species is identified with a set of dyons that are invariant under $\scA$, i.e., with an irreducible representation of $\scA$. Such irreducible representations form quantum double category, with each representation identified with a quantum double label. Hence the natural algebraic theory of dyons is the quantum double category theory.

In the rest of this subsection we expand on the ideas of the previous paragraph.  
 The irreducible representations of tube algebra form a quantum double category. The key structure in the latter is the half braiding. A half-braiding tensor $z$ satisfies the naturality condition
\begin{equation}
\sum_{lrs}\rmd_r\rmd_sz_{lnqr}z_{pmls}G ^{m^*sl^*}_{nr^*t}G^{s^*pm}_{jn^*t}G^{m^*tr^*}_{q^*n^*k}=\delta_{mnj^*}\frac{\delta_{jk}}{\rmd_j}z_{pjqt}.
\label{eq:zzzNaturalityHalfBraiding}
\end{equation}

A minimal solution to 
this equation is associate with 
  a quantum double label $J$. Each $J$ is one-to-one corresponding to an irreducible representation of tube algebra \cite{Mueger}. Denote each minimal solution by $z^J_{pjqt}$.

Quantum double labels classify dyon species.
Each dyon species may carry different charges $q$ just like each spin may carry different magnetic components. In this case, we say these dyons belong to the same dyon species, denoted by a quantum double label $J$. A dyon species is identified by a minimal central projection in $\scA$. Here ``central'' means it commutes with all topological observables in $\scA$. It is a sum
\begin{equation}
\Pi^J=\sum_q \Pi^J_q,
\end{equation}
where $q$ runs over all dyons that belong to $J$.

Each dyon that belongs to species $J$ has the projection $\Pi^J_q$ arising from $z^J$ by
\begin{equation}
\frac{\Pi^J_{qkt}}{\Pi^J_{q0q}}=\frac{\rmd_k\rmd_t}{\rmd_q}z^J_{qkqt}.
\end{equation}
There may be several projections $\Pi_p$, $\Pi_q$, $\dots$, arise from the same $J$, with $p\neq q$.  

In general, each $J$ may carry multiple copies of a charge $q$, denoted by index $\alpha$ in $\Pi^J=\sum_{q,\alpha} \Pi^J_{q,\alpha}$. Throughout this paper for simplicity, we assume each $q$ appears at most once in all $J$.

\subsection{Dyon string operator}\label{SS:stringOperators}

In this subsection we define dyon-pair creation and annihilation operators.   
In the $\Q=1$ subspace, all tails are labeled by the trivial string type $q=0$.  We draw the dotted line to present the trivial label $0$ for convenience.
Fix an edge $e$, and consider a state $\ket{\Psi}$ with no charge at either of the two vertices of $e$.  For such an edge we can define a creation operator.  For example, in the following diagram, we create a pair of dyons across the middle vertical edge labeled by $j_2$.  Define creation operator by
\begin{equation}
	\label{eq:CreateCrossedDyonPair}
	\begin{aligned}
		&W^{J;pq^*}_e\bket{\GroundStateTailedTwoPlaquette}\\
		=\;&\sum_{j'_2}\frac{\rmv_{j'_2}}{\rmv_{j_2}}\overline{z^J_{pj^{\prime}_2qj_2}}
		\bket{\DyonTwoPlaquette}.
	\end{aligned}
\end{equation}

The resulting state $W^{J;pq^*}_e\ket{\Psi}$ has charge $p$ at the lower left vertex, and charge $q^*$ at the upper left vertex. The two plaquettes are occupied by a pair of dyons. 

The generated state is normalized to:
\begin{equation}
	\bra{\Psi}{W^{J;pq^*}_e}^{\dagger}W^{J;pq^*}_e\ket{\Psi}
	=\rmd_p\rmd_q/\rmd_J\langle\Psi\ket{\Psi}.
\end{equation}

Given a ground state, $W^{J;pq^*}_e$ enables us to explicitly write down the elementary excitation wavefunction.

Now we develop the dyon string operator, which creates a pair of dyons at the end of the string. First, we create two dyon-pairs across two edges respectively, e.g., the two labeled by $j_9$ and by $j_2$ as follows. 
\vspace{2cm}
\begin{widetext}
	\begin{equation}
	\begin{aligned}
	&\sum_{q'}W^{J;pq^{\prime*}}_{e_1}W^{J;q'q^*}_{e_2}\bket{\GroundStateTailedThreePlaquette}
	\\
	=\;&\sum_{q'j'_9j'_2}\overline{{z}^J_{pj'_9q'j_9}}\,\overline{z^J_{q'j'_2qj_2}}
	\frac{\rmv_{j'_9}}{\rmv_{j_9}}
	\frac{\rmv_{j'_2}}{\rmv_{j_2}}
	\bket{\TwoDyonThreePlaquette}.
	\end{aligned}
	\label{eq:CreateTwoDyonPairs}
	\end{equation}
\end{widetext}
The resulting state has the first pair of dyons occupying the left and the middle plaquette, and the second pair occupying the middle and the right plaquette. Next we annihilate the two charges in the middle plaquette. By $\hat{T}_1$ followed by a sequence of $\hat{T}_4$ moves, we can move the tail $q'$ to be at the same edge with $q^{\prime*}$. Then we annihilate the charge by
\begin{equation}
	\label{eq:QQW::AnnihilateChargeI}
	\bket{\ChargeAnnihilateA} \mapsto \delta_{k_1k_2}\delta_{qk_1j^*_2}\frac{\rmv_{j}}{\rmv_{k_1}}
	\bket{\ChargeAnnihilateB},
\end{equation}

The desired string operator is the composition of the following: we first create two dyon-pairs as in Eq.\ \eqref{eq:CreateTwoDyonPairs}, second we move the two tails in the middle plaquette to be at the same edge, and third we annihilate the two charges as in Eq.\ \eqref{eq:QQW::AnnihilateChargeI}. This procedure defines a string operator across two edges. We can repeat this procedure to define a string operator along a longer string. 


The process to annihilate charge pairs $q$ and $q^*$ and sum over $q$ at the middle plaquette is called the \textit{contraction} of charges. After the contraction, no nontrivial quasiparticle is left at the middle plaquette. The string operator defined is path independent: \textit{two string operators along two isotopic paths result in the same final state if the final position of the dyon is the same and there is no non-trivial quasiparticle in the area enclosed by the two paths.}

\begin{figure}[t]
\centering
\subfigure[]{\includegraphics[]{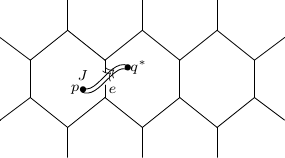}}
\qquad\qquad
\subfigure[]{\includegraphics[]{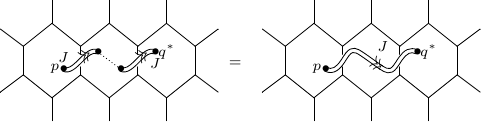}}
\quad\subfigure[]{\includegraphics[]{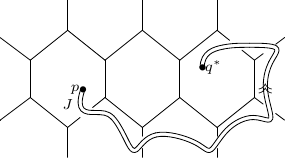}}
\caption{(a) A ribbon presents a pair of dyons created at two plaquettes across edge $e$. (b) Two dyons are created. The dotted line presents the contraction of the charges at the middle plaquette. The resulting state is dyon pair state with two dyons at the two ends of the ribbon string. (c). Creation along two isotopic strings result in the same dyon pair state if no nontrivial quasiparticle exists in the area enclosed by the two strings. The two string operators in (b) and (c) are the same.}
\label{fig:QDstring}
\end{figure}

We use ribbon strings to represent creation and string operators. In Fig.\ref{fig:QDstring}(a), the ribbon string 
represents the creation by 
$W^{J;pq^*}_e$of a dyon pair across an edge $e$. 
There are charges $p$ and $q^*$ at the two ends of this string.   In Fig.\ref{fig:QDstring}(b) the dotted line presents the contraction of charges at the middle plaquette. Here the charge contraction connects two strings to a new one.  The two string operators in Fig.\ref{fig:QDstring}(b) and Fig.\ref{fig:QDstring}(c) are equal, illustrating the path independence of the string operator.

\subsection{Twist and modular $S$ matrix}\label{SS:TwistSmatirx}

In this subsection, we analyze some topological observables in terms of the string operators to characterize the dyon species.

We need a special choice of $z^J$ to write down the creation operator (or equivalently the elementary excitation wavefunction). This amounts to picking a specific representation in which the state transforms under the tube algebra $\scA$. On the other hand, the topological observables that characterize topological properties of dyon excitations do not depend on specific choices of $z^J$. In the following, we explore the topological properties using $\Pi^J$, which are uniquely determined by the $6j$-symbols and does not depend choices of $z^J$.

The simplest invariant is obtained by contraction of the charges at the two ends of a string, leading to a loop operator. If no nontrivial quasiparticle exists in the area enclosed by the loop, the closed string operator gives a multiple of the identity matrix:
\begin{equation}
\bmm\ContractString\emm\,\,=
\,\bmm\LoopString\emm\,
=\rmd_J \mathbf{1},
\end{equation}
where $\rmd_J$ is called the quantum dimension of $J$, defined by
\begin{equation}
\rmd_J=\sum_{q\in J}\rmd_q.
\end{equation}

The next topological observable is the twist: define the \textit{twist} by
\begin{equation}
\Theta=\sum_{q}\rmd_q B_{qq^*q0}.
\label{eq:TwistDyon}
\end{equation}
It commutes with all dyon projections, and hence is a good quantum number of a dyon state. For states with dyon identified by $\Pi^J_q$, the eigenvalue is solved to be
\begin{equation}
\theta_J=\frac{1}{\rmd_q\Pi^J_{q0q}}\sum_{t}\Pi^J_{qqt}.
\label{eq:TwistDyonEigenvalue}
\end{equation} 
The twist $\theta_J$ is the same for all dyons in the same species $J$, even with different charges $q$.

 This scalar  is a $U(1)$ number which we identify with $\theta_J$.  
The definition in Eq.\ \eqref{eq:TwistDyon} has a graphical presentation of a self rotation of the tail. See Fig.\ref{fig:FuseLoop}(c). Hence $\theta$ is identified with the dyon's statistical spin $s$ via $\theta=\exp(2\pi i s)$. 

\begin{figure}[ht]
	\centering
	\subfigure[]{\includegraphics[]{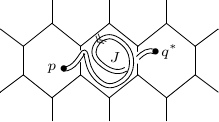}}
	\quad
	\subfigure[]{\includegraphics[]{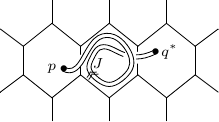}}
	\caption{A dyon twist is presented by a twist of the ribbon string.}
	\label{fig:TwistString}
\end{figure}

If we apply the twist in Fig.\ref{fig:QDstring}(b), we obtain a string operator as in Fig \ref{fig:TwistString}. Twisting either dyon in the middle plaquette of Fig. \ref{fig:QDstring}(b) before the charge contraction leads to the string operators in Fig \ref{fig:TwistString}(a) and Fig \ref{fig:TwistString}(b) separately. Therefore the twist $\theta_J$ of a dyon can be detected by the string operators
\begin{equation}
\label{eq:TwistString}
\bmm\RibbonTwistL\emm=\theta_J\,\,\bmm\RibbonId\emm=\bmm\RibbonTwistR\emm.
\end{equation}
Since the string operators are path independent, it is not important where on the spatial graph we put the string. The crossing matters, which indicates the path order of creation and charge contraction operators. Therefore it is safe to draw only the ribbon strings without mentioning the underlying spatial graph.

\begin{figure}
\centering
\subfigure[]{\includegraphics[]{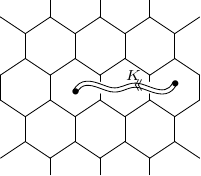}}
\quad
\subfigure[]{\includegraphics[]{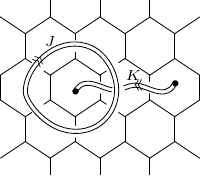}}
\quad
\subfigure[]{\includegraphics[]{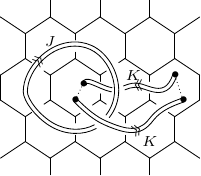}}
\caption{Three steps to evaluate $S$ matrix: (a) create a pair of $K$-dyons; (b)  wound around a $K$-dyon by $J$-dyon closed string operator; (c) contract the two ends of the $K$-string.}
\label{fig:StringSmatrix}
\end{figure}

Another important topological observable is the modular $S$ matrix, as defined as follows. 
First, we create a dyon-pair. Second, we add a closed string operator around on end point of the previously created dyon-pair. Third, we contract the ends of the dyon-pair. 
See Fig.\ref{fig:StringSmatrix}. If no nontrivial quasiparticle exists in the area enclosed both strings, the final operator is a multiple of the identity matrix. Presented in terms of the ribbon strings, they are
\begin{equation}
\label{eq:SmatrixString}
\bmm\SmatrixString\emm=\bmm\SmatrixStringB\emm=S_{JK}\mathbf{1}.
\end{equation}

The $S$ matrix characterizes the holonomy effect of winding dyon $J$ around $K$, or equivalently, exchanging $J$ and $K$ twice.

The $S$ matrix turns out to be independent of choices of $z^J$. It evaluates to be
\begin{equation}
S_{JK}=\sum_{p,q,t}\overline{\,\left(\frac{\Pi^J_{pqt}\Pi^K_{qpt}}{\Pi^J_{p0p}\Pi^K_{q0q}}\right)\,}\frac{1}{\rmd_t}.
\end{equation}

Sometimes it is useful to define the $T$ matrix by
\begin{equation}
T_{JK}=\delta_{JK}\theta_J.
\end{equation}
Therefore, the twist $\theta_J$ is actually an eigenvalue of the T matrix. In our approach, the T matrix is realized as the operator that moves the quasiparticle around the plaquette by one turn. This operator commutes with the Hamiltonian, and thus its eigenvalue is a good quantum number. 

Modular matrices $S$ and $T$ are believed to characterize the quantum double category, and contain all information on the good quantum numbers of the dyon species.

\subsection{Fusion and hopping operators}
\label{subsec:Fusion}

Dyon species are closed under fusion. Here the fusion process is described as follows: when two pairs of dyons of species $J$ and $K$ are created on 
the same 
two plaquettes, the resulting state is a linear combination of ones obtained from the creation of one pair of dyons. 
If the dyon-pair state $L$ appears in this linear combination,
then we define $\delta_{JKL^*}=1$ and  $\delta_{JKL^*}=0$ otherwise.

\begin{figure}[ht!]
\centering
\subfigure[]{\includegraphics[]{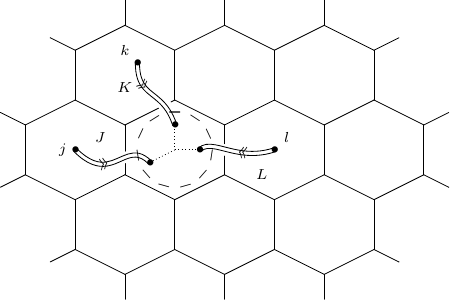}}
\subfigure[]{\includegraphics[]{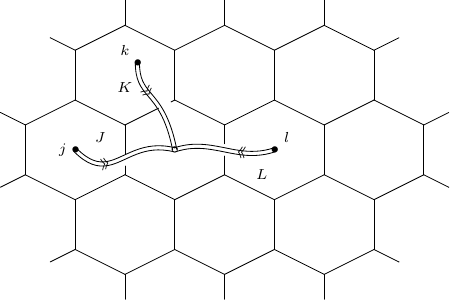}}
\caption{(a). Create three pair of dyons, and fuse three dyons at the middle plaquette by annihilating charges and fluxons. The dotted line presents charge annihilation and dashed line presents fluxon annihilation. (b). If $\delta_{JKL}=1$, the fusion process results in a elementary three-dyon excitation.}
\label{fig:fusionDyon}
\end{figure}

Next we consider another way to describe the fusion process that results in elementary three-dyon states (on the sphere). In terms of string operators, we create three pairs of dyons of species $J,K$ and $L$, with three dyons at one end of each pair meeting at the same plaquette. If we annihilate these three dyons, the resulting state is the zero vector or a nonzero three-dyon state. See Fig.\ref{fig:fusionDyon}(a). We define the fusion rule by $\delta_{JKL}=1$ if we obtain a nonzero three-dyon state and $\delta_{JKL}=0$ otherwise. (In general, there may be more than one fusion channels, but throughout the paper we consider the multiplicity free cases for simplicity.)

Let us consider the later fusion process in more detail.  To do this we define the fusion of charge by
\begin{equation}
	\label{eq:fusionCharge}
	\begin{aligned}
	&\bket{\NewPlaquette}
	\\
	\mapsto&\sum_{q'_1}\rmv_{q_2}\rmv_{j_2}\rmv_{q'_1}G^{q^*_1k_1j^*_2}_{k^{*}_2q^*_2q'_1}
	\bket{\ChargeHoppingNewPlaquette}
	\end{aligned},
\end{equation}
where the charge $q_2$ is moved upward and fused with $q_1$, resulting a linear combination of charge $q'_1$ state.


The fusion process is illustrated in Fig.\ref{fig:fusionDyon}(a) and described as follows.   First, we create three pairs of dyons by $W^{J;pj},W^{K;qk}$ and $W^{L;rl}$, with summation over $p,q$ and $r$. Second, we annihilate the three charges at the middle plaquette by composition of $\hat{T}_1$ moves and the above charge fusion given in Eq.\ \eqref{eq:fusionCharge}.  Then we apply the projection $\Q_v$. Finally, we apply $\B_p$ at the plaquette to annihilate the fluxon. If $\delta_{JKL}=1$, the nonzero resulting state is graphically presented by a riboon three-valent tree structure. See Fig. \ref{fig:fusionDyon}(b).

The fusion rule is completely determined by the $S$ matrix (known as Verlinde formula\cite{Verlinde}):
\begin{equation}
\delta_{JKL}=\frac{1}{D^2}\sum_{N}\frac{S_{JN}S_{KN}S_{LN}}{S_{0N}}.
\end{equation}

\begin{figure}[ht!]
	\centering
	\subfigure[]{\includegraphics[]{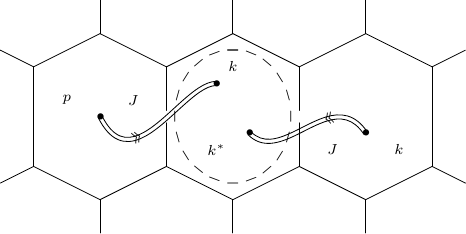}}
	\caption{Hopping Operator: the dash circle represents fluxon annihilation. } 
	\label{fig:TwoDyonFusion}
\end{figure}

We end this subsection by describing the \emph{hopping operator}:  First, 
we create two dyon pairs, both of species $J$, with two particular charges 
$k$  and $k^*$ are at the same plaquette.  See Fig.\ref{fig:TwoDyonFusion}.
Second we annihilate charge using Eq.\ \eqref{eq:QQW::AnnihilateChargeI} and fluxon using $B_p$.  

The hopping operator is equal to the composition of $\frac{\rmd_J}{\rmd_k}W^{J;kk^*}$ with the charge and fluxon annihilation.  We present this equality graphically by
\begin{equation}
\begin{aligned}
&\bmm\TwoDyonFusionEq\emm
\\
\,=\,
&\frac{\rmd_k}{\rmd_J}\,\bmm\TwoDyonContractionEq\emm
\,\,=\,\frac{\rmd_k}{\rmd_J}\,\,\bmm\StringOperatorJ\emm
\end{aligned}
\end{equation}

\section{Excitation Spectrum} 
\label{subsec:Spectrum}

In Sec.\ \ref{subsec:Fusion}, we showed how to generate a three-dyon excitation from a ground state. In this section we study the full spectrum for all excitations.

We consider models on the sphere only. There is only one ground state\cite{GSD}. We claim all excitations can be generated from the ground by string operators and annihilation operators.

\begin{figure}[ht!]
	\centering
	\subfigure[]{\includegraphics[]{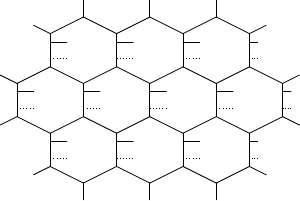}}
	\subfigure[]{\includegraphics[]{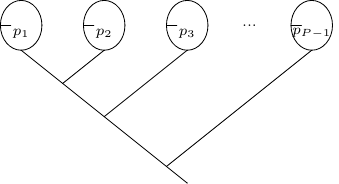}}
	\caption{(a). Each plaquette has one tail that may take nontrivial charge. (b) Unitary $\hat{T}_1$ moves mutate the graph to a tree-like graph, with $P$ the total number of plaquettes.}
	\label{fig:UnitCellGraph}
\end{figure}

To reveal the structure of the spectrum, it is convenient to consider unit cells of one vertex (including one charge) and one plaquette. For simplicity, let us first consider a simplified situation: each plaquette contains at least one tail. At each plaquette, we only consider nontrivial charge at a particular tail. This amounts to enforcing all other tails at each plaquette to be labeled by trivial charge. Then each plaquette together with the tail form a unit cell that supports exactly one dyon. See Fig.\ref{fig:UnitCellGraph}(a) for example.

By applying 
$\hat{T}_1$ moves, one can always mutate the graph to have a tree-like graph as in Fig.\ref{fig:UnitCellGraph}(b). All plaquettes are transformed to $P-1$ bubbles except the last one. (The outside region forms one plaquette on sphere.) These bubbles are the new unit cells that support exactly one dyon. All (good quantum numbers of) dyons at bubbles are preserved during the mutation.

A typical excitation can be generated by $P-1$ pairs of dyons across the bubbles. Let us denote the dyons inside bubbles by $\{J_p,q_p\}_{p\le P-1}$. But the $P-1$ dyons outside the bubbles form a huge multiplicity. Similar analysis based on fusion process in Sec.\ \ref{subsec:Fusion} implies that this multiplicity can be organized by tree structure:
\begin{equation}
\bmm\DyonBasis\emm.
\end{equation}

The fusion channels that occur outside region are diagonalized by $\{K_e\}_{1\le e\le P-3}$. These $K$ degrees of freedom describe how the $P-1$ dyons are fused into linear combination of $J_Pq_{P}$ dyons at the outside plaquette.

Therefore the basis for excitations is
\begin{equation}
\label{eq:DyonBasis}
\begin{aligned}
&\{\ket{\{J_pq_p\}_{1\le p\le P},\{K_e\}_{1\le e\le P-3}}|
\\
&\delta_{J_1J_2K^*_1}\delta_{K_{P-3}J_{P-1}J_P}\prod^{P-4}_{e}\delta_{K_eJ_{e+2}K^*_{e+1}}=1\}.
\end{aligned}
\end{equation}

In the models arising from modular tensor categories, each quantum double label is a pair $i\overline{j}$. See Sec.\ \ref{subsec:ExampleModularCategory}. 
The basis is simplified as
\begin{equation}
	\label{eq:MTCDyonBasis}
	\begin{aligned}
		&\{\ket{\{i_p\overline{j},q_p\}_{1\le p\le P},\{k_e\overline{l}_e\}_{1\le e\le P-3}}|
		\\
		&(\prod_p\delta_{i_pj_pq^*_p})\delta_{i_1i_2k^*_1}\delta_{k_{P-3}i_{P-1}i_P}\prod^{P-4}_{e}\delta_{k_ei_{e+2}k^*_{e+1}}=1\}
	\end{aligned}
\end{equation}


\section{Emergent braiding statistics}

The basis \eqref{eq:DyonBasis} allows us to calculate the fractional exchange statistics of dyons. The transformation of degenerate $N$-dyon states under the exchange of any two dyons can be computed using the hopping operators we have developed in 
Sec.\ \ref{subsec:Fusion}. They form a representation of the Braid group $B_N$, because of the path independence of the hopping operators. 

Consider $N$-dyon excitation states, with $N$ dyons labeled by $\{J_pq_p\}_{1\le p\le N}$ at $N$ fixed unit cells (plaquette together with a tail). The braiding matrix is computed in the $N$-dyon excitations have basis
\begin{equation}
	\label{eq:NDyonBasis}
	\begin{aligned}
		&\{\ket{\{K_e\}_{1\le e\le N-3}}
		\\
		&|\delta_{J_1J_2K^*_1}\delta_{K_{N-3}J_{N-1}J_N}\prod^{N-4}_{e}\delta_{K_eJ_{e+2}K^*_{e+1}}=1\}
	\end{aligned}
\end{equation}
Although they dyons may be different of different species, the braiding matrices form a presentation of the braid group $B_N$. The braiding matrices are independent of charges $q_p$.

In models arising from modular tensor category, any three dyon states has the basis
\begin{equation}
	\ket{i_1\overline{j}_1,q_1;i_2\overline{j}_2,q_2;i_3\overline{j}_3,q_3}.
\end{equation}

The braiding matrices $\sigma_1$ and $\sigma_2$ are diagonal matrices given by diagonal of $R^{i^*_3}_{i_1i_2}/R^{j^*_3}_{j_1j_2}$ and $R^{i^*_1}_{i_3i_2}/R^{j^*_1}_{j_3j_2}$.

Consider the doubled Fibonacci model for example. Consider the four-fluxon states on a sphere. Each fluxon is either a pure fluxon labeled by $(J=\tau\overline{\tau},q=0)$, or carrying a charge $\tau$, labeled by $(\tau\overline{\tau})$. The four-fluxon states have a basis
\begin{equation}
	\bmm\FourFluxonBasis\emm,
\end{equation}
where $K=\mathbf{1},\tau,\overline{\tau},\tau\overline{\tau}$. The four dots at the top label four fluxons, which may or may not carry a charge $\tau$. These charges do not affect the braiding matrices, and are thus not presented in the basis. For simplicity, we choose all four fluxons to be  $(J=\tau\overline{\tau},q=0)$, and the computation is within the usual Levin-Wen Hilbert space.

If we exchange two fluxons in the counterclockwise direction by the hopping operators, we obtain the braiding matrices in the above basis
\begin{align}
	&\sigma _1=\sigma _3=\left(
	\begin{array}{cccc}
		1 & 0 & 0 & 0 \\
		0 & e^{\frac{3 i \pi }{5}} & 0 & 0 \\
		0 & 0 & e^{-\frac{3 i \pi }{5}} & 0 \\
		0 & 0 & 0 & 1 \\
	\end{array}
	\right),\nonumber\\
	&\sigma _2=\left(
	\begin{array}{cccc}
		\phi ^2 & e^{-\frac{3 i \pi }{5}} \phi ^{3/2} & e^{\frac{3 i \pi
			}{5}} \phi ^{3/2} & \phi  \\
		e^{-\frac{3 i \pi }{5}} \phi ^{3/2} & e^{-\frac{i \pi }{5}} \phi ^2
		& \phi  & e^{\frac{2 i \pi }{5}} \phi ^{3/2} \\
		e^{\frac{3 i \pi }{5}} \phi ^{3/2} & \phi  & e^{\frac{i \pi }{5}}
		\phi ^2 & e^{-\frac{2 i \pi }{5}} \phi ^{3/2} \\
		\phi  & e^{\frac{2 i \pi }{5}} \phi ^{3/2} & e^{-\frac{2 i \pi
			}{5}} \phi ^{3/2} & \phi ^2 \\
	\end{array}
	\right),
\end{align}
where $\phi=\frac{\sqrt{5}-1}{2}$. 
$\sigma_1$ exchanges the fluxon 1 and 2, $\sigma_2$ exchanges 2 and 3, and $\sigma_3$ exchanges 3 and 4. They generate the representation of the braid group $B_4$.

The four eigenvalues of $\sigma_1$ and $\sigma_3$ are verified to be $R^{K}_{\tau\overline{\tau},\tau\overline{\tau}}$. $\sigma_2$ can be obtained by a basis transformation in terms of $6j$ symbols.

\section{Braided models}
\label{sec:BraidedModels}
 
Many example models are equipped with $R$ matrix, include the models arising from representations of finite groups and quantum groups. The presence of an $R$-matrix simplifies the disruption of the operators in the model.  In this section we analyze in detail how to characterize the dyons by three good quantum numbers: charge, fluxon, and twist. We study in more detail the creation, annihilation, and string operators in this situation.
 
Let $\{\rmd,\delta,G\}$ be the data discussed in Sec.\ \ref{S:LWModels}.  The $R$ matrix is a map $R:L^3\rightarrow \ds{C}$ that satisfies hexagon equations:
\begin{align}
&\sum_g \rmd_g G^{cad^*}_{be^*g}R^e_{gc}G^{abg^*}_{ce^*f}=R^d_{ac}G^{acd^*}_{be^*f}R^f_{bc}.\\
&\sum_g \rmd_g G^{e^*bd}_{cag}R^e_{ad}G^{e^*ag}_{bcf}=R^d_{ac}G^{e^*bd}_{acf}R^f_{ab}.
\end{align}
The data $\{\rmd,\delta,G,R\}$ is a tensor description of a unitary braided category. Examples include the models arising from representations of finite groups and quantum groups.

\subsection{Good quantum numbers of dyons}

\subsubsection{Charge}

Recall the definition of a charge:  an excited state $\ket{\psi}$ has a charge at vertex $v$, if $\Q_v\ket{\psi}=0$, namely, if the tail label $q_v$ associated to the vertex $v$ is not the trivial $0$. We say $\ket{\psi}$ carries a charge $q_v$. More precisely, define
\begin{equation}
\Q_v^{q}\bket{\NewVertex}=\delta_{q_1,q}\bket{\NewVertex}.
\end{equation}
It commutes with the Hamiltonian \eqref{eq:HamiltonianGaugeModel}, and thus $q$ is good quantum number of $\ket{\psi}$. In particular, $\Q_v^{q=0}=\Q_v$ projects onto trivial charge.

Another good quantum number in the charge excitations is related to the topological spin of the charge. To construct it, we examine how a tail charge is associated to a vertex. There are different choices to associate a tail to vertex to specify the Hilbert space.  In this section, there is no canonical choice better than the others.  For example, if the three edges incoming into one vertex are labeled by $i$, $j$, and $l$, there six possible ways to associate a tail labeled by $q$:
\begin{equation}
\begin{aligned}
&\VertexA \qquad
\VertexB \qquad
\VertexC \\
&\VertexD \qquad
\VertexE \qquad
\VertexF .
\end{aligned}
\label{eq:SixWaysTail}
\end{equation}

All of six choices are equivalently good. These six ways specify a basis of six different Hilbert spaces. Define the basis transformations among them:
\begin{align}
\label{eq:mu::BasisTransformations}
&\mu:\quad
\bket{\VertexA}
\mapsto\;
\sum_{v}\rmv_{u}\rmv_{v}G^{jiu}_{lq^*v^*}
\bket{\VertexB},\\
\label{eq:nu::BasisTransformations}
&\nu:\quad
\bket{\VertexF}
\mapsto\;
\overline{R^{u'}_{q^*l}}\;\bket{\VertexA}.
\end{align}

The transformation $\nu$ moves the tail between the left side and the right side on the same edge, while $\mu$ moves a tail to another edge. Both moves are in the clockwise direction.

In the absence of fluxon at plaquette, the \textit{twist} defined in Eq.\ \eqref{eq:TwistDyon} can be reinterpreted as
\begin{equation}
\label{eq:Theta::TwistCharge}
\Theta_v=\nu\mu\nu\mu\nu\mu,
\end{equation}
which take a tail in the counterclockwise direction around the vertex and finally back to the same position. This process realizes ``self-rotation'' of the charge.

The eigenvalue for charge $q$ is
\begin{equation}
\theta_q=\overline{R^{l^*}_{u^*q^*}R^{u}_{lq^*}},
\end{equation}
which is a $U(1)$ number that depends only on $q$.

\subsubsection{Fluxon}

Suppose there is a nontrivial charge at plaqueete $p$. An excited state $\ket{\psi}$ has a pure fluxon at plaquette $p$, if $B_p\ket{\psi}=0$. To identify fluxons in the presence of nontrivial charges on the tail inside $p$, we extend $\B_p^s$ by
\begin{widetext}
	\begin{equation}
	\label{eq:Bps::Extended}
	\begin{aligned}
	\Bt_p^s\bket{\bmm\NewPlaquette\emm}
	=&\sum_{j'_1j'_2j'_3j'_4j'_5j'_6k'_1k'_3k'_4k'_6}
	\rmv_{j_1}\rmv_{j_2}\rmv_{j_3}\rmv_{j_4}\rmv_{j_5}\rmv_{j_6}\rmv_{k_1}\rmv_{k_3}\rmv_{k_4}\rmv_{k_6}
	\rmv_{j'_1}\rmv_{j'_2}\rmv_{j'_3}\rmv_{j'_4}\rmv_{j'_5}\rmv_{j'_6}\times\\
	&\rmv_{k'_1}\rmv_{k'_3}\rmv_{k'_4}\rmv_{k'_6}G^{l_1k^*_1j_1}_{s^*j'_1k^{\prime*}_1}
	G^{l_2j^*_3k_2}_{s^*k'_2j^{\prime*}_3}
	G^{k_3j^*_4j_3}_{s^*j'_3j^{\prime*}_4}
	G^{l_4k^*_4j_4}_{s^*j'_4k^{\prime*}_4}
	G^{l_5j^*_6k_5}_{s^*k'_5j^{\prime*}_6}
	G^{k_6j^*_1j_6}_{s^*j'_6j^{\prime*}_1}\times\\
	&\left(R^{j'_2}_{q^*k'_1}G^{q_1j^*_2k_1}_{s^*k'_1j^{\prime*}_2}\overline{R^{j_2}_{q^*_1k_1}}\right)
	\left(R^{k'_2}_{q^*_2j'_2}G^{q_2k^*_2j_2}_{s^*j'_2k^{\prime*}_2}\overline{R^{k_2}_{q_2^*j_2}}\right)
	G^{q_4j^*_5k_4}_{s^*k'_4j^{\prime*}_5}
	G^{q_5k^*_5j_5}_{s^*j'_5k^{\prime*}_5}
	\bket{\NewPlaquettePrime}.
	\end{aligned}
	\end{equation}
\end{widetext}
The operator $\Bt_p^s$ is a straightforward extension of the Levin-Wen operator $B_p^s$ in Eq.\ \eqref{eq:Bps::InLW}. It can be obtained as follows. We first apply the basis transformations $\nu$ on all tails that point into the plaquette $p$, i.e., the tails labeled by $q_1$ and $q_2$ in the above example. The resulting graph contains the plaquette with four tails pointing outwards:
\begin{equation}
\label{eq:PlaquetteEffective}
\NewPlaquetteEffective.
\end{equation}
Now the Levin-Wen operator $B_p^s$ is well-defined on this new plaquette, which is treated as having 10 vertices on the boundary with 10 external lines labeled by $l_1$, $l_2$, $\dots$, $l_6$, as well as $q_1$, $q_2$, $q_4$, and $q_5$. After applying $B_p^s$, we move back the two tails $q_1$ and $q_2$ towards right by the inverse transformations $\nu^{-1}$. Loosely speaking, $\Bt_p^s=\nu_1^{-1}\nu_2^{-1}B_p^s\nu_1\nu_2$, where $\nu_{1,2}$ is the half twist on the tail $q_{1,2}$. This results in Eq.\ \eqref{eq:Bps::Extended}. 

The formula in Eq.\ \eqref{eq:Bps::Extended} can be read as follows. The effective plaquette in \eqref{eq:PlaquetteEffective} is treated as having 10 boundary vertices. According to the definition of $B_p^s$ in Eq.\ \eqref{eq:Bps::InLW}, we arrive at 20 of $v$'s and 10 of $6j$-symbols in the Eq.\ \eqref{eq:Bps::InGaugeModel}. The four copies of the half twist $\nu$ gives rise to four of $R$ tensors (or their complex conjugates $\overline{R}$), in the two brackets, for two tails $q_1$ and $q_2$ respectively.

All local operators $\Bt_p^s$ and $\Q_v^q$ are mutually commuting with any other. According to the interpretation in Eq.\ \eqref{eq:PlaquetteEffective}, this is a direct consequence of the property that $Q_v$ and $B_p$ are mutually commuting projection operators. 
The operator $\B_p^s$ can be recovered from  
$\Bt_p^s$ 
by $\B_p^s=\Bt_p^s\left(\prod_{v\text{ around }p}\Q_v\right)$.

In the following we use $\Bt_p^s$ to identify fluxons at $p$. The ocal operators $\Bt_p^s$ forms the \textit{fusion algebra}
\begin{equation}
\label{eq:BpsAlgebra::GaugeModel}
\Bt_p^r\Bt_p^s=\sum_{t}\delta_{rst^*}\Bt_p^t.
\end{equation}
with multiplication obeying the fusion rule. 

In the following we derive a set of orthonormal projection operators in the fusion algebra to identify particle species of the fluxons.
The braided model equipped with $R$ matrix has $\delta_{ijk}=\delta_{jik}$. The algebra \eqref{eq:BpsAlgebra::GaugeModel} is now abelian, and it is uniquely determines a $N{\times}N$ matrix $X^A_j$, called the \textit{fusion characters}, satisfying
\begin{align}
\label{eq:XmatrixDefiningEquation1}
&X^A_{j^*}=\overline{X^A_j}\\
\label{eq:XmatrixDefiningEquation2}
&X^A_iX^A_j=\sum_{k}\delta_{ijk^*}X^A_kX^A_0\\
\label{eq:XmatrixDefiningEquation3}
&\sum_{j}X^A_j\,\overline{X^B_j}=\delta_{A,B},
\sum_{A}X^A_i\,\overline{X^A_j}=\delta_{i,j}.
\end{align}
The matrix $X^A_j$ is unique up to the relabeling of $A=0,1,\dots,N-1$. $X^A_j$ can be viewed as normalized one-dimensional irreducible representations of the fusion algebra, as observed in Eq \eqref{eq:XmatrixDefiningEquation2}. The factor $X^A_0$ on the RHS of Eq \eqref{eq:XmatrixDefiningEquation2} normalizes $X^A_j$ to satisfy Eq.\  \eqref{eq:XmatrixDefiningEquation3}.

The matrix $X^A_j$ determines a set of orthonormal projections operators $n^A_p$ at $p$:
\begin{align}
\label{eq:nA:definition}
n^A_p:=\sum_{s}\overline{X^A_sX^A_0}\Bt_p^s,
\end{align}
satisfying
\begin{equation}
\label{eq:nA:OrthonormalConditions}
n^A_pn^B_p=\delta_{A,B}n^A_p,\quad\sum_{A}n^A=\mathds{1}.
\end{equation}

These projection operators identify the particle species $A$ of the fluxons at $p$. Each $n^A$ projects onto the states with $A$-type fluxon at $p$. There is a special fluxon type, denoted by $A=0$, corresponding to quantum dimensions by $X^0_j=d_j/\sqrt{D}$. For $A=0$, $n^0_p=\Bt_p$ projects onto states without any nontirvial fluxon at $p$, and thus we say $A=0$ is the trivial type. Each $A$ comes with a conjugate $A^*$ such that $X^{A^*}_j=\overline{X^A_j}$, and we say $A^*$-type fluxon is the antiparticle of $A$-type fluxon.

\subsubsection{Twist}

In additional to the charge $q$ and the fluxon $A$, there is another good quantum number, the twist, which arises from exchange between $q$ and $A$. For example, in the $\Z_2$ gauge theory (toric code model), there are four types of elementary quasiparticles: the trivial one $\unit$, the $\Z_2$ charge $e$, the $\Z_2$ flux $m$, and the charge-flux composite $em$. The twist of $em$ is $-1$ because the wave function acquires the Arharonov-Bohm phase $-1$ by exchanging $e$ and $m$ twice(or equivalently, by winding $e$ around $m$ once), which renders $em$ the fermionic statistics.

\begin{figure}[ht]
\centering
\vspace{0.5cm}
\subfigure[]{\includegraphics[]{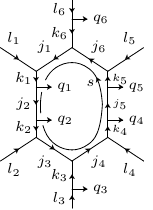}\label{subfig:WilsonLoopPlaquette}}
\hspace{1mm}
\subfigure[]{\includegraphics[]{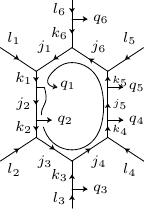}\label{subfig:TwistFluxonPlaquette}}
\caption{(a) $\Bt_p^s$ creates a Wilson loop labeled by $s$; (b) the twist $\Theta_{vp}$ of the charge $q_1$ moves $q_1$ around the plaquette.}
\label{fig:TwistDyon}
\end{figure}
 
We define twist as follows. Take the unit cell of a vertex $v$ and the plaquette $p$, and consider a dyon that carries charge $q$ and fluxon $A$. The idea of the twist is to wind the charge $q$ around the fluxon $A$ at plaquette.
 
Start with the plaquette as in Fig.\ref{subfig:TailedPlaquette}, and consider the dyon living at the unit cell of the vertex of $q_1$ and the plaquette. Suppose it carries the charge $q_1$ and the fluxon $A$. We shall construct the twist operator to move the charge $q_1$ around the plaquette. This operation is similar to $\Bt_p^s$ which creates a Wilson loop labeled by $s$, see Fig.\ref{subfig:WilsonLoopPlaquette}. However, the twist differ by moving $q_1$ along an open path labeled by $q_1$, with the open end treated as the new tail, see Fig.\ref{subfig:TwistFluxonPlaquette}. 

Define the twist by
\begin{widetext}
\begin{equation}
\label{eq:hpq::Holonomy}
\begin{aligned}
&\Theta_{vp}\bket{\bmm\NewPlaquette\emm}
=\sum_{j'_1j'_3j'_4j'_5j'_6k'_1k'_3k'_4k'_6}
\rmv_{j_1}\rmv_{j_2}\rmv_{j_3}\rmv_{j_4}\rmv_{j_5}\rmv_{j_6}\rmv_{k_3}\rmv_{k_4}\rmv_{k_6}
\rmv_{j'_1}\rmv_{j'_3}\rmv_{j'_4}\rmv_{j'_5}\rmv_{j'_6}\rmv_{k'_1}\rmv_{k'_3}\rmv_{k'_4}\rmv_{k'_6}\times
\\
&G^{l_2j^*_3k_2}_{q_1^*k'_2j^{\prime*}_3}
G^{k_3j^*_4j_3}_{q_1^*j'_3j^{\prime*}_4}
G^{l_4k^*_4j_4}_{q_1^*j'_4k^{\prime*}_4}
G^{l_5j^*_6k_5}_{q_1^*k'_5j^{\prime*}_6}
G^{k_6j^*_1j_6}_{q_1^*j'_6j^{\prime*}_1}
G^{q_1j^*_2k_1}_{s^*k'_1j^{\prime*}_2}
\left(R^{k'_2}_{q^*_2k_1}G^{q_2k^*_2j_2}_{q^*_1k_1k^{\prime*}_2}\overline{R^{k_2}_{q_2^*j_2}}\right)
G^{q_4j^*_5k_4}_{q_1^*k'_4j^{\prime*}_5}
G^{q_5k^*_5j_5}_{q_1^*j'_5k^{\prime*}_5}
\bket{\NewPlaquettePrimePrime}
\end{aligned}.
\end{equation}
\end{widetext}
When the fluxon is trivial at plaquette $p$, $\Theta_{vp}$ becomes the special twist $\Theta_v$ of the pure charge at $v$ in Eq.\ \eqref{eq:Theta::TwistCharge}. Therefore $\Theta_{vp}$ takes the dyon around the entire unit cell $vp$.


It turns out that twist $\Theta_{vp}$ is commutes with both $\Q_v^q$ and $\B_p^s$ and yields a good quantum number. A dyon in an excitation $\ket{\psi_J}$ is characterized by the three good quantum numbers, the charge $q$, the fluxon $A$, and the twist $\theta_J$, if
\begin{equation}
\begin{aligned}
&\Q_v^q\ket{\psi_J}=\ket{\psi_J},\\
&n_p^A\ket{\psi_J}=\ket{\psi_J},\\
&\Theta_{vp}\ket{\psi_J}=\theta_J\ket{\psi_J}.
\end{aligned}
\label{eq:q:A:theta::DyonThreeQunaumtNumbers}
\end{equation}

Two dyons that carry the same charge $q$ and the same fluxon $A$ may have different twists, because the twist measures more information than the Wilson loop, see Fig.\ref{subfig:WilsonLoopPlaquette} and Fig.\ref{subfig:TwistFluxonPlaquette}. The Wilson loop $B_p^s$ acting of the dyons is completely determined by the test charge $s$ and the fluxon $A$ of the dyon. It creates a pair of charge $s$ and $s^*$, winds $s$ around the fluxon $A$ and then annihilate $s$ and $s^*$. The entire process gives rise to the Arhoranov-Bohm phase $X^A_s/X^A_0$ according to Eq.\ \eqref{eq:nA:definition}. The twist, on the other hand, exchanges the charge $q$ and the fluxon $A$ twice, and measures more information about the states that is characterized by the twist $\theta_J$.

Although our discussion is restricted at a particular plaquette in Fig.\ref{subfig:TailedPlaquette}, the definition of the twist $\Theta_{vp}$ is valid on any unit cell of a vertex $v$ and a plaquette $p$ (with $v$ on the boundary of $p$). 

\subsection{Dyon-pair state}\label{sec:DyonPairState}

In this subsection, we study the lowest excitation on sphere, the dyon-pair states. We shall study the three good quantum numbers qualitatively as well as quantitatively.

\begin{figure}[!t]
\centering
\vspace{0.5cm}
\subfigure[]{\includegraphics[]{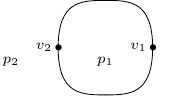}\label{subfig:BareCircle}}
\hspace{1cm}
\subfigure[]{\includegraphics[]{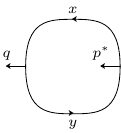}\label{subfig:TailedCircle}}
\caption{(a) A circle with two two-valent vertices on the sphere, with two vertices $v_1,v_2$ and two plaquettes $p_1,p_2$. (b) Circle with two tails.}
\label{fig:CircleOnSphere}
\end{figure}

Start with a circle on sphere with two two-valent vertices, see Fig.\ref{subfig:BareCircle}. The Hilbert space is spanned by the for degrees of freedom on the circle with two tails, denoted by $p^*$, $q$, $x$, and $y$, see Fig.\ref{subfig:TailedCircle}. We use $p^*$ for the future convenience.

We divide the space into two unit cells $v_1p_1$ and $v_2p_2$. The excitations are classified by the two dyons living at these two unit cells. The two dyons are always paired. 

We shall explore the following properties of these dyon-pair states in the section:

\begin{itemize}
\item[1.]  No single (nontrivial) dyon exists on a sphere.
\item[2.]  If there is no (nontrivial) fluxon, the charges at $v_1$ and $v_2$ are opposite to each other.
\item[3.]  Two fluxons at $p_1$ and $p_2$ are opposite to each other.
\item[4.]  Two dyons have the same twist.
\end{itemize}

The charge projections at the two vertices $v_1,v_2$ are
\begin{equation}
\label{eq:Q::Circle}
\Q_{v_1}^{q_1}=\delta_{p^*,q_1},
\quad
\Q_{v_2}^{q_2}=\delta_{q,q_2}.
\end{equation}
The fluxon projections at two plaquettes $p_1,p_2$ are
\begin{equation}
n_{p_{1,2}}^A=\sum_{s}\overline{X^A_sX^A_0}\;\Bt_{p_{1,2}}^s,
\end{equation}
with $\Bt_{p_{1,2}}^s$ being
\begin{equation}
\begin{aligned}
&\Bt_{p_1}^s\bket{\EqTailedCircle}\\
=&\sum_{x'y'}\rmv_{x}\rmv_{y}\rmv_{x'}\rmv_{y'}
G^{q^*y^*x}_{s^*x'y^{\prime*}}\times\\
\quad&\left(R^{x'}_{py'}G^{px^*y}_{s^*y'x^{\prime*}}\overline{R^{x}_{py}}\right)
\bket{\EqTailedCirclePrime},
\end{aligned}
\label{eq:Bps1::Circle}
\end{equation}
and
\begin{equation}
\begin{aligned}
&\Bt_{p_2}^s\bket{\EqTailedCircle}\\
=&\sum_{x'y'}\rmv_{x}\rmv_{y}\rmv_{x'}\rmv_{y'}
G^{pyx^*}_{s^*x^{\prime*}y'}\times\\
\quad&\left(R^{x^{\prime*}}_{q^*y^{\prime*}}G^{q^*xy^*}_{s^*y^{\prime*}x'}\overline{R^{x^*}_{q^*y^*}}\right)
\bket{\EqTailedCirclePrime}.
\end{aligned}
\label{eq:Bps2::Circle}
\end{equation}

Let us fix the fluxon $A$ at $p_1$, and consider the dyon-pair states with charges $p^*$ at $v_1$ and $q$ at $v_2$ fixed. Such dyon-pair states are  $n_{p_1}^A\Q_{v_1}^{p^*}\Q_{v_2}^q=1$ eigenstates. The total number of distinguished dyon pair states is then
\begin{equation}
\begin{aligned}
N^A_{p,q}
=&\mathrm{tr}(n_{p_1}^A\Q_{v_1}^p\Q_{v_2}^q)\\
=&\sum_{xys}\rmd_{x}\rmd_{y}\overline{X^A_sX^A_0}
G^{q^*y^*x}_{s^*xy^*}
G^{px^*y}_{s^*yx^*},
\end{aligned}
\label{eq:TotalDyonNumber}
\end{equation}
where in the second equality we used property that $R^x_{py}\overline{R^x_{py}}=\delta_{p^*xy^*}$.

In non-abelian models, i.e., with $\left|\rmd_j\right|>1$ for some string type $j$, the charges $p^*$ and $q$ may not match exactly as $p=q$. However, there is some selection rule to pair $p^*$ and $q$, depending on the fluxon $A$. 

We prove the properties as follows. When the fluxon at $p_1$ is trivial: $A=0$, the dyon excitations are $n_{p_1}^{A=0}=1$ eigenstates. If we fix the charge $p$ at $v_1$ and $q$ at $v_2$, from Eq.\  \eqref{eq:TotalDyonNumber}, the total number of possible states is
\begin{equation}
\begin{aligned}
\mathrm{tr}(n_{p_1}^{A=0}\Q_{v_1}^p\Q_{v_2}^q)
=&\frac{1}{D}\sum_{xys}\rmd_{x}\rmd_{y}\rmd_s
G^{q^*y^*x}_{s^*xy^*}
G^{px^*y}_{s^*yx^*}\\
=&\delta_{p,q},
\end{aligned}
\end{equation}
where in the second equality we used Eqs.\ \eqref{eq:6jcond} and \eqref{eq:dimcond}. If the charge $p^*$ at $v_1$ is fixed, there exists one (and only one) dyon excitations with charge charge at $v_2$ being $p$. This proves the property 2. In the particular case when $p=0$, then the only allowed state is the ground state, and hence property 1 is proved.

Next we prove property 3. By definition of $R$ matrix, the expression in the bracket in Eq.\ \eqref{eq:Bps1::Circle} can be expressed as
\begin{equation}
\left(R^{x'}_{py'}G^{px^*y}_{s^*y'x^{\prime*}}\overline{R^{x}_{py}}\right)
=G^{pyx^*}_{sx^{\prime*}y'}\overline{R^{y^{\prime*}}_{y^*s^*}}R^{x'}_{xs}.
\end{equation}
We also rewrite the expression in the bracket in Eq.\ \eqref{eq:Bps2::Circle} similarly, followed by the substitution $R^{x^{\prime*}}_{x^*s}=R^{x'}_{xs^*}$ and $R^{y'}_{ys^*}=R^{y^{\prime*}}_{y^*s}$. After substituting the formula in the brackets, we find
\begin{equation}
\Bt_{p_1}^s=\Bt_{p_2}^{s^*},
\end{equation}
which implies
\begin{equation}
n_{p_1}^A=n_{p_2}^{A^*}.
\end{equation}
If the dyon at $v_1p_1$ has fluxon $A$, then the fluxon of the dyon at $v_2p_2$ must be the its anti-fluxon $A^*$.

It can be also proved that
\begin{equation}
\Theta_{v_1p_1}=\Theta_{v_2p_2}.
\end{equation}
with
\begin{equation}
\begin{aligned}
&\Theta_{v_{1,2}p_{1,2}}\bket{\EqTailedCircle}\\
=&\sum_{y'}\rmv_{x}\rmv_{y'}
G^{q^*y^*x}_{pyy^{\prime*}}
\bket{\EqTailedCircleTwistPrime}.
\end{aligned}
\label{eq:Twist1=Twist2::Circle}
\end{equation}
The two dyons carry the same twists.

Elementary excitations are dyon-pair states. In each pair the two dyons have the same twists, opposite fluxons. The charges of two dyons may not match exactly, but satisfy some constraint that depends on the fluxon.

An elementary excitation is given by an simultaneous eigenvector $\psi$ of $\Q^{p^*}_1$, $\Q^q_2$, $n^A_{1,2}$ and $\Theta_{1,2}$ with eigenvalues being the quantum numbers $p^*,q,A,\theta$. They are given by the half-braiding tensors
\begin{equation}
\label{eq:DyonTensor}
\psi\bpm\EqTailedCirclePrimeC\epm=\rmv_x\rmv_y \overline{z^J_{p^*xq^*y}}\delta_{pp'}\delta_{qq'},
\end{equation}
for some quantum double label $J$, where $p^*$ and $q$ are fixed charges that two dyons carry.

Each quantum double label $J$ is parameterized by fluxon type $A$ and the twist $\theta$.

In a special case, the ground state is
\begin{equation}
\label{eq:GroundStateCircle}
\ket{\Phi}=\sum_{x}\frac{\rmd_x}{\sqrt{D}}\bket{\GroundStateTailedCircle}.
\end{equation}

\subsection{Creation, annihilation, and string operators}

The properties of dyon-pair states analyzed in the previous subsection hold on a generic graph.
To see this, here we shall 
consider the creation and annihilation operators of dyons and string operators, in the setting of an $R$-matrix. This will enable us to generate all elementary excitations from a ground state.

\subsubsection{Charge string operators}

We first study the pure charge case.  Recall the ground states are the simultaneous $\Q_v=1$ and $\B_p=1$ eigenstates. In the $\Q=1$ subspace, all tails are labeled by the trivial string type $q=0$. As above, we draw the dotted line to present the trivial label $0$ for convenience. In the following we  
give an explicit formula for creation operator that in terms of the $R$ matrix.

In the $\Q=1$ subspace  the creation operator that creates a pair of charges at the two ending vertices of an edge $e$ by
\begin{widetext}
\begin{equation}
\label{eq:CreateChargePair::Q=1space}
\begin{aligned}
W^q_e\bket{\GroundStatePlaquette}
=\;&\sum_{j'_2}\frac{\rmv_{j'_2}}{\rmv_{j_2}}
\bket{\ChargePairPlaquette}.
\end{aligned}
\end{equation}
\end{widetext}
Here $e$ denote the left boundary edge of the plaquette. The resulting state is an pair of charges, $q$ at the top vertex, and $q^*$ at the bottom vertex. 

The operator $W^q_e$ is normalized as follows:
\begin{equation}
\bra{\psi}{W^q_e}^{\dagger}W^q_e\ket{\psi}=\rmd_q \left\langle\psi|\psi\right\rangle.
\end{equation}

If the two tails on the edge $e$ are not on the same side, $W^q_e$ is defined up to a basis transformation $\mu$ or $\nu$ acting on Eq.\ \eqref{eq:CreateChargePair::Q=1space}. For example, 
\begin{widetext}
\begin{equation}
\label{eq:CreateCrossedChargePair}
\begin{aligned}
W^q_e\bket{\CrossedGroundStatePlaquette}
=\sum_{j'_2}\frac{\rmv_{j'_2}}{\rmv_{j_2}}\overline{R^{j_2}_{q^*j^{\prime*}_2}}
\bket{\CrossedChargePairPlaquette}.
\end{aligned}
\end{equation}
\end{widetext}


In cases where $R$-matrix exists, each charge $q_0$ itself forms a dyon species with trivial fluxon type. The corresponding half braiding tensor is
\begin{equation}
z^{J=q_0}_{pkqt}=\delta_{p,q_0}\delta_{q,q_0}R^{t}_{q_0k}.
\end{equation}
Eq.\ \eqref{eq:CreateCrossedChargePair} is a special case of Eq.\ \eqref{eq:CreateCrossedDyonPair}.
In general, without a $R$ matrix, a charge does not form a dyon species.

If there are nontrivial charge already present inside the plaquette, $W^q_e$ is defined by
\begin{widetext}
\begin{equation}
\label{eq:CreateChargePairGeneral}
W^q_e\bket{\NewPlaquette}
=\sum_{j'_2q'_1q'_2}\rmv_{j_2}
\rmv_{j'_2}\rmv_{q'_1}\rmv_{q'_2}G^{q^*_1k_1j^*_2}_{j^{\prime*}_2q^*q'_1}G^{qj'_2j^*_2}_{k^*_2q^*_2q'_2}
\bket{\ChargePairNewPlaquette}
\end{equation}
\end{widetext}
which creates two charges: $q$ at the upper left vertex and $q^*$ at the lower left vertex. The operator $W^q_e$ in Eq.\ \eqref{eq:CreateChargePairGeneral} is unitary. One sees that when $q_1=0$ and $q_2=0$, Eq.\ \eqref{eq:CreateChargePair::Q=1space} is recovered.

The operator $W^q_e$ can be used to recover some of the previously defined operators.  In particular, the charge fusion operator define in Eq.\ \eqref{eq:fusionCharge} is equal to $\Q_2\sum_q \rmd_{q}W^{q}_e$.  Also, the special case when $q_1=q_2^*=q$ in Eq.\ \eqref{eq:CreateChargePairGeneral} then  $\Q_1\Q_2\sum_{q'} \rmd_{q'}W^{q'}_e$ is equal to charge annihilation operator define in Eq.\ \eqref{eq:QQW::AnnihilateChargeI}.

The above shows that the hopping operator defined in Sec.\ \ref{subsec:Fusion} can be express in terms of the $R$ matrix.   Also, the operators in this subsection allow us to define the string operators completely in terms of $R$ matrix. 

\subsubsection{Fluxon string operator}

In this subsection, we study the string operators for pure fluxons that carry no charge. We restrict to the $\Q=1$ subspace.

We define the creation operator $W_e^A$ on an edge $e$ by
\begin{equation}
\label{eq:CreationOperator}
W^A_e \bket{\bmm\,\TwoPlaquettes\emm}
:=\frac{\overline{X^A_0}\,\overline{X^A_{j_e}}}{X^0_0X^0_{j_e}}\bket{\bmm\,\TwoPlaquettes\emm}.
\end{equation}
It is diagonal in the matrix form. Only two plaquettes are shown, assuming the rest  of the graph is unaffected. The definition holds for arbitrary shaped plaquettes.

The operator $W^A_e$ generate a fluxon-pair state from any ground state $\ket{\Phi}$, with fluxon $A^*$ on $p_1$ and $A$ on $p_2$,
where $p_1$ is plaquette left to the edge $e$ and $p_2$ right to $e$:
\begin{align}
&n^B_{p_1}W^A_e\ket{\Phi}=\delta_{A^*,B}W^A_e\ket{\Phi}\nonumber\\
&n^B_{p_2}W^A_e\ket{\Phi}=\delta_{A,B}W^A_e\ket{\Phi}\nonumber\\
&n^B_{p'}W^A_e\ket{\Phi}=\delta_{B,0}W^A_e\ket{\Phi}.
\end{align}
These properties can be proved using the conditions  \eqref{eq:6jcond} on $6j$-symbols.

\begin{figure}[t]
\centering
\subfigure[$W^A_e\ket{\Phi}$]{\includegraphics[]{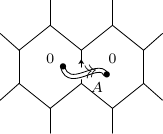}}
\quad
\subfigure[$W^{A^*}_{e^{-1}}$]{\includegraphics[]{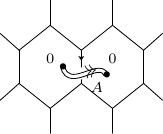}}
\caption{Fluxon-pair state $W^A_e\ket{\Phi}$ generated from a ground state $\ket{\Phi}$. The creation operator does not depend on the edge direction. The fluxon-pair state $W^A_e\ket{\Phi}$ in (a) is the same as $W^{A^*}_{e^{-1}}\ket{\Phi}$ in (b).}
\label{fig:WPhi}
\end{figure}

The definition of $W^A_e$ does not depend on the direction of the edge $e$. In fact, if we reverse the direction of $e$, $j_e$ in Eq.\  \eqref{eq:CreationOperator} is replaced by $j_e^*$. $X^A_{j_e^*}=X^{A^*}_{j_e}$ implies $W^A_e=W^{A^*}_{e^{-1}}$, where $e$ and $e^{-1}$ are the same edge with opposite direction. Both $W^A_e$ and $W^{A^*}_{e^{-1}}$ create the same fluxon pairs across the edge, see Fig.\ref{fig:WPhi}.

From Eq.\ \eqref{eq:CreationOperator}, $W^0_e$ is the identity operator when $A=0$, as creating a trivial fluxon pair does nothing. The hermitian of $W^A_e$ creates a conjugate pair of fluxons because $X^{A^*}_j=\overline{X^A_j}$:
\begin{equation}
  W_e^{A^*}={W_e^A}^{\dagger}.
\end{equation}

In general (non-abelian case, i.e., with $|\rmd_j|>1$ for some $j$ in the input data), even a pure fluxon carries charges. The operator $W^A_e$ is a special case of a generic fluxon creation operator $W^{J=A;00}_e$ with quantum double label $J=A$ and with trivial charges at both ends.

In the following we show how to annihilate and hop fluxons in the absence of charge at the plaquette. 

\begin{figure}[t]
	\centering
	\subfigure[]{\includegraphics[]{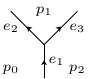}}
	\qquad\qquad
	\subfigure[]{\includegraphics[]{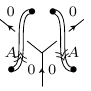}}
	\qquad\qquad
	\subfigure[]{\includegraphics[]{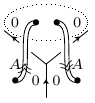}}
	\quad
	\subfigure[]{\includegraphics[]{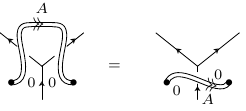}}
	\caption{(a) Three neighboring plaquettes around a trivalent vertex. (b) Create two fluxon pairs across the edge $e_2$ and $e_3$. (c) Annihilate fluxons at $p_1$ by $n^0_{p_1}$. (d) The final fluxon-pair state in (c) is equal to that obtained by directly creating a fluxon pair across edge $e_1$. This implies $n^0_{p_1}W^A_{e_2}$ is path independent, and thus is a hopping operator of fluxon $A$ at $p_1$. }
	\label{fig:nWW}
\end{figure}

Let us start with a ground state $\ket{\Phi}$, and consider a trivalent vertex and its three neighboring plaquettes $p_0$, $p_1$ and $p_2$, see Fig.\ref{fig:nWW}(a). In the following, we suppress $W^A_{e_i}$ by $W^A_i$ for $i=1,2,3$.

In Fig.\ref{fig:nWW}(b), $W^A_2$ creates a $A^*$--$A$ fluxon pair at $p_0$ and $p_1$, while $W^A_3$ creates a $A^*$--$A$ fluxon pair at $p_1$ and $p_2$. Now $p_1$ is occupied by two fluxons, $A$ from $W^A_2$, and $A^*$ from $W^{A}_3$. The resulting state may no longer be an eigenstate of certain $n^B_{p_1}$, because $A$ and $A^*$ may couple to more than one types of fluxons. The operator $W^A_3W^A_2\ket{\Phi}$ can be decomposed by the orthonormal projections $n^B_{p_1}$. The operator $n^B_{p_1}$ projects onto the state $n^B_{p_1}W^A_3W^A_2\ket{\Phi}$ with only $B$-fluxon at $p_1$. 

Particularly, $n^0_{p_1}$ kills any nontrivial fluxon at $p_1$. In the above example, $n^0_{p_1}$ projects onto a fluxon-pair state, with $A^*$ at $p_0$ and $A$ at $p_2$. In this killing process, $n^0_{p_1}$ plays the role of annihilation operator. The annihilation can occur only if the two fluxons at $p_1$ are antiparticles of each other.

The above process is also a hopping process, in which the hopping operator $n^0_{p_1}W^A_3$ moves the fluxon $A$ from $p_1$ to $p_2$. In this process, a $A$-fluxon is created at $p_2$ while a $A$-fluxon is annihilated at $p_1$.

The hopping operator must satisfy some topological property: hopping along two homotopic paths (without any nontrivial quasiparticle enclosed by the two paths) leads to the same final state. Consider again the above example. We apply the hopping operator $n^0_{p_1}W^A_3$ to the fluxon pair state $W^A_2\ket{\Phi}$, and obtain a fluxon pair state. The \textit{path independence} requires 
\begin{equation}
\label{eq:PathIndependenceHopping}
n^0_{p_1}W^A_3W^A_2\ket{\Phi}=W^A_1\ket{\Phi},
\end{equation}
around any trivalent vertex. This property can be verified by using the conditions \eqref{eq:6jcond} on $6j$-symbols.

The hopping operators induce a string operator that creates a pair of fluxons far apart. We choose a path along plaquettes $p_1, p_2,\dots$, and $p_{n+1}$, going across edges $e_1,e_2,...$, and $e_{n}$, as illustrated below:
\newcommand{\VertEdge}[1]
{\quad
	\bmm\xy
	(0,0)*{};(0,10)*{} **\dir{-} ?(0.6)*\dir{>}+(2,-5) *{\scriptstyle #1}
	\endxy\emm\quad
}
\begin{equation}
p_1\VertEdge{e_1}p_2\VertEdge{e_2}\dots\VertEdge{e_{n}}p_{n+1}.
\end{equation}
This is a string consisting of plaquettes. First we create a fluxon pair on the neighboring plaquettes across $e_1$, with $A^*$-fluxon at $p_1$ and $A$-fluxon at $p_2$. Then we move the $A$-fluxon to $p_{n}$ by a sequence of hopping operators, and the final state is
\begin{equation}
\label{eq:StringOperator}
n^0_{p_{n}}W^A_{e_{n}}\dots n^0_{p_2}W^A_{e_2}W^A_{e_1}\ket{\Phi}.
\end{equation}
The two fluxons are at the starting plaquette $p_1$ and the ending plaquette $p_{n+1}$ of the string. The string operator in Eq.\  \eqref{eq:StringOperator} only depends on the two ends of the string because of the path independence of the hopping operator.


\section{Examples}

There are many examples of input data for the models considered in this paper.  Including examples related to the representations of finite groups, the group algebra of finite groups, and the representations of $q$-deformed universal enveloping algebra of Lie algebras.  
 In this section, we discuss some typical examples of these three classes.

\subsection{From finite group representations}
\label{subsec:FiniteGroupRep}

Given a finite group $H$, Levin-Wen models admit two different types of input data: from representations 
of $H$, with labels identified as irreducible representations; and from the group itself, with labels identified as group elements. We call the former the $Rep_H$ model and the latter the $Vec_H$ model.

In this subsection we consider several examples of models arising from representations of a finite group $H$.  
To this end we now discuss a few general features in this context.

The models are based on a 
tensor description of the representation category $Rep_{H}$ of $H$. String types $j$ are (representatives of) irreducible representations $(\rho_j,V_j)$.  Quantum dimensions $\rmd_j=\alpha_j \mathrm{dim}(V_j)$ are equal to the dimension of the representation space, multiplied by the Frobenius-Schur indicator $\alpha_j$, which is 1 if the representation $j$ is real or complex, and $-1$ if pseudo-real.

The fluxons are classified by the conjugacy classes. Since the number of conjugacy classes is equal to the number of irreducible representations, the number of fluxons is equal to the number of charges, as expected from the analysis in previous section.

Let  $\{C^A\}_A$ be the set of conjugacy classes of $H$ indexed by labels $A$. The fusion characters $X^A_j$ are just the usual characters $\chi_j(A)$ for $H$ (up to normalization factors):
\begin{equation}
X^A_j=\sqrt{\frac{|C^A|}{|H|}}\chi_j(A)\alpha_j,
\end{equation}
where $|H|$ is the order of $H$, and $|C^A|$ is the cardinality of $C^A$. 
Note that $X^0_0X^0_j=\alpha_j\mathrm{dim}(V_j)=\rmd_j$. 
The orthogonality relations \eqref{eq:XmatrixDefiningEquation3} for $X^A_j$ result from those for character functions.

The quantum double labels are classified by pairs $(A,\mu)$, 
where $A$ labels a conjugacy class of $H$, and $\mu$ is an irreducible representation of the centralizer $Z_A=\{g\in H | gh_A=h_Ag\}$. Here $h_A$ is a arbitrary representative element in $C^A$ but fixed once and for all.

\subsubsection{Abelian group}

Consider an abelian group $H$. All irreducible representations are $1$-dimensional, and hence $\rmd_j=1$. The $6j$-symbol is given by
\begin{equation}
G^{{i}{j}{m}}_{{k}{l}{n}}=
\delta_{{i}{j}{m}}
\delta_{{k}{l}\dual{{m}}}
\delta_{{j}{k}\dual{{n}}}
\delta_{{i}{n}{l}}.
\end{equation}

Each group element is itself a conjugacy class, so the quantum double labels are pairs $(g,\mu)$ of group elements and irreducible representations of $H$. Each dyon is a charge-fluxon composite. 

For example, let $H=\Z_N$, the quantum double charges are $(g,\mu)$ for $g,\mu=0,1,\dots,N-1$ and the $z$ tensors are
\begin{align}
\label{eq:ZNztensor}
z^{(g,\mu)}_{pjqt}=\delta_{p,\mu}\delta_{q,\mu}\exp(2\pi i g/N)\delta_{pjt^*}\delta_{jqt^*},
\end{align}
where $\delta_{pjt^*}=1$ if $p+j-t=0$ mod $N$ and 0 otherwise.

\subsubsection{$Rep_{S_3}$ model.} 

Consider the model arising from representations of $S_3$. The string types are the three irreducible representations of the symmetry group $S_3$, denoted by $L=\{0,1,2\}$. All labels are self-dual, i.e., $j^*=j$. The fusion rules are given by $\delta_{000}=\delta_{011}=\delta_{022}=\delta_{122}=\delta_{222}=1$.

The quantum dimension $\rmd_j$ is the dimension of the representation space $V_j$: $d_0=d_1=1$ and $d_2=2$. The independent nonzero symmetrized $6j$-symbols are
\begin{align}
G_{000}^{000}=1,G_{111}^{000}=1,
G_{222}^{000}=\frac{1}{\sqrt{2}},G_{011}^{011}=1,
G_{222}^{011}=\frac{1}{\sqrt{2}},
\nonumber\\
G_{022}^{022}=\frac{1}{2},
G_{122}^{022}=\frac{1}{2},G_{222}^{022}=\frac{1}{2},
G_{122}^{122}=\frac{1}{2},G_{222}^{122}=-\frac{1}{2}.
\end{align}
All other nonzero $6j$-symbols are obtained through the tetrahedral symmetry in Eq.\ \eqref{eq:6jcond}.

The nontrivial $R$ matrix is $R^1_{22}=-1$.

There are three conjugacy classes, 
labeled by $A=0,1,2$, with $|C^A|=1,2,3$ respectively. The fluxons are classified by the three conjugacy classes, with the character table presented in Table \ref{tab:CharacterTableS3}.

\begin{table}[ht]
	\caption{Character table of $H=S_3$.}
	\label{tab:CharacterTableS3}
	\begin{center}
		\begin{tabular}{c|ccc}
			\hline
			$\chi_j(A)$     & $C^{A=0}$ & $C^{A=1}$ & $C^{A=2}$ \\
			\hline
			$\chi_{j=0}$ & 1 & 1 & 1  \\
			$\chi_{j=1}$ & 1 & 1 & $-1$  \\
			$\chi_{j=2}$ & 2 & $-1$ & 0  \\
			\hline
		\end{tabular}
	\end{center}
\end{table}

There are 8 quantum double labels. Indeed, the centralizers for the three conjugacy classes are $Z_{A=0}=S_3$, $Z_{A=1}\cong\Z_3$, and $Z_{A=2}\cong\Z_2$. 
In total 
there are 8 irreducible representations of $Z_A$. We denote 8 quantum double labels by $J=1,2,\dots, 8$.

We present dyon-pairs graphically by a string $\dyonpair{A}{p}{q}$, with fluxon $A$.  All distinguished dyon-pair states and the corresponding twists are enumerated in Table \ref{tab:DyonPairStatesS3}.

The properties developed in previous section can be verified, e.g., the total number of dyon-pair states for fixed $A,p,q$ obey the counting formula in Eq.\ \eqref{eq:TotalDyonNumber}.

\begin{widetext}
\begin{table}[ht]
	\begin{center}
		\begin{tabular}{c|cc|cc|cc}
			\hline
			$A=0$     &  $\dyonpair{A=0}{0}{0}$ & $\theta_1=1$ & $\dyonpair{A=0}{1}{1}$ & $\theta_2=1$ & $\dyonpair{A=0}{2}{2}$ & $\theta_3=1$\\
			\hline
			$A=1$ & $\begin{array}{cc}\dyonpair{A=1}{0}{0}&\dyonpair{A=1}{0}{1}\\ \dyonpair{A=1}{1}{0}&\dyonpair{A=1}{1}{1}\\\end{array}$ & $\theta_4=1$ & $\dyonpair{A=1}{2}{2}$ & $\theta_5=\exp(\frac{2\pi i}{3})$ & $\quad\dyonpair{A=1}{2}{2}\quad$&   $\theta_6=\exp(-\frac{2\pi i}{3})$\\
			\hline
			$A=2$ & $\begin{array}{cc}\dyonpair{A=2}{0}{0} &\dyonpair{A=2}{0}{2} \\	\dyonpair{A=2}{2}{0} &\dyonpair{A=2}{2}{2}\\\end{array}$& $\theta_7=1$ & $\begin{array}{cc}\dyonpair{A=2}{1}{1} &
			\dyonpair{A=2}{1}{2} \\
			\dyonpair{A=2}{2}{1} &
			\dyonpair{A=2}{2}{2} \\\end{array}$ & $\theta_8=-1$&  &\\
			\hline
		\end{tabular}
	\end{center}
	\caption{17 dyon-pair states in $Rep_{S_3}$ model.}
	\label{tab:DyonPairStatesS3}
\end{table}

The explicit wavefunction for each dyon-pair is specified by the half-braiding tensors $z$:

\begin{itemize}
	\item[] $z^{1}_{pjqt}=\delta_{p,0}\delta_{q,0}\delta_{j,t}$,
	\item[]  $z^{2}_{pjqt}=\delta_{p,1}\delta_{q,1}\left(
	\begin{array}{ccc}
	0 & 1 & 0 \\
	1 & 0 & 0 \\
	0 & 0 & -1 \\
	\end{array}
	\right)_{jt}$,
	\item[]  $z^{3}_{pjqt}=\delta_{p,2}\delta_{q,2}\left(
	\begin{array}{ccc}
	0 & 0 & 1 \\
	0 & 0 & -1 \\
	1 & -1 & 1 \\
	\end{array}
	\right)_{jt}$,
	\item[]  $z^{4}_{pjqt}=\delta_{p,0}\delta_{q,0}\left(
	\begin{array}{ccc}
	1 & 0 & 0 \\
	0 & 1 & 0 \\
	0 & 0 & -\frac{1}{2} \\
	\end{array}
	\right)_{jt}+\delta_{p,1}\delta_{q,1}\left(
	\begin{array}{ccc}
	0 & 1 & 0 \\
	1 & 0 & 0 \\
	0 & 0 & \frac{1}{2} \\
	\end{array}
	\right)_{jt}-\frac{\sqrt{3}}{2}i\delta_{p,0}\delta_{q,1}\delta_{j,3}\delta_{t,3}+\frac{\sqrt{3}}{2}i\delta_{p,1}\delta_{q,0}\delta_{j,3}\delta_{t,3}$,
	\item[]  $z^{5}_{pjqt}=\delta_{p,2}\delta_{q,2}\left(
	\begin{array}{ccc}
	0 & 0 & 1 \\
	0 & 0 & -1 \\
	e^{-\frac{2 \ii \pi }{3}} & e^{\frac{\ii \pi }{3}} & e^{\frac{2 \ii \pi }{3}}
	\\
	\end{array}
	\right)_{jt}$,
	\item[]  $z^{6}_{pjqt}=\delta_{p,2}\delta_{q,2}\left(
	\begin{array}{ccc}
	0 & 0 & 1 \\
	0 & 0 & -1 \\
	e^{\frac{2 \ii \pi }{3}} & e^{-\frac{\ii \pi }{3}} & e^{-\frac{2 \ii \pi }{3}}
	\\
	\end{array}
	\right)_{jt}$,
	\item[]  $z^{7}_{pjqt}=\delta_{p,0}\delta_{q,0}\left(
	\begin{array}{ccc}
	1 & 0 & 0 \\
	0 & -1 & 0 \\
	0 & 0 & 0 \\
	\end{array}
	\right)_{jt}+\delta_{p,2}\delta_{q,2}\left(
	\begin{array}{ccc}
	0 & 0 & 1 \\
	0 & 0 & 1 \\
	1 & 1 & 0 \\
	\end{array}
	\right)_{jt}+\delta_{p,0}\delta_{q,2}\delta_{j,3}\delta_{t,3}+\delta_{p,2}\delta_{q,0}\delta_{j,3}\delta_{t,3}$,
	\item[]  $z^{8}_{pjqt}=\delta_{p,1}\delta_{q,1}\left(
	\begin{array}{ccc}
	0 & 1 & 0 \\
	-1 & 0 & 0 \\
	0 & 0 & 0 \\
	\end{array}
	\right)_{jt}+\delta_{p,2}\delta_{q,2}\left(
	\begin{array}{ccc}
	0 & 0 & 1 \\
	0 & 0 & 1 \\
	-1 & -1 & 0 \\
	\end{array}
	\right)_{jt}+i\delta_{p,1}\delta_{q,2}\delta_{j,3}\delta_{t,3}+i\delta_{p,2}\delta_{q,1}\delta_{j,3}\delta_{t,3}$.
\end{itemize}
\end{widetext}

\subsection{Kitaev's quantum double model}

Here we consider 
the Levin-Wen model arising from a finite group $H$ itself. Set the string types to be group elements: $I=\{h\}_{h\in H}$, with $h^*=h^{-1}$. Set 
 $\rmd_h=1$, for all $h\in H$ 
and $\delta_{ghk}=1$ if $ghk=1$ and 0 otherwise. Set
\begin{equation}
G^{{i}{j}{m}}_{{k}{l}{n}}=
\delta_{{i}{j}{m}}
\delta_{{k}{l}\dual{{m}}}
\delta_{{j}{k}\dual{{n}}}
\delta_{{i}{n}{l}}.
\end{equation}
Hence $\rmv_h=1$.

The model is identified with Kitaev's quantum double model on the dual triangulation graph. The local operators form a quantum double algebra $D(H)$ of $H$.

Let $b^q_k=B_{k^{-1}qk,k^{-1},k^{-1}q}$ for $q,k\in H$. The tube algebra has the multiplication rule:
\begin{equation}
b^p_r b^q_s=\frac{1}{\sqrt{D}}\delta_{p,rbr^{-1}}b^p_{rs},
\end{equation}
which recovers $D(H)$.

The dyons in elementary excitations are determined by solutions of Eq.\ \eqref{eq:PiPi=Pi}. Fix $q$ at the tail, set $\pi^q_k=\Pi_{q,k^{-1},k^{-1}q}$ for $k\in Z_q=\{t\in H|tq=qt\}$. The equation $\pi^q_k=\sum_{m\in Z_q}\pi^q_m\pi^q_{m^{-1}k}$ has the solution $\pi^q_k=\frac{\mathrm{dim}(\alpha)}{|H|}\chi^q_{\alpha}(k)$ given by the character of all irreducible representations $\alpha$ of $Z_q$.

The dyon species are identified by a pair $(A,\alpha)$ 
where $A$ is a label of a conjugacy class $C^A$ of $H$ and $\alpha\in \mathrm{Irrep}(Z_q)$ 
for a representative element $q$ in $C^A$. 
The modular matrices are
\begin{align}
&S_{(A,\alpha),(B,\beta)}=\frac{1}{|H|}\sum_{\substack{g\in A,h\in B\\ gh=hg}}\overline{\chi^g_{\alpha}(h)\chi^h_{\beta}(g)},\\
&T_{(A,\alpha),(B,\beta)}=\delta_{AB}\delta_{\alpha\beta}\frac{\chi^g_{\alpha}(g)}{\mathrm{dim}_{\alpha}}, \text{ for any } g\in A.
\end{align}

The above procedure also applies to the twisted quantum double case. For finite group $H$ and a 3-cocycle $\omega$ in $H^3(H,U(1))$, setting the $6$j-symbol to be $\omega$ identifies the corresponding LW model with the twisted quantum double model\cite{HWW}. (The tetrahedral symmetry of $6j$-symbol may be violated, which could be fixed by introducing an ordering of triangulation.) The tube algebra becomes the twisted quantum double algebra $D^{\omega}(H)$. We will not discuss the details in this paper.

Both $Rep_H$ model and the quantum double model have excitations classified by the same quantum double labels $(A,\alpha)$. This reveals an electric-magnetic (EM) duality: the former support quasiparticles of charges at vertices and fluxons at plaquettes while the latter support charges at plaquettes and fluxons at vertices. We will discuss EM duality in section \ref{sec:EMduality} in more detail.

\subsection{From modular category}
\label{subsec:ExampleModularCategory}

For a braided model with input data $\{\rmd,\delta,G,R\}$, 
as in Sec.\ \ref{sec:BraidedModels},
 defines the $S$ matrix 
\begin{equation}
S_{ab}=\sum_c \rmd_c R^c_{ab}R^c_{ba}.
\end{equation}
If $S$ matrix is invertible, the input data is a tensor description of a unitary modular category. The quantum double classification  
is quite simple in this case.

The quantum double labels are pairs denoted by $i\overline{j}$, with quantum dimension
\begin{equation}
\rmd_{i\overline{j}}=\rmd_i \rmd_j.
\end{equation}
The fluxon type of $i\overline{j}$ is $j$. 
In particular, 
pure fluxons are $j\overline{j}$.

The half-braiding tensors are
\begin{equation}
z^{i\overline{j}}_{pjqt}
=\sum_{ab}\rmd_a\rmd_b R^a_{ik}R^b_{jk}G^{a^*ik}_{bj^*t}G^{ijq^*}_{t^*ka^*}G^{ibt^*}_{k^*p^*j^*}.
\end{equation}

The $S$ matrix for the quantum double labels are
\begin{equation}
S_{i\overline{j},k\overline{l}}=S_{ik}\overline{S_{jl}},
\end{equation}
and the twist is
\begin{equation}
\theta_{i\overline{j}}=R^0_{jj^*}/R^0_{ii^*}.
\end{equation}

Modular categories can be derived from representations of the quantum group $\mathcal{U}_q\left(\mathfrak{su}(2)\right)$ (called the quantum universal enveloping algebra of $\mathfrak{su}(2)$). When $q$ is taken to be a primitive root of unity, $U_q(\mathfrak{su}(2))$ has 
finitely many irreducible representations with nonzero quantum dimensions, 
which lead to 
symmetric $6j$-symbols. An efficient way to construct this data is through the Jone-Wentzl projectors in Temperley-Lieb algebra (see ref \cite{Wang} for example). Examples include semion, Fibonacci, and Ising data,
which we discuss now.

\subsubsection{Doubled semion model}

Semion data can be obtained at the $q$-deformation parameter $q=\exp(\pi \ii/3)$.  String types are $L=\{0,1\}$ (sometimes denoted by $\{\mathbf{1},s\}$), with quantum dimensions $\rmd_0=1$ and $\rmd_1=-1$. It has the same fusion rule $\delta_{110}=1$ as that of the group $\mathbb{Z}_2$ representation theory.

The nonzero symmetric $6j$-symbols are
\begin{equation}
G_{000}^{000}=1,
G_{111}^{000}=\ii,
G_{011}^{011}=-1.
\end{equation}
The other nonzero $6j$-symbols are obtained through the tetrahedral symmetry.

The nontrivial $R$ matrix is 
 $R^0_{11}=\ii$.

There are four quantum double labels: $0\overline{0},0\overline{1},1\overline{0},1\overline{1}$, called boson, semion, anti-semion, and doubled semion. The $S$ matrix is
\begin{equation}
S=\left(
\begin{array}{cccc}
1 & -1 & -1 & 1 \\
-1 & -1 & 1 & 1 \\
-1 & 1 & -1 & 1 \\
1 & 1 & 1 & 1 \\
\end{array}
\right).
\end{equation}

The twists are 
$1,\ii,-\ii,1$.

\subsubsection{Double Fibonacci model}

Fibonacci data can be obtained at the $q$-deformation parameter $q=-\exp(\pi \ii/5)$.

The string types are $L=\{0,2\}$, 
sometimes denoted by $\{\mathbf{1},\tau\}$. Let $\phi=\frac{1+\sqrt{5}}{2}$ be the golden ratio. The quantum dimensions of $0,2$ are $\rmd_0=1$ and $\rmd_2=\phi$.

The fusion rules are 
\begin{equation}
\label{Fibbranchingrule}
\delta_{000}=\delta_{022}=\delta_{222}=1,\delta_{002}=0
\end{equation}
and the nonzero $6j$-symbols $G$ are given by
\begin{align}
\label{Fib6js}
G^{000}_{000}=1,
G^{022}_{022}=G^{022}_{222}=1/\phi,
\nonumber\\
G^{000}_{222}=1/\sqrt{\phi},
G^{222}_{222}=-1/{{\phi}^2}.
\end{align}

The other nonzero symmetrized $6j$-symbols are obtained through the tetrahedral symmetry. The nontrivial $R$ matrices are $R^0_{22}=\exp(-4\pi i/5)$ and $R^2_{22}=\exp(3\pi i/5)$.

The four quantum double labels are  $0\overline{0},0\overline{2},2\overline{0},2\overline{2}$. The $S$ matrix is
\begin{equation}
S=\left(
\begin{array}{cccc}
1 & \phi  & \phi  & \phi ^2 \\
\phi  & -1 & \phi ^2 & -\phi  \\
\phi  & \phi ^2 & -1 & -\phi  \\
\phi ^2 & -\phi  & -\phi  & 1 \\
\end{array}
\right).
\end{equation}
The twists  
for the labels $0\overline{0}, 0\overline{2}, 2\overline{0}$ and $2\overline{2}$
are $1, \exp(4\pi i/5), \exp(-4\pi i/5)$ and $1$, 
respectively.

\subsubsection{Doubled Ising model}

Ising data can be obtained at the $q$-deformation parameter $q=\exp(3\pi \ii/4)$.

The string types are $L=\{0,1,2\}$, 
sometimes denoted by $\{\mathbf{1},\sigma,\psi\}$. The quantum dimensions  are $\rmd_0=1,\rmd_1=\sqrt{2}$ and $\rmd_2=1$.

The fusion rules are 
\begin{equation}
\label{eq:Isingbranchingrule}
\delta_{000} = 1,
\delta_{011} = 1, 
\delta_{022} = 1,
\delta_{112} = 1, 
\end{equation}
and the nonzero $6j$-symbols $G$ are given by
\begin{align}
\label{eq:IsingG}
\begin{aligned}
&G_{000}^{000}=1,G_{111}^{000}=\frac{1}{\sqrt[4]{2}},
	G_{222}^{000}=1,G_{011}^{011}=\frac{1}{\sqrt{2}},\\
&G_{122}^{011}=\frac{1}{\sqrt[4]{2}},G_{211}^{011}=\frac{1}{\sqrt{2}},G_{022}
^{022}=1,G_{112}^{112}=-\frac{1}{\sqrt{2}}.
\end{aligned}
\end{align}

The other nonzero symmetrized $6j$-symbols are obtained through the tetrahedral symmetry. 
The nontrivial $R$ matrices are 
\begin{equation}
R^{0}_{22}=-1,
 R^{1}_{21}=-\ii,R^0_{11}=\exp(-\pi \ii/8),R^2_{11}=\exp(3\pi \ii/8).
\end{equation}

There are nine quantum double labels: $0\overline{0}$, $0\overline{1}$, $0\overline{2}$, $1\overline{0}$, $1\overline{1}$, $1\overline{2}$, $2\overline{0}$, $2\overline{1}$, and $2\overline{2}$. The $S$ matrix is
\begin{equation}
\left(
\begin{array}{ccccccccc}
1 & \sqrt{2} & 1 & \sqrt{2} & 2 & \sqrt{2} & 1 & \sqrt{2} & 1 \\
\sqrt{2} & 0 & -\sqrt{2} & 2 & 0 & -2 & \sqrt{2} & 0 & -\sqrt{2} \\
1 & -\sqrt{2} & 1 & \sqrt{2} & -2 & \sqrt{2} & 1 & -\sqrt{2} & 1 \\
\sqrt{2} & 2 & \sqrt{2} & 0 & 0 & 0 & -\sqrt{2} & -2 & -\sqrt{2} \\
2 & 0 & -2 & 0 & 0 & 0 & -2 & 0 & 2 \\
\sqrt{2} & -2 & \sqrt{2} & 0 & 0 & 0 & -\sqrt{2} & 2 & -\sqrt{2} \\
1 & \sqrt{2} & 1 & -\sqrt{2} & -2 & -\sqrt{2} & 1 & \sqrt{2} & 1 \\
\sqrt{2} & 0 & -\sqrt{2} & -2 & 0 & 2 & \sqrt{2} & 0 & -\sqrt{2} \\
1 & -\sqrt{2} & 1 & -\sqrt{2} & 2 & -\sqrt{2} & 1 & -\sqrt{2} & 1 \\
\end{array}
\right).
\end{equation}
The twist is determined by $R$ matrix.

\section{Electric-magnetic duality in topological theory with finite gauge groups}
\label{sec:EMduality}

Here we give a consequence of our results in the context of 
 electric-magnetic duality in topological gauge theory with a finite (gauge) group $H$, which has been proposed  and 
 studied in \cite{Aquado,Kong}.   

For an arbitrary finite group $H$, there is a procedure\cite{HW15} to derive its unitary symmetric 
$6j$-symbols equipped with an $R$ matrix, which can be used as the LW input data to construct the $Rep_H$ 
model, see Sec.\ \ref{subsec:FiniteGroupRep}. On the other hand, we can define a $Vec_H$ model on the same 
trivalent graph, denoted by $\Gamma$. This $Vec_H$ can be identified with Kitaev's quantum double model 
with $H$ defined on a triangulation -- the dual graph of $\Gamma$. As discussed in the previous section, in 
the $Rep_H$ model, fluxons at plaquettes of $\Gamma$ are labeled by conjugacy classes $A$ of $H$, and 
charges at vertices (of $\Gamma$) by irreducible representations of $H$. In the $Vec_H$ model, charges at 
triangular plaquettes of the triangulation are labeled by conjugacy classes $A$, while fluxons at vertices 
(of the triangulation) by irreducible representations of $H$. 
 In terms of local operators, $\Bt_p$ (and $\Q_v$) in the $Rep_H$ model can be identified with $\Q_v$ (and  $\B_p$, rep.) in the $Vec_H$ model.

This gives rise to an electric-magnetic transformation (EMT) between these two models 
\cite{Aquado}. The electric-magnetic duality asserts that the two models connected by the EMT are actually 
equivalent to each other \cite{Kong}. 

Since the existence of a transformation in general does not imply the existence of a corresponding symmetry 
or invariance, the validity of EMD is much stronger than the existence of EMT. Well-known examples in 
quantum field theory include spontaneous symmetry breaking and non-abelian gauge anomaly in quantized 
chiral gauge theory \cite{ZWZ}. 
Namely, one needs to check that the EMD in topological gauge theory is not 
violated by symmetry breaking or global excitations. Even if sometimes the arguments for the EMD is 
intuitively simple, the concrete checks for exact duality may be highly nontrivial. Here we provide two 
concrete checks for the EMD between the $Rep_H$ model and the $Vec_H$ model.                    

\begin{figure}
\centering
\subfigure[]{\includegraphics[]{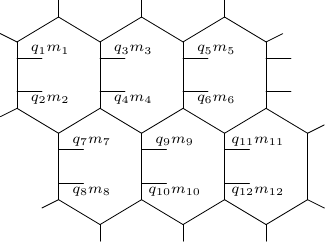}}
\quad
\subfigure[]{\includegraphics[]{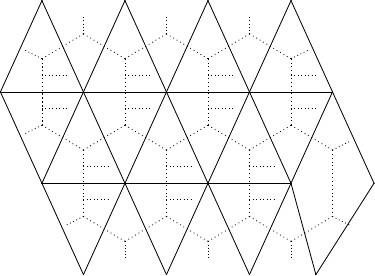}}
\caption{(a). The Hilbert space for $Rep_H$ model on trivalent graph $\Gamma$, with each tail labeled by $q_v, m_v$. (b) The Hilbert space for $Vec_H$ model on the triangulation (solid line) dual to $\Gamma$, with local Hilbert space $\ds{C}[H]$ on each edge.}
\label{fig:EMdualityGraph}
\end{figure}

Our first check is to verify that the Hilbert space of the two models connected by the EMT has the same 
dimension. Certainly this is a necessary condition for the two models to be equivalent to each other.  
We slightly extend the $Rep_H$ model by enriching its Hilbert space again at each tail. To each vertex $v$ 
we associate a tail labeled by $q_v$ and a matrix index $m_v$ of representation $q_v$, which take values 
$1,2,\dots,\mathrm{dim}_{q_v}$. (Recall $\mathrm{dim}(q_v)=\alpha_{q_v}\rmd_{q_v}$.)
 Define the Hamiltonian by
\begin{equation}
H=-\sum_{v}\Q_v-\sum_p\Bt_p,
\end{equation}
where $\Bt_p$ is defined in Eq.\ \eqref{eq:Bps::Extended}. The operators $\Q_v$ and $\Bt$ will not affect 
$m_v$. We still call this slightly extended model a $Rep_H$ model. The Hilbert space is illustrated in 
Fig.\ref{fig:EMdualityGraph}(a).

The $Vec_H$ model is identified with Kitaev's quantum double model with $H$. The Hilbert space is spanned by group elements in $H$ at all edges in the triangulation. See Fig.\ref{fig:EMdualityGraph}(b).

The Hilbert space of two models has the same dimension. To see this, we look at local Hilbert space $\mathcal{H}_v$ at each vertex. It has basis $\{|i,j,k,l,q,m\rangle\}$, labeling the following diagram
\[
\VertexTail,
\]
where $i,j,k$ are labels on three incoming edges, and $l,q,m$ the enriched charge degrees of freedom. Note that there is exactly one tail to each vertex. The dimension of $\mathcal{H}_v$ is
\begin{equation}
\mathrm{dim}(\mathcal{H}_v)=\sum_{l,q\in L}\delta_{jkl}\delta_{il^*q^*}\rmd_q=\rmd_i\rmd_j\rmd_k.
\end{equation}
Therefore, effectively the local Hilbert space at each edge $e$ labeled by $j_e$ has dimension $\rmd_{j_e}^2$. By a theorem for the order (number of elements, or dimension of the group algebra) of a finite group, the local Hilbert space at each edge has dimension $\sum_j\rmd_j^2=|H|$. This gives the same dimension of the Hilbert space of the $Vec_H$ model on the triangulation. In fact, there is a duality transformation between the two models
\begin{equation}
\label{eq:EMduality}
\begin{aligned}
&\mathcal{H}^{Rep_H}
\\
=&\bigoplus_{\{j\}}\bigotimes_v\mathcal{H}_v
\\
=&\bigoplus_{\{j\}}\bigotimes_v\left[Hom\left(\bigotimes_{\substack{e:\text{into }v\\e':\text{out of }v}}V_{j_e}\otimes V_{j_{e'}},V_{q_v}\right)\otimes V_{q_v}\right]
\\
\rightarrow&\bigoplus_{\{j\}}\bigotimes_v\left(\bigotimes_{\substack{e:\text{into }v\\e':\text{out of }v}}V_{j_e}\otimes V_{j_{e'}}\right)
\\
=&\bigoplus_{\{j\}}\bigotimes_{e}\left(V_{j_e}\otimes V_{j_e}^*\right)
\\
=&\bigotimes_{e}\bigoplus_{j_e}\left(V_{j_e}\otimes V_{j_e}^*\right)
\\
\rightarrow&\bigotimes_e \ds{C}[H]=\scH^{Vec_H}.
\end{aligned}
\end{equation}
with the summation $\{j\}$ over labels on all edges on graph $\Gamma$ (or on the triangulation), and $V_j$ 
for representation space of $j$. The map on the 4th line is an isomorphism composing two Wigner's 
$3j$-symbols or the Clebsch-Gordan coefficients that decomposes the tensor product of two representations. 
The map on the last line is a generalized Fourier transformation (i.e. the Peter-Weyl theorem) between two 
bases of $\ds{C}[H]$, from $\ket{\rho^j_{\alpha\beta}}_{\alpha\beta=1,\dots,\rmd_j}$ to $\ket{h}_{h\in H}$.

Our second check is about the spectrum of the models. The equality of the ground state degeneracy in the 
$Rep_H$ and $Vec_H$ models has been verified in, e.g., \cite{GSD}. With our quantum double classification 
of quasiparticle excitations in the $Rep_H$ model, we are able to check the equivalence of the two models 
at the level of quasiparticle excitation species. Previously it is known that in the $Vec_H$ model 
(Kitaev's quantum double model) it is the quantum double of the finite group H that classifies the 
elementary excitations \cite{Kitaev}. In this paper we have shown that the $Rep_H$ model (or the LW model 
with input data from finite group H) accommodates dyon excitations classified by the quantum double of $H$. 
So the two models have the same excitation spectrum in their quantum numbers and energy levels. This is 
certainly a highly nontrivial check for the electric-magnetic duality between the two models.

\section{Relation to topological quantum field theory}

Levin-Wen model is viewed as a Hamiltonian approach to Turaev-Viro topological quantum field theory (TQFT). The topological observables in the former are related to topological invariant of 3-manifolds.

\subsection{GSD and Turaev-Viro TQFT}

We first consider zero temperature case. We denote the input data $\{\rmd,\delta,G\}$ by a unitary fusion category $\scC$ that derives them. The zero temperature partition function of Levin-Wen models on surface $\Sigma$ equals the ground state degeneracy (GSD). The GSD is is related to Turaev-Viro invariant -- a topological invariant for 3-manifold defined below -- by

\begin{equation}
\mathrm{GSD}_{\scC}(\Sigma)=\tau^{\mathrm{TV}}_{\scC}(\Sigma\times S^1).
\label{eq:GSD=tauTV}
\end{equation}

We first define $\tau^{\mathrm{TV}}_{\scC}$ and then sketch the proof.

Given $\mathcal{\scC}$ and a compact oriented 3-manifold $M$, we construct the number $\tau^{\mathrm{TV}}_{\scC}(M)$ as follows. Any 3-manifold $M$ has a triangulation, i.e., can be discretized into tetrahedral. We choose an arbitrary one, and the desired number will be triangulation independent.

\begin{enumerate}
	\item Assign labels to all edges.
	\item Assign $6j$-symbols to all tetrahedral as follows. Due to tetrahedral symmetry in Eq.\ \eqref{eq:6jcond}, such $6j$-symbols do not change under rotation of tetrahedral.
	\begin{equation}
	\bmm \tetrahedron{i}{j}{m}{k}{l}{n} \emm \quad\Rightarrow\quad G^{ijm}_{kln}.
	\end{equation}
	\item Assign quantum dimensions $\rmd_j$ to all edges labeled by $j$.
	\item Assign $1/D$ to each vertex (in the triangulation).
	\item Multiply all the quantities in step 2 - 4, and take the product over tetrahedral, edges, and vertices of these numbers.
	\item Sum over all labels.
\end{enumerate}

We get
\begin{equation}\label{E:DefTVinvariant}
\tau^{\mathrm{TV}}_{\scC}(M)=\sum_{\text{labels}}\prod_{\text{vertices}}\frac{1}{D}\prod_{\text{edges}}\rmd\prod_{\text{tetrehedra}}(6j\text{-symbols}).
\end{equation}

This number does not depend on choices of triangulation.

Now we sketch the proof in Eq.\ \eqref{eq:GSD=tauTV}. The ground states are $\prod_pB_p=1$ eigenvectors. Hence $\text{GSD}=\text{tr}(\prod_{p}B_p)$. To relate the trace to $\tau^{\mathrm{TV}}_{\scC}$, we first write down the dual triangulation of the trivalent graph as follows. 
\begin{equation}
\TriangleYpart{j_4}{j_1}{j_2}{j_3}{j_5}{j_6}{>}{<}{>}{<}{<}{<}
\quad
\Rightarrow
\quad
\TriangulationYpartLarge{j_4}{j_1}{j_2}{j_3}{j_5}{j_6}{>}{<}{>}{<}{<}{<}
\end{equation}

Eq.\ \eqref{eq:Bps::InLW} becomes
\begin{align}
\label{eq:Bps::DualTriangulation}
&\Biggl\langle
\TriangulationYpart{j_4}{j^{\prime}_1}{j^{\prime}_2}{j^{\prime}_3}{j_5}{j_6}{>}{<}{>}{<}{<}{<}
\Biggr|
B_p^s
\Biggl|\TriangulationYpart{j_4}{j_1}{j_2}{j_3}{j_5}{j_6}{>}{<}{>}{<}{<}{<}\Biggr\rangle\nonumber\\
=&
\rmv_{j_1}\rmv_{j_2}\rmv_{j_3}\rmv_{j'_1}\rmv_{j'_2}\rmv_{j'_3}
G^{j_5j^*_1j_3}_{sj'_3j^{\prime*}_1}G^{j_4j^*_2j_1}_{sj'_1j^{\prime*}_2}G^{j_6j^*_3j_2}_{sj'_2j^{\prime*}_3}.
\end{align}
It can presented by three tetrahedral as follows. The top three triangles are those in the bra of above equation and the three bottom ones are in the ket.
\[
\CorbodismYpart{j_4}{j_1}{j_2}{j_3}{j_5}{j_6}{j'_1}{j'_2}{j'_3}{>}{<}{>}{<}{<}{<}
\]

This identification with $B_p^s$ leads to $\text{tr}(\prod_pB_p^s)=\tau^{\mathrm{TV}}_{\scC}(\Sigma\times S^1)$.

\subsection{Excitations and the extended Turaev-Viro invariant}


In this subsection we will explain how the excitation in the above model are related to an extension of the Turaev-Viro invariant to manifolds containing links defined by Turaev and Virelizier\cite{TVir}.

Here we continue using the conventions of the last subsection.  Let $Z(\scC)$ be the quantum double category associated to $\scC$.  As mentioned above a minimal solution to Equation \eqref{eq:zzzNaturalityHalfBraiding} is identified with a quantum double element $J$.   Let $L$ be a link in $\Sigma\times S^1$ whose components are labeled with such quantum double elements.    

In \cite{TVir}, the Turaev-Viro invariant is extended to $Z(\scC)$-colored links and in particular, defines an invariant of the  pair $(\Sigma\times S^1,L)$.  In this work the invariant is defined using skeleton which are 2-polyhedron with certain properties.  Taking the dual of a triangulation gives a skeleton.  A link in a skeleton is a collection of loops immersed in the 2-dimensional simplices of the skeleton with certain transversality conditions.  A quantum double label $J$ and its associated half braiding $z^J$ can be used to define a new symbol similar to a $6j$-symbol, see \cite{TVir}.   The extended T-V invariant is defined in a similar way to the T-V invariant outlined above:  First, a skeleton of $(\Sigma\times S^1,L)$ can be decomposed into building blocks (which are analogous to tetrahedron).  Each face is assigned a label (here the a face of the skeleton is dual to an edge in the triangulation).  Each building block correspond to a $6j$-symbol or a new symbol coming from the quantum double labels of the link.  As in Eq.\ \eqref{E:DefTVinvariant}, the extended invariant is obtained by taking a weighted sum over all possible labelings of the product of these symbols.    

The model given in this paper is a Hamiltonian realization of the extended T-V invariant.  As outlined below, the link $L$ in $\Sigma\times S^1$ can be associated with an operator $f_L$ which is  a composition of certain operators given in Section \ref{sec:ElementaryExcitations}. (Note $f_L$ is not unique.)

\begin{figure}[!t]
	\centering
	\subfigure[]{\includegraphics[]{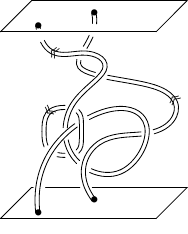}}
	\hspace{3cm}
	\subfigure[]{\includegraphics[]{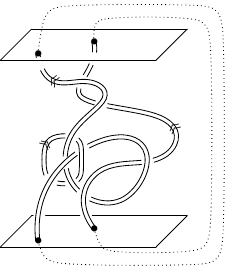}}
	\quad
	\subfigure[]{\includegraphics[]{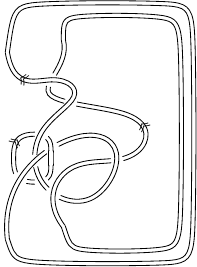}}
	\caption{(a) A tangle representing $f_L$, which is the creation of  a dyon-pair from a ground state. (b) Illustration of gluing dots along $S^1$. (c) Closure of underlying topological object is $L$.}
	\label{fig:trfllll}
\end{figure}

For simplicity, let us first describe the situation when $\Sigma=S^2$.  
  In Fig.\ref{fig:trfllll}(a), the bottom plate presents an initial state with two dots presenting dyons. The top plate presents a final state. The operator $f_L$ is defined by composing  certain  creation and annihilation operators and charge contractions, with the composition order coinciding with the time direction in the figure and determined by the topology of $L$.  Since the string operator is path independent, $f_L$ only depends on the topology of the tangle, i.e. the portion of the link between the two plates. The operator $f_L$ is parameterized by charges at the ends of strings on top and bottom plates.

We require that the closer along $S^1$ of the tangle underlying $f_L$ is the link $L$, see Fig.\ref{fig:trfllll}(c).  When the link is contained in a 3-ball the closer is trivial and the tangle can be chosen to be the link.  For example, when $L$ is the Hopf link $H_{J,K}$ in a 3-ball in $\Sigma\times S^1$ whose components are labeled with $J$ and $K$ then $f_L$ is the composition of the operators described in Figure \ref{fig:StringSmatrix}.  

In general, $\Sigma \neq S^2$ and $L$ may not be in a 3-ball.  The description of $f_L$ in such a situation involves the topology of $\Sigma$ and/or a nontrivial closer of the tangle along $S^1$.  However, since the operators are all local, $f_L$ can be described by a similar process as above. 

The extended Turaev-Viro  invariant $\tau^{\mathrm{TV}}_{\scC}$ associated to $\scC$ is equal to the Reshitkhin-Turaev invariant $\tau^{\mathrm{RT}}_{Z(\scC)}(\Sigma\times S^1,L)$ associated to the quantum double category $Z(\scC)$, see \cite{TVir}.  The R-T invariant is defined by representing the 3-manifold by surgery on some framed link $K$ in $S^3$ then applying certain quantum invariants coming from $Z(\scC)$ to the link  $K\cup L$.  Thus in the definition of the R-T invariant one does not have to work with trivalent graphs but the evaluation of certain link invariants.

 An argument similar to the one in the last section shows that 
\begin{equation}\label{E:RTTVOp}
\text{tr}(f_L)=\tau^{\mathrm{TV}}_{\scC}(\Sigma\times S^1,L)=\tau^{\mathrm{RT}}_{Z(\scC)}(\Sigma\times S^1,L).
\end{equation}
The above trace is taken in the Hilbert space and is the sum over all charges on the open strings.  This trace can be viewed as a charge contraction in the time direction, and connects all open strings in Fig.\ref{fig:trfllll}(a) to a closed link in Fig.\ref{fig:trfllll}(c).  In the case of ground states i.e. when $L$ is trivial Eq.\ \eqref{E:RTTVOp}  is just Eq.\   \eqref{eq:GSD=tauTV}.

In particular, when the link $L$ is trivial, $f_L=\prod_p\B_p$ is the ground state projection operator, and Eq. \eqref{E:RTTVOp} recovers Eq. \eqref{eq:GSD=tauTV} as a special example.

\vspace{3cm} 

\section{Conclusions and Discussions} 

In this paper we have studied how to describe the full spectrum of dyon excitations in the extended 
Levin-Wen models. Previously it was known that in the LW models, fusion of two pure fluxons generally may lead to the appearance of charge d.o.f. To incorporate the latter explicitly, we enlarge the Hilbert space by introducing a tail (labeled by a string type) at one of the edges of each vertex, and modified the LW Hamiltonian accordingly. Though we have to deal with new configurations with an extra tail at each vertex, in this approach we have been able to achieve the following:  

(1) In our extended Hilbert space with enriched d.o.f. for charge at vertices, we are able to study the 
properties of charge and fluxon-type of dyon excitations, and in particular their interplay through the 
twist operation. We have shown that one needs three quantum numbers -- charge, fluxon-type and twist -- to 
describe the dyon species for elementary excitations, or the total dyonic quantum numbers of excitations 
localized in a region ``surrounded by vacuum". We emphasize the necessity of introducing the twist, as the 
third quantum number beyond the charge and fluxon-type, for a complete description of a dyon species.

(2) The above conclusions are obtained by studying the operator algebra formed by local operators and its 
irreducible representations (simple modules). We have shown that 
all local plaquette operators
\textit{preserving 
	topological symmetry}, i.e.\ invariant under 
Pachner moves 
form the so-called Tube algebra. The latter is a generalization of usual $B_p^s$ operators in the LW models; in fact, the operators $B_p^s$ form a subalgebra of the Tube algebra. 

(3) String operators can be realized as linear maps on the extended Hilbert space. Irreducible 
representations (simple modules) of the Tube algebra are shown to be in one-to-one correspondence 
with the half-braiding tensors that are used to define string operators \cite{Mueger,LW14}. 
In this way, we establish that the Tube 
algebra and string operators are dual to each other by a (generalized) Fourier transformation.
On the other hand, half-braiding tensors are ingredients 
to define the quantum double (the center) category of the input unitary fusion category. So we conclude 
that the quantum numbers of dyon excitations are organized by irreducible representations of the Tube 
algebra, or equivalently by the quantum double category, as the center of the input unitary fusion category 
of the LW model. Twist is a property necessarily associated with the quantum double category.

(4) Realizing string operators, 
as linear maps over the extended Hilbert space enables us to obtain not only 
the $S$, $T$ matrices, but also the braid group representations for dyons. This knowledge is important for 
describing emergent braid statistics of dyon excitations \cite{Wu1984} and, therefore, will play a crucial 
role in designing quantum computation codes that exploit manipulation of excitations in the topological 
phases.       

(5) We can systematically construct explicit states/wavefunctions with given quantum numbers. This enables 
one to study more physically interesting quantities, such as entropy, entanglement entropy etc, and to 
design quantum computation algorithm based on manipulation of the non-abelian anyonic 
quasiparticles.                       

(6) A consequence of our results is that the Kitaev quantum double model (the Toric Code model) associated 
with a finite group on a triangulation and the (extended) LW model with input data from the same finite group has the exactly same dyon excitation spectrum, characterized by the same quantum double category.   
This provides a strong check/test/eveidence for the electric-magnetic duality between the two models, not only for ground states but also at the level of the full excitation spectrum. 

As for the physical consequences, one may naturally ask whether our extended string-net models, with the enlarged Hilbert space and modified Hamiltonian,  could give rise to new topological phases? To answer this question we note that when all the tails (labeled by a string type), that we have added at one of the edges of each vertex, are labeled by the trivial type $0$, the states in our extended Hilbert space are restricted to the un-enlarged Levin-Wen Hilbert space, and our modified Hamiltonian reduces to the LW Hamiltonian as well. So the subspace of degenerate ground states in our extended string-net model are the same as that in the LW model. Therefore we assert that {\it at zero temperature}, our extended string-net model does {\it not} give rise to new classes of topological phases beyond the quantum double model or the LW string-net model. On the other hand, with our extension of the string-net models we have been able to achieve a proper and complete treatment of the excited states, resulting in a better understanding of the excitation spectrum, especially of the charged or dyonic excitations, above the string-net ground states. Such a treatment is lack and improbable in the original LW model, because it lacks the labels for charged degrees of freedom at the vertices. Hence our extension could give a different perspective, from the original LW model, about the properties, phases and, possibly, phase transitions of the system at finite temperatures involving charged or dyonic excitations. We would like to come back to address these issues in the future. 

There are some future directions. One is how to develop similar approach to solve discrete 3+1D models\cite{JM,WW,WWH} for topological phases. The observable algebra (of local operators that commute with the Hamiltonian, which is the Tube algebra in 2+1D case) will be expanded due to the extra dimension. Another direction is experimental design of quantum simulations\cite{NMM,LWH} of string-net excitations using the anyon manipulation operators proposed in this paper.

Finally we want to emphasize the following point. One may add more terms into the LW Hamiltonian, which may 
not commute with the existing two terms. When the coupling strengths of these additional terms are sufficiently small, we expect that the model remains in the same topological phase, with the 
energy levels of the many-body states getting shifted, provided there is no level crossing between the 
ground states and excited states. With such more general Hamiltonians, we believe that the quantum double 
category or the pertinent Tube algebra of local operators 
we have obtained for the LW model still provides a ``complete basis" for many-body excitation states in the 
enlarged Hilbert space and, therefore, could still be useful. For example, we may use this ``basis" to 
formulate/compute perturbation theory corrections for elementary excitations. 

\section{Acknowledgments}  

Y. Hu thanks Department of Physics and Astronomy, College of Science, University of Utah for their partial support. The research of N. Geer was partially supported by NSF grants DMS-1007197 and DMS-1308196. The 
work of Y. S. Wu was supported in part by US NSF grant PHY-1068558.

\appendix

\vspace{1cm}

\section{Some properties of quantum double}
\label{app:QuantumDouble}

The quantum double category is characterized by the half-braiding tensors $z$. We list some properties and symmetry conditions on $z$.

Orthonormal relation:
\begin{align}
\label{eq:OrthonormalZ}
\sum_{l}z^J_{ljqt}\,\overline{z^J_{ljpt}}=\delta_{pq}\delta_{jpt^*},\nonumber\\
\sum_{l}z^J_{qjlt}\,\overline{z^J_{pjlt}}=\delta_{pq}\delta_{pjt^*},
\end{align}

$z^J_{pjqt}$ satisfies the symmetry conditions
\begin{align}
\label{eq:SymmetryZ1}
&z^J_{pjqt}=\sum_{r}\rmd_rG^{j^*pr^*}_{jq^*t}\,\overline{z^J_{qj^*pr}}\,\\
\label{eq:SymmetryZ2}
&\overline{z^J_{qj^*pr}}\,=\sum_t\rmd_tG^{jrp^*}_{jt^*q}z^J_{pjqt},
\end{align}
where the second condition is a consequence of the first one together wit the orthogonality relation \eqref{eq:6jcond}.


\begin{thebibliography}{100}
\bibitem{Ocneanu} Ocneanu first introduced the Tube algebra in his lectures, but 
    never published it. Later it was published in \cite{EK95}.
\end{thebibliography}
\end{document}